\documentclass[11pt,a4paper]{article}
\pdfoutput=1
\usepackage{jheppub}

\usepackage[utf8]{inputenc}
\usepackage{amsmath}
\usepackage{amsfonts}
\usepackage{amssymb}
\usepackage{slashed}
\usepackage{graphicx}
\usepackage{dsfont}
\usepackage{subcaption}
\usepackage{xcolor}
\usepackage[colorinlistoftodos,textwidth=45mm,textsize=small]{todonotes}

\graphicspath{{img/}}

\newcommand{\cc}{c}
\newcommand{\ct}{t}
\newcommand{\co}{o}
\newcommand{\cn}{n}
\newcommand{\cz}{z}
\newcommand{\PL}{{\text P}_{\text L}}
\newcommand{\PR}{{\text P}_{\text R}}
\newcommand{\TL}{{\text{L}}}
\newcommand{\TR}{{\text{R}}}
\newcommand{\Acharg}{A _1}
\newcommand{\Adip}{A _2}
\newcommand{\AZ}{A _Z}
\newcommand{\Abox}{A _{\text{box}}}
\newcommand{\fAchargn}{F_0 ^n}
\newcommand{\fAdipan}{F _1 ^n}
\newcommand{\fAdipbn}{F _2 ^n}
\newcommand{\fAchargc}{F _0 ^c}
\newcommand{\fAdipac}{F _1 ^c}
\newcommand{\fAdipbc}{F _2 ^c}
\newcommand{\fAZa}{F _{1} ^{Z}}
\newcommand{\fAZb}{F _{2} ^{Z}}
\newcommand{\fAZbf}{f _{2} ^{Z}}
\newcommand{\fIfour}{I _4}
\newcommand{\fJfour}{J _4}
\newcommand{\ffa}{F _a}
\newcommand{\ffb}{F _b}
\newcommand{\neut}{\chi ^0}
\newcommand{\charg}{\chi ^-}
\newcommand{\slep}{{\tilde{l}}}
\newcommand{\sneu}{{\tilde{\nu}}}
\newcommand{\squark}{{\tilde{q}}}
\newcommand{\hc}{{\text{H.c.}}}
\newcommand{\muu}[1]{\mu_u^{\text{eff,}#1}}
\newcommand{\mud}[1]{\mu_d^{\text{eff,}#1}}
\newcommand{\msusy}{m_{\text{SUSY}}}
\newcommand{\mLOSP}{m_{\text{LOSP}}}
\newcommand{\Rmue}{R(\text{Al})}
\newcommand{\Rmuedipole}{R^{\text{only dip.}}(\text{Al})}

\newcommand{\ALKitano}{\mathbb{A}_L}
\newcommand{\ARKitano}{\mathbb{A}_R}

\title{Low-energy lepton physics in the MRSSM:\\ $(g-2)_\mu$, $\mu\to
  e\gamma$ and $\mu\to e$ conversion}

\author[1]{Wojciech Kotlarski\note{Corresponding author.},}
\author{Dominik St\"ockinger,}
\author{and Hyejung St\"ockinger-Kim}
\affiliation{
Institut f\"ur Kern- und Teilchenphysik, TU Dresden, 01069 Dresden, Germany
}
\emailAdd{wojciech.kotlarski@tu-dresden.de}
\emailAdd{dominik.stoeckinger@tu-dresden.de}
\emailAdd{hyejung.stoeckinger-kim@tu-dresden.de}

\abstract{

Low-energy lepton observables are discussed in the Minimal
R-symmetric Supersymmetric Standard Model.  
We present comprehensive numerical analyses and the analytic one-loop results
for  $(g-2)_\mu$, $\mu \to e \gamma$, and $\mu \to e$ conversion.
The interplay between the three observables is investigated as well as
the parameter regions with large $g-2$.  
A striking difference to the MSSM is the absence of $\tan\beta$
enhancements; however we find smaller enhancements governed by
MRSSM-specific R-Higgsino couplings $\lambda_d$ and $\Lambda_d$.
As a
result we find significant contributions to $g-2$ only in a small
parameter space with several SUSY masses below $200$ GeV, compressed spectra
and large $\lambda_d$, $\Lambda_d$. In this parameter space there is a
correlation between all three considered observables. In the
parameter region with small $(g-2)_\mu$ the SUSY masses can be larger
and the correlation between $\mu\to e\gamma$ and $\mu\to e$ conversion
is weak. Therefore already COMET Phase 1 has a promising sensitivity to the
MRSSM.

}

\begin{document}
\maketitle

\section{Introduction}

Low-energy lepton physics is an area which could lead to
fundamental discoveries in the forthcoming years, and intriguing
anomalies and deviations from Standard Model (SM) predictions have
accumulated in observables related to leptons. In particular, there currently 
is a $3$--$4\sigma$ discrepancy in the muon anomalous magnetic moment
$a_\mu$. 
Future measurements of $a_\mu$ are ongoing at
Fermilab \cite{Grange:2015fou} and planned at J-PARC
\cite{Iinuma:2011zz,Mibe:2011zz}, with the prospect of a 
significant reduction of the uncertainty and the potential to
firmly establish the existence of physics beyond the SM.

In addition, several measurements of charged lepton flavour violating (CLFV) processes in 
$\mu\to e$ transitions are planned. An upgrade of the MEG experiment
\cite{TheMEG:2016wtm} will
increase the sensitivity for 
the $\mu\to e\gamma$ decay by an order of magnitude \cite{Baldini:2013ke,Galli:2018bsc},
and the planned Mu3e experiment~\cite{Blondel:2013ia} 
promises four orders of magnitude improvement on the upper limit for $\mu \to eee$.
Likewise the planned COMET and Mu2e
experiments at J-PARC and Fermilab are expected to improve
the current sensitivity~\cite{Bertl:2006up} to $\mu\to e$ conversion in muonic atoms
by four orders of magnitude~\cite{Cui:2009zz,Kuno:2013mha,Adamov:2018vin,Abrams:2012er}.
The progress of these experiments is accompanied by high-precision calculations
of background processes~\cite{Czarnecki:2011mx,Szafron:2015kja,Pruna:2016spf,Pruna:2017upz}.

In preparation of
the planned experiments it is timely to study the range of
possible predictions for these observables in candidate alternatives
to the SM.
See e.g.\ Ref.~\cite{Lindner:2016bgg} for a recent summary focusing on simple models.

Supersymmetry (SUSY) remains one of the best motivated ideas for
physics beyond the SM. However SUSY might not be realized in its
minimal form, the MSSM. In recent years, the minimal R-symmetric SUSY
standard model (MRSSM) has been put forward as a viable and attractive
alternative \cite{Kribs:2007ac}. It is based on a continuous
unbroken U(1)$_\text{R}$ symmetry under which the superparticles are
charged. It involves Dirac gauginos, $N=2$ SUSY multiplets in the
gauge and Higgs sectors, and supersoft SUSY breaking \cite{Fox:2002bu}. In contrast to
many other non-minimal SUSY models, it has no MSSM limit and thus
constitutes a separate, alternative realization of SUSY.

One of the original motivations was the observation \cite{Kribs:2007ac} that
large flavour violating mixing is viable in the sfermion sector. The
consequences for $\mu\to e\gamma$, $\mu\to e$ conversion and the
$\mu\to eee$ decay have  first been studied in Ref.\ \cite{Fok:2010vk};
further flavour physics observables have also been studied in
Refs.\ \cite{Dudas:2013gga,Sun:2019wii}. The result of Ref.\ \cite{Fok:2010vk} was
that particularly significant effects in $\mu\to e$ conversion can be
possible in the MRSSM.

Here we provide an extensive analysis of the three observables: $a_\mu$,
$\mu\to e\gamma$ and $\mu\to e$ and their correlations in the
MRSSM. This is the first MRSSM study of $a_\mu$  and the first MRSSM
study of lepton flavour violation where the role of the MRSSM-specific
superpotential
parameters $\lambda$, $\Lambda$ is analysed. These parameters were
already very important in phenomenological studies of electroweak
observables in the MRSSM \cite{Bertuzzo:2014bwa,Diessner:2014ksa,Diessner:2015iln}. As we will show, they have a similar
influence as $\tan\beta$ in the ordinary MSSM.

Our study can be compared to similar studies in the MSSM. With respect
to $a_\mu$ it is well known that the MSSM can provide a very natural
explanation of the currently observed deviation, see
Refs.\ \cite{Czarnecki:2001pv,Martin:2001st,DSreview} for reviews, and
very detailed studies have been performed including higher-order
corrections \cite{Fargnoli:2013zda,Fargnoli:2013zia,Athron:2015rva}. The striking property of the MSSM prediction for
$a_\mu$ is the enhancement proportional to $\tan\beta$
\cite{Moroi:1995yh,DSreview}. A similar enhancement is present in the
amplitudes relevant for $\mu\to e\gamma$ and $\mu\to e$ conversion \cite{Hisano:1995cp}. As
a result, the observables are strongly correlated. The correlation
between $a_\mu$ and $\mu\to e\gamma$ has been studied in
Refs.\ \cite{Chacko:2001xd,Isidori:2007jw,Kersten:2014xaa}, the correlation between the
lepton flavour violating observables has been studied in
Refs.\ \cite{Hisano:1995cp,Ilakovac:2012sh} and  more recently, in the light of LHC
data, in Ref.\ \cite{CalibbiShadmi}. 

We will show here that the MRSSM has very different properties: There
is no $\tan\beta$ enhancement for any of the observables; $a_\mu$ can
only be accommodated in a very small parameter space, and there is an
interesting non-correlation between $\mu\to e\gamma$ and $\mu\to e$,
which implies that $\mu\to e$ conversion places important
complementary bounds on the MRSSM flavour structure.

We remark that in parallel to model specific studies there has been
significant recent progress
on model-independent effective field theoretical (EFT) approaches to CLFV:
Loop corrections to $\mu \to e \gamma$ have been evaluated in an EFT with dimension--6 operators~\cite{Crivellin:2013hpa,Pruna:2014asa};
higher-order corrections to $\mu \to e \gamma$, $\mu \to eee$ and $\mu \to e$ from running below the weak scale have been evaluated~\cite{Davidson:2016edt,Davidson:2016utf,Crivellin:2017rmk} and
disentanglement of different Wilson coefficients by experimental observables has been studied~\cite{Davidson:2017nrp,Davidson:2018kud}. 

The paper is structured as follows.
In section \ref{sec:model} we provide relevant properties of the MRSSM
including mass matrices and Feynman rules. Section \ref{sec:observables}
presents the theory of the three observables in general
and the specific analytical results in the MRSSM. In section
\ref{sec:numerics} we analyse all three observables in detail,
exploring their numerical behaviour in all
relevant corners of parameter space and highlighting the
(non-)correlations between the observables. 
The Appendix contains a list of Feynman rules.

\section{Details of the MRSSM}
\label{sec:model}
\subsection{Model definition}

In this section we provide the definition and relevant properties of
the minimal R-symmetric supersymmetric standard model (MRSSM),
originally introduced in Ref.\ \cite{Kribs:2007ac}. The
MRSSM is a supersymmetric (SUSY) extension of the SM with a continuous
unbroken R-symmetry --- a global U(1)$_R$ invariance under which
SM-like fields and their superpartners transform differently.
Our notation and presentation extends the one of
Refs.\ \cite{Diessner:2014ksa,Staub:2013tta,Staub:2012pb,Staub:2008uz}. 
The derivations of the following formulas has been done both using
SARAH \cite{Staub:2013tta,Staub:2012pb,Staub:2008uz} based on a model
file developed for Ref.\ \cite{Diessner:2014ksa} as well as by hand;
the relevant formulas have been implemented in FlexibleSUSY
\cite{Athron:2014yba,Athron:2017fvs} and in a dedicated mathematica
code, allowing cross checks. 

The
R-charge of all SM-like fields is chosen as zero; the R-charge of all
superpartner fields is then fixed by the SUSY algebra. The requirement
of U(1)$_R$ invariance forbids the usual MSSM-like Majorana mass terms
for gauginos and the Higgsino-mass $\mu$-parameter.  In the MRSSM,
gauginos and Higgsinos obtain Dirac-like masses involving new
superfields which have no MSSM counterparts. The gauginos of each
gauge group require an additional chiral superfield in the adjoint
representation: $\hat{\cal O}$ (octino, octet), $\hat{T}$ (triplino,
triplet), $\hat{S}$ (singlino, singlet); the adjoint scalar components
have R-charge 0. The Higgsinos require two new SU(2)$_L$ doublets:
${\hat{R}}_{d,u}$ (R-Higgsinos, R-Higgs fields); the R-Higgs fields
have R-charge $+2$.

The MRSSM superpotential reads
\begin{align}
\nonumber W = & \mu_d\,\hat{R}_d \cdot
\hat{H}_d\,+\mu_u\,\hat{R}_u\cdot\hat{H}_u
\nonumber\\
&{}+\Lambda_d\,\hat{R}_d\cdot \hat{T}\,\hat{H}_d\,+\Lambda_u\,\hat{R}_u\cdot\hat{T}\,\hat{H}_u\,
+\lambda_d\,\hat{S}\,\hat{R}_d\cdot\hat{H}_d\,+\lambda_u\,\hat{S}\,\hat{R}_u\cdot\hat{H}_u
\nonumber\\
&{} - Y_d \,\hat{d}\,\hat{q}\cdot\hat{H}_d\,- Y_e \,\hat{e}\,\hat{l}\cdot\hat{H}_d\, +Y_u\,\hat{u}\,\hat{q}\cdot\hat{H}_u\, ,
\label{eq:superpot}
 \end{align} 
where the dot denotes $\epsilon$ contraction with $\epsilon_{12} = +1$ and where the triplet is defined as
\begin{equation}
\hat{T}=
\begin{pmatrix}
\hat{T}_0/\sqrt{2}&\hat{T}_+\\
\hat{T}_-&-\hat{T}_0/\sqrt{2}\\
\end{pmatrix}
\,.
\end{equation}
The MSSM-like superfields appearing here are the Higgs doublets
$\hat{H}_{u,d}$, the quark and lepton doublets $\hat{q}$, $\hat{l}$
and singlets $\hat{u}$, $\hat{d}$, $\hat{e}$. The terms in the last
line are the usual Yukawa couplings as in the MSSM. In the present
paper the quark Yukawa couplings 
are not relevant, so we neglect the CKM matrix and assume
generation-diagonal Yukawa coupling matrices in the quark and lepton
sectors. We denote quarks and leptons
of the three generations as
\begin{align}
\nu _g &=(\nu _e, \nu _\mu, \nu _\tau),\,
&l _g &= (e, \mu, \tau),\,
&u _g &= (u, c, t),\,
&d _g &= (d, s, b),
\end{align}
with a generation index $g\in\{1,2,3\}$. If no ambiguities
can arise, such as in Eq.\ (\ref{eq:superpot}),
we drop the index $g$. Accordingly, we denote the
diagonal entries of the Yukawa couplings as
$Y_e=\text{diag}(Y_{l_1},Y_{l_2},Y_{l_3})
$, etc.

\begin{table}[th]
\begin{center}
\begin{tabular}{c|l|l||l|l|l|l}
\multicolumn{1}{c}{Field} & \multicolumn{2}{c}{Superfield} &
                              \multicolumn{2}{c}{Boson} &
                              \multicolumn{2}{c}{Fermion} \\
\hline 
 \phantom{\rule{0cm}{5mm}}Gauge Vector    &\, $\hat{g},\hat{W},\hat{B}$        \,& \, $\;\,$ 0 \,
          &\, $g,W,B$                 \,& \, $\;\,$ 0 \,
          &\, $\tilde{g},\tilde{W}\tilde{B}$             \,& \, +1 \,  \\
Matter   &\, $\hat{l}, \hat{e}$                    \,& \,\;+1 \,
          &\, $\tilde{l},\tilde{e}^*_R$                 \,& \, +1 \,
          &\, $l,e^*_R$                                 \,& $\;\;\,$\,\;0 \,    \\
          &\, $\hat{q},{\hat{d}},{\hat{u}}$       \,& \,\;+1 \,
          &\, $\tilde{q},{\tilde{d}}^*_R,{\tilde{u}}^*_R$ \,& \, +1 \,
          &\, $q,d^*_R,u^*_R$                             \,& $\;\;\,$\,\;0 \,    \\
 $H$-Higgs    &\, ${\hat{H}}_{d,u}$   \,& $\;\;\,$\, 0 \,
          &\, $H_{d,u}$               \,& $\;\;\,$\, 0 \,
          &\, ${\tilde{H}}_{d,u}$     \,& \, $-$1 \, \\ \hline
\phantom{\rule{0cm}{5mm}} R-Higgs    &\, ${\hat{R}}_{d,u}$   \,& \, +2 \,
          &\, $R_{d,u}$               \,& \, +2 \,
          &\, ${\tilde{R}}_{d,u}$     \,& \, +1 \, \\
  Adjoint Chiral  &\, $\hat{\cal O},\hat{T},\hat{S}$     \,& \, $\;\,$ 0 \,
          &\, $O,T,S$                \,& \, $\;\,$ 0 \,
          &\, $\tilde{O},\tilde{T},\tilde{S}$          \,& \, $-$1 \,  \\
\end{tabular}
\end{center}
\caption{The R-charges of the superfields and the corresponding bosonic and
             fermionic components. From Ref.\ \cite{Diessner:2014ksa}.
        }
\label{tab:Rcharges}
\end{table}

The $\mu_{u,d}$-terms in the superpotential provide Higgsino masses
which, in contrast to the MSSM, do not involve a transition between
up- and down-sectors. The $\Lambda_{u,d}$- and $\lambda_{u,d}$-terms
are MRSSM-specific new interaction terms. The structure of these terms
resembles the one of the Yukawa couplings. As discussed in
Ref.\ \cite{Diessner:2014ksa} the $\Lambda_{u,d}$- and
$\lambda_{u,d}$-terms are very important for phenomenology and can
influence the $\rho$-parameter and Higgs mass calculations in a
similar way as the top/bottom Yukawa couplings.

The soft SUSY-breaking Lagrangian has been defined in
Ref.\ \cite{Kribs:2007ac}, based on the discussion of supersoft
supersymmetry breaking in Ref.\ \cite{Fox:2002bu}. It contains scalar mass terms for the
MSSM-like squarks and sleptons  with parameters
\begin{align}
  \text{$(m_{\tilde{q}}^{2})_{ij}$,
    $(m_{\tilde{u}}^{2})_{ij}$,
    $(m_{\tilde{d}}^{2})_{ij}$,
    $(m_{\tilde{l}}^{2})_{ij}$,
    $(m_{\tilde{e}}^{2})_{ij}$}.
\end{align}
Here $i,j=1,2,3$ are generation indices. It also
contains scalar mass terms for the Higgs fields and the new scalar
fields which however are not required for the present paper. Finally,
there are non-MSSM-like soft SUSY-breaking terms which give Dirac
masses to the gauginos. These can be generated from
spurions $W'_\alpha=\theta_\alpha D$ from a hidden sector $U(1)'$ that
acquires a D-term  \cite{Fox:2002bu}, in the form
$
\int d^2\theta\frac{\hat{W'_\alpha}}{M} \hat{W}_i^\alpha \hat{\Phi}_i 
$,
where $\hat{W}_i^\alpha$ and $\hat{\Phi}_i$ are the field strength
superfields and the new adjoint chiral superfields for each gauge
group. This construction leads to the Lagrangian  (see also the
discussion in Ref.\ \cite{Diessner:2014ksa})
\begin{align}
  {\cal L}_{\text{soft}}\ni
   \, & -M_B^D (\tilde{B}\,\tilde{S}-\sqrt{2} \mathcal{D}_B\, S)-
   M_W^D(\tilde{W}^a\tilde{T}^a-\sqrt{2}\mathcal{D}_W^a T^a)
   \nonumber\\
   &{}
   -M_g^D(\tilde{g}^a\tilde{O}^a-\sqrt{2}\mathcal{D}_g^a O^a)
   + \mbox{h.c.}\,,
\end{align}
which describes Dirac mass terms for the gauginos and interaction
terms between the adjoint scalars and the auxiliary
$D$-fields of the corresponding gauge multiplet.

\subsection{Masses and mixings}

For the purpose of the present paper we need the Feynman rules for the
lepton and quark interactions with the photon and $Z$ bosons and with
sleptons/squarks and charginos and neutralinos.
These in turn require an understanding of masses and mixings in the
MRSSM.

We begin with basic tree-level relations between couplings, vacuum
expectation values and masses
of the SM $W$ and $Z$ gauge bosons:
\begin{align}
  e&= g_2\sin\theta_{\text{W}}=g_1\cos\theta_{\text{W}} \,,
  &g_Z&=\frac{e}{\sin\theta_{\text{W}}\cos\theta_{\text{W}}} \,,\\
  m_Z^2 &= \frac{g^2_1+g^2_2}{4} v^2 \,,
  &m_W^2 &= \frac{g^2_2}{4} (v^2 + 4 v_T^2) \,,
  \label{eq:bosonmasses}
  \\
  v^2&= v_u^2+v_d^2 \,,
  &
  \tan\beta&=\frac{v_u}{v_d} \,.
\end{align}
In the following we abbreviate $s_{\text{W}} \equiv \sin \theta _{\text{W}}$ and $c_{\text{W}} \equiv \cos\theta _{\text{W}}$. 
The vacuum expectation values are defined and normalized such that
$H_{u,d}^0=\frac{1}{\sqrt2}v_{u,d}+\ldots$,
$T^0=\frac{1}{\sqrt2}v_T+\ldots$, $S=\frac{1}{\sqrt2}v_S+\ldots$ and
$v\approx250$ GeV.
The relations between the quark and lepton masses and the Yukawa couplings are 
\begin{align}
  Y_{l_g,d_g}&=\frac{\sqrt{2}\,m_{l_g,d_g}}{v_d}=\frac{m_{l_g,d_g} \, e}{\sqrt{2} m_Z\cos\beta
    \sin\theta_W\cos\theta_W} \,,
  \\
  Y_{u_g}&=\frac{\sqrt{2}\,m_{u_g}}{v_u}=\frac{m_{u_g} \, e}{\sqrt{2}m_Z\sin\beta 
    \sin\theta_W\cos\theta_W} \,.
\end{align}
The SU(2)$_L\times$U(1)$_Y$ gauge covariant derivative is defined as
$D_\mu=\partial_\mu+ig_2 T^a W^a_\mu+ig_1 Y B_\mu$ with generators
$T^a$ and $Y=Q-T^3$ with the electric charge $Q$; hence the
interaction Lagrangian for fermions and 
photon/$Z$ reads
\begin{align}
  \label{eq:LintEW}
  {\cal L}_{\text{int}}\ni
  -e Q_fA_\mu \, \bar{f}\gamma^\mu f
  -g_Z Z_\mu \, \bar{f}\gamma^\mu
  \left(Z^{\TL}_f \PL+ Z^{\TR}_f \PR\right) f \,,
\end{align}
with left- and right-handed projectors
$\text{P}_{\TL,\TR}=\frac{1}{2}\left(1\mp\gamma_5\right)$ and
\begin{align}
  Z^{\TL}_f&=T^3_f-Q_f\sin^2\theta_W \,,
  &Z^{\TR}_f&= -Q_f\sin^2\theta_W \,,
\label{ZLRcoefficients}
\end{align}
with the electric charge $Q_l=-1$, $Q_u=+2/3$, $Q_d=-1/3$ and the weak
isospin $T^3_{l,d}=-1/2$, $T^3_u=+1/2$.

The interaction eigenstate sleptons and squarks are
$\tilde{\nu}_{g\TL}$, $\tilde{l}_{g\TL/\TR}$, $\tilde{d}_{g\TL/\TR}$,
$\tilde{u}_{g\TL/\TR}$, where the generation index $g=1,2,3$. Since the
left-handed and right-handed sfermions have opposite
R-charges, there is no left-right mixing. This is an important
distinction to the MSSM and at the heart of the modified flavour
properties of the MRSSM \cite{Kribs:2007ac}. Still it
is useful to define the mass matrices and 
the diagonalization in the following, MSSM-like way. For the interaction eigenstate sfermions $\tilde{f} ^{\text{I}}$ of any type
we write the mass term in the Lagrangian as
$ {{\tilde{f}}^{\text{I}\dagger}} {\mathcal{M}} _{\tilde{f}} ^2 \tilde{f} ^{\text{I}}$ in terms of the
basis of chirality and generation eigenstates $\tilde{f} ^{\text{I}}=
  ({\tilde{f}}_{1\TL}, {\tilde{f}}_{2\TL}, {\tilde{f}}_{3\TL},
  {\tilde{f}}_{1\TR}, {\tilde{f}} _{2\TR}, {\tilde{f}} _{3\TR})
  ^{\text{T}} .$ The mass matrix $  {\mathcal{M}} _{\tilde{f}}^2$
 is a $6\times6$ block
  matrix, and it can be written and diagonalized as
\begin{align}
  {\mathcal{M}} _{\tilde{f}} ^2 &=
  \begin{pmatrix}
    {\mathcal{M}} ^2 _{\tilde{f} _\TL} & 0
    \\
    0 & {\mathcal{M}} ^2 _{\tilde{f} _\TR}
  \end{pmatrix} \,,
&
  U ^{\tilde{f}} {\mathcal{M}} ^2 _{\tilde{f}} U ^{\tilde{f} \dagger}&=
  {\mathcal{M}} ^2 _{\tilde{f}\,\text{diag}}
  \,,
  &
 \tilde{f} ^{\text{I}}_i &=\sum_{j=1}^6 U ^{\tilde{f}\,*}_{ji}  \tilde{f} _j \,,
\end{align}
by introducing a unitary matrix $U ^{\tilde{f}}$ and  mass eigenstate
fields $\tilde{f}_j$. The values of the 
$3\times3$ block mass matrices depend on the sfermion type. For the
sneutrinos and charged sleptons  they are given by\footnote{We do not consider neutrino masses and right-handed neutrinos in the present paper. To make the equations and the corresponding implementation more uniform we nevertheless introduce six sneutrino fields and Eq.\ (\ref{eq:right_handed_sneutrino_mass}); the additional terms do not appear in physical calculations.}
\begin{align}
  ({\mathcal{M}} ^2 _{\tilde{\nu} _\TL})_{ij} &=
  (m ^2 _{\slep}) _{ij}
  +
  \delta _{ij}
  \left(
  -\frac{g _1 ^2}{8}(v _u ^2 - v _d ^2)
  -
  \frac{g _2 ^2}{8}(v _u ^2 - v _d ^2)
  -
  g _1 M _{B} ^{D} v _S
  +
  g _2 M _{W} ^{D} v _T
  \right) \,,
  \\
\label{eq:right_handed_sneutrino_mass}
  ({\mathcal{M}} ^2 _{\tilde{\nu} _\TR})_{ij} &=0\,,
  \\
  ({\mathcal{M}} ^2 _{\slep _\TL})_{ij} &=
  (m ^2 _{\slep}) _{ij}
  +
  \delta _{ij}
  \left(
  m ^2 _{l _i}
  -
  \frac{g _1 ^2}{8}(v _u ^2 - v _d ^2)
  +
  \frac{g _2 ^2}{8}(v _u ^2 - v _d ^2)
  -
  g _1 M _{B} ^{D} v _S
  -
  g _2 M _{W} ^{D} v _T
  \right) \,,
  \\
  ({\mathcal{M}} ^2 _{\slep _\TR}) _{ij} &=
  (m ^2 _{\tilde{e}}) _{ij}
  +
  \delta _{ij}
  \left(
  m ^2 _{l_i}
  +
  \frac{g _1 ^2}{4}(v _u ^2 - v _d ^2)
  +
  2 g _1 M _B ^{D} v _S
  \right)\,.
\end{align}
The formulas for squark masses are as follows:
\begin{align}
  ({\mathcal{M}} ^2 _{\tilde{u} _\TL})_{ij} &=
  (m ^2 _{\tilde{q}_\TL}) _{ij}
  +
  \delta _{ij}
  \left(
  m ^2 _{u _i}
  +
  \frac{g _1 ^2}{24}(v _u ^2 - v _d ^2)
  -
  \frac{g _2 ^2}{8}(v _u ^2 - v _d ^2)
  +
  \frac{1}{3}g _1 M _{B} ^{D} v _S
  +
  g _2 M _{W} ^{D} v _T
  \right) \,,
  \\
  ({\mathcal{M}} ^2 _{\tilde{u} _\TR}) _{ij} &=
  (m ^2 _{\tilde{u}_\TR}) _{ij}
  +
  \delta _{ij}
  \left(
  m ^2 _{u_i}
  -
  \frac{g _1 ^2}{6}(v _u ^2 - v _d ^2)
  -
  \frac{4}{3} g _1 M _B ^{D} v _S
  \right) \,,
  \\
  ({\mathcal{M}} ^2 _{\tilde{d} _\TL})_{ij} &=
  (m ^2 _{\tilde{q}_\TL}) _{ij}
  +
  \delta _{ij}
  \left(
  m ^2 _{d _i}
  +
  \frac{g _1 ^2}{24}(v _u ^2 - v _d ^2)
  +
  \frac{g _2 ^2}{8}(v _u ^2 - v _d ^2)
  +
  \frac{1}{3}g _1 M _{B} ^{D} v _S
  -
  g _2 M _{W} ^{D} v _T
  \right) \,,
  \\
  ({\mathcal{M}} ^2 _{\tilde{d} _\TR}) _{ij} &=
  (m ^2 _{\tilde{d}_\TR}) _{ij}
  +
  \delta _{ij}
  \left(
  m ^2 _{d_i}
  +
  \frac{g _1 ^2}{12}(v _u ^2 - v _d ^2)
  +
  \frac{2}{3} g _1 M _B ^{D} v _S
  \right)\,.
\end{align}
The MRSSM neutralinos are Dirac fermions with twice as many degrees of
freedom as in the MSSM. The two-component basis fields $ {\xi}_i=({\tilde{B}},
\tilde{W}^0, \tilde{R}_d^0, \tilde{R}_u^0)$ have R-charge $+1$, the
two-component basis fields  $\zeta_i=(\tilde{S}, \tilde{T}^0,
\tilde{H}_d^0, \tilde{H}_u^0) $ have R-charge $-1$.
In terms of this basis, the mass matrix can be written as
\begin{equation} 
\label{eq:neut-massmatrix}
m_{{\chi}} =
\left( 
\begin{array}{cccc}
  M^{D}_B &0 &-\frac{1}{2} g_1 v_d  &\frac{1}{2} g_1 v_u \\ 
  0 &M^{D}_W &\frac{1}{2} g_2 v_d  &-\frac{1}{2} g_2 v_u \\ 
  - \frac{1}{\sqrt{2}} \lambda_d v_d  &-\frac{1}{2} \Lambda_d v_d  & - \mud{+} &0\\ 
  \frac{1}{\sqrt{2}} \lambda_u v_u  &-\frac{1}{2} \Lambda_u v_u  &0 & \muu{-}
\end{array} 
\right) \,,
\end{equation} 
with $\mu_i^{\text{eff,}\pm} = \mu_i+\frac{\lambda_iv_S}{\sqrt2} \pm\frac{\Lambda_iv_T}{2}$.
The mass matrix is diagonalized with two unitary mixing matrices
$N^1$ and $N^2$   and mass eigenstates $\kappa_i$ and $\psi_i$ are
defined as
\begin{align}
{N^1} ^\ast m_{{\chi}} {N^2} ^\dagger &= {m_\chi}_{\text{diag}} \,,
&
{\xi}_i &=\sum_{j=1}^4 {N_{ji}^1}^\ast {\kappa}_j\,,
&
\zeta_i=\sum_{j=1}^4 {N_{ji}^2}^\ast{\psi}_j\,,
\end{align}
and physical four-component Dirac neutralinos are constructed as
\begin{equation}
{\chi}^0_i=\left(\begin{array}{c}
\kappa_i\\
{\psi}^{\ast}_i\end{array}\right)\,,\qquad i=1,2,3,4\,.
\end{equation}

The MRSSM charginos also involve twice as many degrees of freedom as
in the MSSM and can be grouped according to their R-charge. The
$\chi$-charginos have R-charge$=$electric charge; the $\rho$-charginos
have R-charge$=$($-$electric charge). The $\chi$-charginos are defined
in terms of the basis \( (\tilde{T}^-, \tilde{H}_d^-), (\tilde{W}^+,
\tilde{R}_d^+) \). The mass matrix and the diagonalization procedure
are defined as 
\begin{align} 
  m_{{\chi}^+} &=
  \left( 
  \begin{array}{cc}
    g_2 v_T  + M^{D}_W \; &\frac{1}{\sqrt{2}} \Lambda_d v_d \\ 
    \frac{1}{\sqrt{2}} g_2 v_d \; &+ \mud{-}
  \end{array} 
  \right) \,,
  &
  {U^1}^* m_{{\chi}^+} {V^1}^\dagger &=
  {m_{\chi^+}}_{\text{diag}} \,,
  \label{eq:cha1-massmatrix}
\end{align}
\begin{align}
  \tilde{T}^- &= \sum_{j=1}^2 {U_{j1}^1}^\ast \lambda^-_{{j}}\,,& 
  \tilde{H}_d^- &= \sum_{j=1}^2 {U_{j2}^{1}}^\ast \lambda^-_{{j}}\,,\\ 
  \tilde{W}^+ &= \sum_{j=1}^2 {V_{j1}^1}^\ast \lambda^+_{{j}}\,,&
  \tilde{R}_d^+ &= \sum_{j=1}^2 {V_{j2}^1}^\ast \lambda^+_{{j}} \,,
\end{align} 
with two unitary matrices  \(U^1\) and \(V^1\) and mass-eigenstate
spinors $\lambda^{\pm}_j$. The corresponding
physical four-component charginos are constructed as  
\begin{equation}
{\chi}^+_i=\left(\begin{array}{c}
\lambda^+_i\\
\lambda^{-*}_i\end{array}\right)\,,\qquad i=1,2 \,.
\end{equation}

The $\rho$-charginos are defined in terms of the basis $ (\tilde{W}^-,
\tilde{R}_u^-)$, $(\tilde{T}^+, \tilde{H}_u^+) $. The mass matrix and
diagonalization procedure read
\begin{align} 
  m_{{\rho}^-} &=
  \left( 
  \begin{array}{cc}
    - g_2 v_T  + M^{D}_W \;&\frac{1}{\sqrt{2}} g_2 v_u \\ 
    - \frac{1}{\sqrt{2}} \Lambda_u v_u \; &  - \muu{+} \end{array} 
  \right)
  &
  {U^2}^\ast m_{{\rho}^-} {V^2} ^\dagger & = {m_{\rho^-}}_{\text{diag}} \,,
  \label{eq:cha2-massmatrix}
\end{align}
\begin{align} 
  \tilde{W}^- & = \sum_{j=1}^2 {U_{j1}^2}^\ast \eta^-_{{j}}\,, &
  \tilde{R}_u^- & = \sum_{j=1}^2 {U_{j2}^2}^\ast \eta^-_{{j}}\,,\\ 
  \tilde{T}^+ & = \sum_{j=1}^2 {V_{j1}^2}^\ast \eta^+_{{j}}\,, &
  \tilde{H}_u^+ & = \sum_{j=1}^2 {V_{j2}^2}^\ast \eta^+_{{j}}\,,
\end{align} 
with two unitary matrices \(U^2\) and \(V^2\) and mass-eigenstate
spinors $\eta^{\pm}_j$.
The corresponding physical four-component charginos are constructed as  
\begin{equation}
  {\rho}^-_i=\left(\begin{array}{c}
    \eta^-_i\\
    \eta^{+*}_i\end{array}\right)\,,\qquad i=1,2\,.
\end{equation}
For reference, the R-charges of the mass eigenstate sfermions,
charginos and neutralinos are collected in table
\ref{tab:Rmasseigenstates}.
\begin{table}[t]
  \begin{center}
    \begin{tabular}{|r|l|r|l|r|l|}
      \multicolumn{2}{|c|}{left-/right-handed sfermions} &
      \multicolumn{2}{c|}{(anti-)neutralinos} &
      \multicolumn{2}{c|}{$\chi^-$/$\rho^-$-charginos}\\
      \hline
      $\tilde{f}_L$        & $+1$
      & $\chi^{0c}_i$       & $-1$
      & $\chi^-_i$         & $-1$
      \\
      $\tilde{f}_R$        & $-1$
      & $\chi^0_i$         & $+1$
      & $\rho^-_i$         & $+1$
      \\
    \end{tabular}
  \end{center}
  \caption{The R-charges of the squarks and sleptons,
    neutralinos, antineutralinos and the  $\chi$- and $\rho$-charginos.
    This table also shows which  pairs of particles can couple to
    leptons or quarks.       }
  \label{tab:Rmasseigenstates}
\end{table}

\subsection{Feynman rules}
\label{sec:Feynmanrules}

Using the definitions of the mass eigenstates we can specify the
interaction Lagrangians relevant for the required Feynman rules. 
The interaction Lagrangian between the $Z$ boson and sleptons, $\chi$-charginos\footnote{We 
omit the $\rho$-chargino terms as these do not contribute due to the R-charge conservation.}
and neutralinos can be written as
\begin{align}
  {\cal L}_{\text{int}}\ni&{}
  -g_Z Z_\mu \overline{\chi^0_A}\gamma^\mu\left(
  \cz^{\TL\chi^0}_{AB}\PL+\cz^{\TR\chi^0}_{AB}\PR\right)
  \chi^0_B
  \\
  &{}
  -g_Z Z^\mu \overline{\chi^-_A}\gamma^\mu\left(
  \cz^{\TL\chi^-}_{AB}\PL+\cz^{\TR\chi^-}_{AB}\PR\right)
  \chi^-_B
  \\
  &{}
  -g_Z Z_\mu
  \tilde{f}_X^\dagger\left(i\overrightarrow{\partial}^\mu-i\overleftarrow{\partial}^\mu\right)
  \cz^{\tilde{f}}_{XY}\tilde{f}_Y \,. 
\end{align}
The values of the $Z$ boson coupling coefficients $\cz^i_j$ and the
resulting Feynman rules can be found in the Appendix.

The interaction Lagrangian between fermions, sfermions and
neutralinos/charginos can be written as
\begin{align}
  {\cal{L}}_{\text{int}} &\ni
  \sum_{(f',{\tilde{f}}),g}
  \Big\{
      {\overline{\tilde{\chi} ^{-} _{A}}}
      (\cc ^{\text{L} (f')} _{g AX} \PL +
      \cc ^{\text{R} (f')} _{g AX} \PR)
      f' _{g} {\tilde{f}} ^{\dagger} _{X}
      +
      {\overline{\tilde{\rho} ^{-} _{A}}}
      (\ct ^{\text{L} (f')} _{g AX} \PL +
      \ct ^{\text{R} (f')} _{g AX} \PR)
      f'_{g} {\tilde{f}} ^{\dagger} _{X}
  \Big\}
  \nonumber\\
  &+ \sum _{(f,{\tilde{f}}),g}
  \Big\{
      {\overline{\tilde{\chi} ^{0} _{A}}}
      (\cn ^{\text{L} (f)} _{g AX} \PL +
      \cn ^{\text{R} (f)} _{g AX} \PR)
      f _g {\tilde{f}}^{\dagger} _{X}
      +
      {\overline{\tilde{\chi} ^{0 c} _{A}}}
      (\co ^{\text{L} (f)} _{g AX} \PL +
      \co ^{\text{R} (f)} _{g AX} \PR)
      f _g {\tilde{f}}^{\dagger} _{X}
  \Big\}
  \nonumber\\
  &+ \hc \,.
\end{align}
In the Lagrangian the sums extend over the fermion/sfermion pairs
$(f ,{\tilde{f} })\in \{ (\nu , {\tilde{\nu}}),\,$ $ (l,
{\tilde{l}}),$ $ (u, {\tilde{u}}),$ $ (d, {\tilde{d}})\}$
and $(f',{\tilde{f} }) \in \{(l,\tilde{\nu}),\, (\nu^c, {\tilde{l}}),\, (d, {\tilde{u}}),\, (u^c, {\tilde{d}})\}$  and over the generation index 
 $g \in \{1,2,3\}$. We have written the Lagrangian in a
form 
analogous to the one of Ref.\ \cite{Fargnoli:2013zia} for easy comparison to
the MSSM case. While in the MSSM there are only two types of
couplings, the MRSSM needs four types of couplings $\cn,\co,\cc,\ct$
for the interactions with neutralinos, antineutralinos,
$\chi$-charginos and $\rho$-charginos. The indices of the couplings
correspond  to the chirality, type and generation of the respective
quark or lepton and to the neutralino/chargino and sfermion indices.

The gaugino and Higgsino couplings are well separated in $\cn$, $\co$,
$\cc$, and $\ct$. The coefficients $\cn ^\TR$, $\co ^\TL$, $\cc ^\TL$,
and $\ct ^\TR$ are from the gaugino interactions whereas $\cn ^\TL$,
$\co ^\TR$, $\cc ^\TR$, and $\ct ^\TL$ are from Higgsino interactions and suppressed by small Yukawa couplings. 
The values of the coupling coefficients and the
resulting Feynman rules can be found in the Appendix.

\section{Theory of $a_\mu$, $\mu\to e\gamma$ and $\mu\to e$ in the
  MRSSM}
\label{sec:observables}

In the present paper we consider three muon observables in the MRSSM:
the flavour-conserving anomalous magnetic dipole moment $a_\mu$ and
the lepton-flavour violating decay $\mu\to e\gamma $ and $\mu\to e $
conversion.
In this section we collect definitions of these observables and
provide explicit formulas valid in the MRSSM. The formulas are written
in a way that facilitates comparisons to the MSSM and generalization
to other models. We begin with $a_\mu$ and $\mu\to e\gamma$ which rely
on the lepton--photon three-point interaction only, and then turn to
$\mu\to e $ conversion, which is based on effective 4-fermion
interactions.

\subsection{$a_\mu$ and $\mu\to e\gamma$}

The muon magnetic moment and the decay $\mu\to e\gamma$ are related to
the lepton--photon interaction. We define the three-point vertex 
$i\Gamma_{\bar{l}_il_j\gamma}$ as the sum of all Feynman diagrams%
\footnote{With this definition, $\Gamma$ corresponds to the loop-corrected effective action in momentum space.}
with incoming lepton $l_j$ with incoming momentum $p$ and outgoing lepton $l_i$ and
outgoing off-shell photon with outgoing momenta $(p-q)$ and $q$ respectively. For the flavour-conserving and CP-conserving case the effective
interaction is commonly written as \cite{Jegerlehner:2017gek,Jegerlehner:2009ry}
\begin{align}
  \label{GammallgammaFC}
  i\Gamma_{\bar{l}l\gamma}^\mu&=
  -i e Q_l\, \bar{u}_{l}(p-q)
  \left[
    \gamma^\mu F_{\text{E}}^l(q^2)
    + \frac{i\sigma^{\mu\nu}(-q_\nu)}{2m_{l}} F_{\text{M}}^l(q^2)
    \right] u_{l}(p)\,,
\end{align}
where the overall sign reflects our choice of the gauge covariant derivative and the corresponding interaction Lagrangian in Eq.~(\ref{eq:LintEW}) and $Q_l = -1$. 
The two form factors $F_{\text{E,M}}(q^2)$ depend on the lepton
generation.
The electric form factor satisfies $F_{\text{E}}(0)=1$ if $e$ is on-shell
renormalized, and the magnetic dipole form factor describes the anomalous
magnetic dipole moment. Specializing to the case of the muon, we have
\begin{align}
  \label{amudef}
  a_\mu&=\frac{(g-2)_\mu}{2} =
  F_{\text{M}}^\mu(0)\,.
\end{align}
The contributions of the MRSSM need to be compared with the
experimental value and the SM prediction for $a_\mu$. Taking the most
recent SM theory evaluation of the KNT collaboration
\cite{Keshavarzi:2018mgv}, the deviation from the Brookhaven measurement
\cite{Bennett:2006fi} is
given by
\begin{align}
  \label{eq:Deltaamu}
  \Delta a_\mu^{\text{Exp}-\text{SM}}&=
  (27.06\pm7.26)\times10^{-10} \,.
\end{align}
Other recent evaluations \cite{Jegerlehner:2018zrj,Davier:2016udg} find similar
deviations with a significance between $3.6$--$4.0\sigma$. It is
noteworthy that in recent years tremendous progress has been made on
consolidating and improving the accuracy of the SM hadronic
contributions using lattice QCD \cite{Chakraborty:2016mwy,DellaMorte:2017dyu,Borsanyi:2017zdw,Blum:2015gfa,Gerardin:2016cqj,Blum:2018mom}, dispersion relations \cite{Pauk:2014rfa,Colangelo:2015ama,Colangelo:2017qdm,Colangelo:2017fiz,Hoferichter:2018dmo,Hoferichter:2018kwz}
and  $e^+e^-\to\pi\pi$ data
\cite{Ablikim:2015orh,Anastasi:2017eio,Xiao:2017dqv}; for further
progress on hadronic, QED and weak contributions see
Refs.\ \cite{Colangelo:2014qya,Kurz:2014wya,Gnendiger:2013pva,Lee:2013sx,Kurz:2013exa,Kurz:2015bia,Kurz:2016bau,Laporta:2017okg}
and the recent reviews \cite{Benayoun:2014tra,Jegerlehner:2018zrj}.

For the flavour-violating case relevant for $\mu\to e\gamma$ and
$\mu\to e$, where the momenta are small and the mass $m_{l_i}$ can be
neglected, the effective interaction of off-shell photon with on-shell fermions
can be written as \cite{Nowakowski:2004cv,Ilakovac:2012sh}
\begin{align}
  \label{GammallgammaFV}
  i\Gamma_{\bar{l}_il_j\gamma}^\mu =
  -ieQ_{l_j}\, \bar{u}_{l_i}(p-q)
  \Big[&
    \left(q^2 \gamma^\mu
    -q^\mu\slashed{q}\right)\left(A_1^{\bar{l}_il_j\TL}\PL+A_1^{\bar{l}_il_j\TR}
    \PR\right)
    \nonumber\\&
    +m_{l_j} i\sigma^{\mu\nu}q_\nu \left(A_2^{\bar{l}_il_j\TL}\PL+A_2^{\bar{l}_il_j\TR}\PR\right)
    \Big] u_{l_j}(p) \,,
\end{align}
with constants $A_{1,2}$, which depend on the lepton generations
and the chirality, as indicated. The constants $A_2$ describe the
photon dipole interaction, and there is a strong similarity but no
equality between $(-m_{\mu}^2(A_2^{{\bar{e}\mu\TL}}\PL+A_2^{\bar{e}\mu\TR}\PR))$ and $F_{\text{M}}^\mu$ because
of the neglected $m_{l_i}$-terms in Eq.\ (\ref{GammallgammaFV}). 
The lepton-flavour violating decay $\mu\to e\gamma$ is then described
in terms of $A_2 ^{\bar{e}\mu \, \TL/\TR}$. The decay rate and the
branching ratio are given by (see e.g.\ \cite{Hisano:1995cp})
\begin{align}
  \Gamma(\mu \to e \gamma) &=
  \frac{\alpha}{4} m_{\mu}^5
  \left( \left|A_2^{\bar{e}\mu \TL}\right|^2
  + \left|A_2^{\bar{e}\mu \TR}\right|^2
  \right) \,,
  \\
  \label{eq:BRmutoeg}
  B_{\mu \to e \gamma} &= \frac{\Gamma(\mu \to e \gamma)}{\Gamma(\mu \to e \nu_\mu {\bar\nu}_e)}
  =\frac{48 \pi ^3 \alpha }{G_{\text{F}} ^{2}}
  \left(
  \left|A_2^{\bar{e}\mu \TL}\right|^2
  + \left|A_2^{\bar{e}\mu \TR}\right|^2
  \right) \,.
\end{align}
The currently best experimental upper limit has been obtained by the MEG
experiment \cite{TheMEG:2016wtm},
\begin{align}
  B_{\mu^+\to e^+\gamma}&<4.2\times10^{-13} \ (90\%\text{ C.L.})\,.
  \label{eq:mu2eglimit}
\end{align}
An upgrade to MEG-II is planned \cite{Baldini:2013ke,Galli:2018bsc},
with a foreseen increase in sensitivity by an order of magnitude.

\begin{figure}
  \begin{center}
    \begin{subfigure}[c]{.23\textwidth}
      \begin{center}
        \includegraphics[scale=.45]{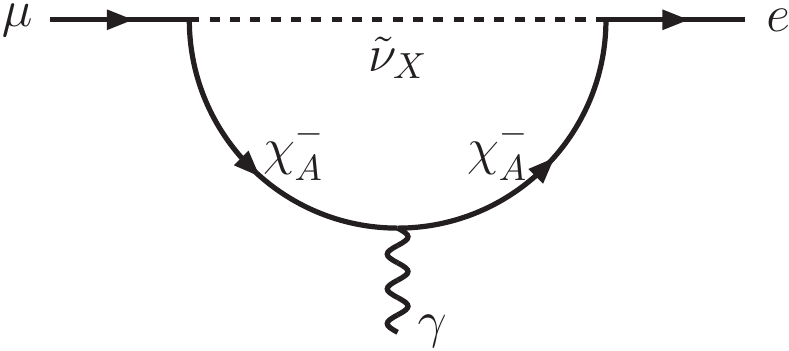}
      \end{center}
    \end{subfigure}
    \,
    \begin{subfigure}[c]{.23\textwidth}
      \begin{center}
        \includegraphics[scale=.45]{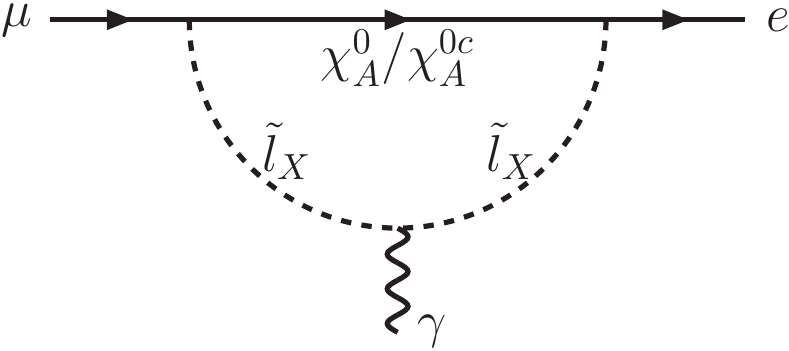}
      \end{center}
    \end{subfigure}
    \,
    \begin{subfigure}[c]{.23\textwidth}
      \begin{center}
        \includegraphics[scale=.45]{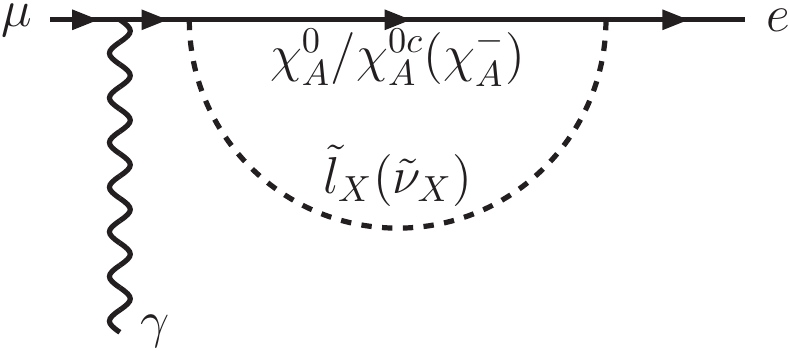}
      \end{center}
    \end{subfigure}
    \,
    \begin{subfigure}[c]{.23\textwidth}
      \begin{center}
        \includegraphics[scale=.45]{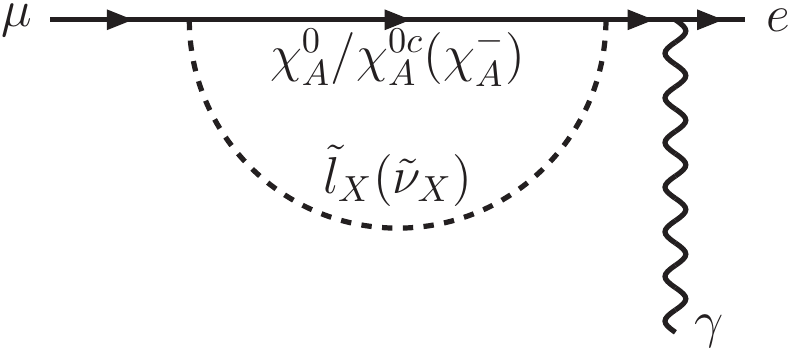}
      \end{center}
    \end{subfigure}
  \end{center}
  \caption{The four types of diagrams contributing to the photon
    interaction (\ref{GammallgammaFV}). All four diagrams contribute
    to the charge radius form factors $\Acharg^{\bar{e}\mu \TL/\TR}$;
    the first two also contribute to the dipole form factors
    $\Adip^{\bar{e}\mu \TL/\TR}$.  }
  \label{fig:muegDiagrams}
\end{figure}
We now discuss the MRSSM one-loop contributions to the required form
factors and focus on the restrictions imposed by R-symmetry. The
Feynman diagrams relevant for $\mu\to e\gamma$ are shown in Fig.\ \ref{fig:muegDiagrams}; 
the diagrams for $a_\mu$ are analogous to the diagrams (a,b) with the replacement $e\to\mu$.
We refer to table \ref{tab:Rmasseigenstates} for the pairs of
particles which can couple to leptons without violation of R-charge
conservation. 
Diagram (a) involves the exchange of a left-handed sneutrino and
$\chi$-chargino, which involves the components $\tilde{W}$ and
$\tilde{H}_d$. The couplings to the leptons involve the
coupling coefficients $\cc$, which contain the
gauge coupling $g_2$ and the lepton Yukawa couplings. Since there is no
right-handed sneutrino, there is no Feynman diagram involving
$\rho$-charginos.
Diagram (b) in Fig.\ \ref{fig:muegDiagrams} involves the
exchange of a slepton and a neutralino or an an antineutralino. In
case of an antineutralino $\chi^{0c}_A$,the slepton must always be
left-handed, see also
Tab.\ \ref{tab:Rmasseigenstates}. So the underlying physics of the antineutralino diagram
is similar to the one of the chargino diagram; the involved couplings
are $\co$, which contain the gauge couplings $g_2$ and $g_1$ and the
lepton Yukawa couplings. Diagram (b) with neutralino exchange  involves the exchange
of a right-handed slepton. Here the
involved coupling coefficients are $\cn$, containing only $g_1$ and
lepton Yukawa couplings. Diagrams (c,d) have a similar behaviour.

The MRSSM results for the constants $A_{1,2} ^{\bar{e}\mu \, \TL/\TR}$ are decomposed as
\begin{align}
  A_1^{\bar{e}\mu \TL}&=\Acharg ^{n\TL} + \Acharg ^{c\TL} \,,\\
  A_2 ^{\bar{e}\mu \TL}&=\Adip ^{n\TL} + \Adip ^{c\TL} \,,\\
  A_1^{\bar{e}\mu \TR}&=\Acharg ^{n\TR} + \Acharg ^{c\TR} \,,\\
  A_2 ^{\bar{e}\mu \TR}&=\Adip ^{n\TR} + \Adip ^{c\TR} \,,
\end{align}
into neutralino and chargino contributions. Using the dimensionless variables
$x_{AX} ^{n} \equiv m_{\neut _{A}} ^{2}/m _{\slep _{X}} ^{2}$ and
$x _{AX} ^{c} \equiv m ^{2} _{\charg _{A}} /m ^{2} _{{\sneu} ^{} _{X}}$
and omitting terms which are suppressed by the electron Yukawa coupling,
the charge radius contributions are given as
\begin{align}
  \label{eq:AchargnL}
  \Acharg ^{n\TL} &=
  \frac{1}{576 \pi ^2}
  \sum_{A, X}
  \Big(
  \co ^{\TL (l) \ast} _{1AX}
  \co ^{\TL (l)} _{2AX}
  +
  \cn ^{\TL (l) \ast} _{1AX}
  \cn ^{\TL (l)} _{2AX}
  \Big)
  \frac{1}{m _{\slep _{X}} ^2}
  \fAchargn(x_{AX} ^n)\,,
  \\
  \label{eq:AchargnR}
  \Acharg ^{n\TR} &=
  \frac{1}{576 \pi ^2}
  \sum_{A,X}
  \Big(
  \cn ^{\TR (l) \ast} _{1AX}
  \cn ^{\TR (l)} _{2AX}
  +
  \co ^{\TR (l) \ast} _{1AX}
  \co ^{\TR (l)} _{2AX}
  \Big)
  \frac{1}{m_{\slep _{X}} ^2}
  \fAchargn(x_{AX} ^n)\,,
\end{align}
and
\begin{align}
  \label{eq:AchargcL}
  \Acharg ^{c\TL} &=
  -\frac{1}{576 \pi ^2}
  \sum_{A,X}
  \cc ^{\TL (l) \ast} _{1AX} \cc ^{\TL (l)} _{2AX}
  \frac{1}{m ^{2} _{\sneu ^{} _{X}}}
  \fAchargc (x _{AX} ^c)\,,\\
  \label{eq:AchargcR}
  \Acharg ^{c\TR} &=
  -\frac{1}{576 \pi ^2}
  \sum _{A, X}
  \cc ^{\TR (l) \ast} _{1AX} \cc ^{\TR (l)} _{2AX}
  \frac{1}{m ^{2} _{\sneu ^{} _{X}}}
  \fAchargc (x _{AX} ^c)\,.
\end{align}
As shortly mentioned at the end of Sec.~\ref{sec:model} the leading
terms in Eqs.~(\ref{eq:AchargnL})--(\ref{eq:AchargcR}) are  
the $\cn ^\TR$-, $\co ^\TL$-, and $\cc ^\TL$-terms, which are from
gaugino interactions; the other terms are suppressed by Yukawa
couplings. 
Similar comments apply to the following results.

The dipole contributions from neutralinos/antineutralinos are given as
\begin{align}
  \Adip ^{n\TL} &=
  \frac{1}{32 \pi ^2}
  \sum _{A, X}
  \frac{1}{m_{\slep _{X}} ^2}
  \Big\{
  \cn ^{\TR (l) \ast} _{1AX}
  \cn ^{\TR (l)} _{2AX}
  \fAdipan(x_{AX} ^n)
  +
  \cn ^{\TR (l) \ast} _{1AX}
  \cn ^{\TL (l)} _{2AX}
  \frac{m_{\neut _{A}}}{m_\mu}
  \fAdipbn (x_{AX} ^n)
  \Big\}\nonumber\\
  &+
  \frac{1}{32 \pi ^2}
  \sum _{A, X}
  \frac{1}{m_{\slep _{X}} ^2}
  \Big\{
  \co ^{\TR (l) \ast} _{1AX}
  \co ^{\TR (l)} _{2AX}
  \fAdipan(x_{AX} ^n)
  +
  \co ^{\TR (l) \ast} _{1AX}
  \co ^{\TL (l)} _{2AX}
  \frac{m_{\neut _{A}}}{m_\mu}
  \fAdipbn (x_{AX} ^n)
  \Big\},
  \\
  \Adip ^{n\TR} &=
  \frac{1}{32 \pi ^2}
  \sum _{A, X}
  \frac{1}{m_{\slep _{X}} ^2}
  \Big\{
  \co ^{\TL (l) \ast} _{1AX}
  \co ^{\TL (l)} _{2AX}
  \fAdipan (x_{AX} ^n)
  +
  \co ^{\TL (l) \ast} _{1AX}
  \co ^{\TR (l)} _{2AX}
  \frac{m_{\neut _{A}}}{m_\mu}
  \fAdipbn (x_{AX} ^n)
  \Big\}\nonumber\\
  &+
  \frac{1}{32 \pi ^2}
  \sum _{A, X}
  \frac{1}{m_{\slep _{X}} ^2}
  \Big\{
  \cn ^{\TL (l) \ast} _{1AX}
  \cn ^{\TL (l)} _{2AX}
  \fAdipan (x_{AX} ^n)
  +
  \cn ^{\TL (l) \ast} _{1AX}
  \cn ^{\TR (l)} _{2AX}
  \frac{m_{\neut _{A}}}{m_\mu}
  \fAdipbn (x_{AX} ^n)
  \Big\}\,.
\end{align}
The dipole contributions from charginos are given as
\begin{align}
  \Adip ^{c\TL} &=
  -\frac{1}{32 \pi ^2}
  \sum _{A, X}
  \frac{1}{m ^{2} _{\sneu ^{} _{X}}}
  \Big\{
  \cc ^{\TR (l) \ast} _{1AX} \cc ^{\TR (l)} _{2AX}
  \fAdipac (x _{AX} ^c)
  +
  \cc ^{\TR (l) \ast} _{1AX} \cc ^{\TL (l)} _{2AX}
  \frac{m_{\charg _{A}}}{m_\mu}
  \fAdipbc (x _{AX} ^c)
  \Big\} \,,\\
  \Adip ^{c\TR} &=
  -\frac{1}{32 \pi ^2}
  \sum _{A, X}
  \frac{1}{m ^{2} _{\sneu ^{} _{X}}}
  \Big\{
  \cc ^{\TL (l) \ast} _{1AX} \cc ^{\TL (l)} _{2AX}
  \fAdipac (x _{AX} ^c)
  +
  \cc ^{\TL (l) \ast} _{1AX} \cc ^{\TR (l)} _{2AX}
  \frac{m_{\charg _{A}}}{m_\mu}
  \fAdipbc (x _{AX} ^c)
  \Big\} \,.
\end{align}

The MRSSM results for the contributions to $a_\mu$ are
\begin{align}
  a _\mu=a_\mu ^{{\tilde{\chi}} ^{-}} +
  a_\mu ^{\tilde{\chi} ^{0}} + a_\mu ^{\tilde{\chi} ^{0 c}} \,, 
  \label{eq:amuanalytical}
\end{align}
with
\begin{subequations}
\label{amuMRSSM}
  \begin{align}
  a _\mu ^{{\tilde{\chi}} ^{-}} =
  \frac{1}{16 \pi ^2}
  \frac{m _\mu ^2}{m_{{\tilde{\nu}}_{X}} ^2}
  \Big\{&
  (
  \cc ^{\TL} _{2 AX} \cc ^{\TL \ast} _{2 AX} +
  \cc ^{\TR} _{2 AX} \cc ^{\TR \ast} _{2 AX}
  )
 \fAdipac (x ^c _{AX})
\nonumber\\&  +
  \frac{m_{\tilde{\chi} ^{-} _{A}}}{2 m_\mu}
  (
  \cc ^{\TL} _{2 AX} \cc ^{\TR \ast} _{2 AX} +
  \cc ^{\TR} _{2 AX} \cc ^{\TL \ast} _{2 AX}
  )
\fAdipbc (x ^c _{AX})
  \Big\}
  \,,\\
  a_\mu ^{\tilde{\chi} ^{0}} =
  \frac{-1}{16 \pi ^2}
  \frac{m _\mu ^2}{m_{{\tilde{\mu}}_{X}} ^2}
  \Big\{&
  (
  \cn ^{\TL} _{2 AX} \cn ^{\TL \ast} _{2 AX} +
  \cn ^{\TR} _{2 AX} \cn ^{\TR \ast} _{2 AX}
  )
 \fAdipan(x ^n _{AX})
\nonumber\\&  +
  \frac{m_{\tilde{\chi} ^{0} _{A}}}{2 m_\mu}
  (
  \cn ^{\TL} _{2 AX} \cn ^{\TR \ast} _{2 AX} +
  \cn ^{\TR} _{2 AX} \cn ^{\TL \ast} _{2 AX}
  )
\fAdipbn(x ^n _{AX})
  \Big\}
  \,,\\
  \label{eq:amuanalyticalcharg}
  a_\mu ^{\tilde{\chi} ^{0 c}} =
  \frac{-1}{16 \pi ^2}
  \frac{m _\mu ^2}{m_{{\tilde{\mu}}_{X}} ^2}
  \Big\{&
  (
  \co ^{\TL} _{2 AX} \co ^{\TL \ast} _{2 AX} +
  \co ^{\TR} _{2 AX} \co ^{\TR \ast} _{2 AX}
  )
  \fAdipan(x ^n _{AX})
\nonumber\\&  +
  \frac{m_{\tilde{\chi} ^{0} _{A}}}{2 m_\mu}
  (
  \co ^{\TL} _{2 AX} \co ^{\TR \ast} _{2 AX} +
  \co ^{\TR} _{2 AX} \co ^{\TL \ast} _{2 AX}
  )
 \fAdipbn(x ^n _{AX})
 \Big\} \,.
  \end{align}
\end{subequations}

The appearing loop functions are defined as\footnote{Note that the
  loop functions for the dipole contributions have a different
  normalization from the loop functions $F_{1,2}^{C,N}$ introduced e.g.\ in
  Refs.\ \cite{Martin:2001st,DSreview,Fargnoli:2013zia}: $F_{1}^C(x)= {12}\fAdipac
  (x)$,  $F_{1}^N(x)= {12}\fAdipan
  (x)$, $F_{2}^C(x)=\frac{3}{2}\fAdipbc (x)$, $F_{2}^N(x)=3\fAdipbn(x)$. The loop functions $F_{0,1,2} ^{n,c}$ are normalized as:
  $\frac{2}{3}\fAchargn(1) = -\frac{2}{3}\fAchargc(1)=1$, $12\fAdipan (1) = 12\fAdipac (1) = 1$, $3\fAdipbn (1) = \frac{3}{2}\fAdipbc (1) =1$.}
\begin{subequations}
  \label{fADef}
  \begin{align}
  \fAchargn (x) &=
  \frac{1}{(1-x)^4}
  (2 - 9x + 18x^2 - 11x^3 + 6x^3\ln x) \,,\\
  \fAdipan (x) &=
  \frac{1}{6(1-x)^4}
  (1 -6x + 3x^2 + 2x^3 -6x^2\ln x) \,,\\
  \fAdipbn (x) &=
  \frac{1}{(1-x)^3}
  (1 - x^2 + 2x\ln x) \,,\\
  \fAchargc (x) &=
  \frac{1}{(1-x)^4}
  (16 - 45 x + 36 x^2 - 7 x^3 + 6(2-3x) \ln x) \,,\\
  \fAdipac (x) &=
  \frac{1}{6(1-x)^4}
  (2 + 3 x - 6 x^2 + x^3 + 6 x \ln x) \,,\\
  \fAdipbc (x) &=
  \frac{1}{(1-x)^3}
  (-3 + 4 x - x^2 - 2 \ln x) \,.
  \end{align}
\end{subequations}

\subsection{$\mu\to e$ conversion}

The coherent $\mu\to e$ conversion in a muonic atom is related to the
photon dipole interaction $A_2$ and the effective 4-fermion
interaction between $\mu$--$e$ and the quarks $q$ in the respective
nucleus. We use the precise evaluation of the $\mu\to e$ conversion
rate of Ref.\ \cite{Kitano:2002mt}, which defines the general effective
Lagrangian
\begin{eqnarray}
  {\cal L}_{\rm int}^{\text{Ref.\ \cite{Kitano:2002mt}}} &=&
    + \frac{4 G_{\rm F}}{\sqrt{2}}
    \left[
    m_\mu  \bar{e} \sigma^{\mu \nu} \left(\ALKitano^*\PL+\ARKitano^*\PR\right) \mu F_{\mu \nu}
    + \hc
    \right] \nonumber \\
    &&
    - \frac{G_{\rm F}}{\sqrt{2}} \sum_{q = u,d,s}
    \left[ {\rule[-3mm]{0mm}{10mm}\ } \right.
    \left(
    g_{LS(q)} \bar{e} \PR \mu + g_{RS(q)} \bar{e} \PL \mu
    \right) \bar{q} q  \nonumber \\
    &&   \hspace*{2.4cm}
    +
    \left(
    g_{LP(q)} \bar{e} \PR \mu + g_{RP(q)} \bar{e} \PL \mu
    \right) \bar{q} \gamma_5 q
    \label{LeffKitano}
\\
    &&   \hspace*{2.4cm}
    +
    \left(
    g_{LV(q)} \bar{e} \gamma^{\mu} \PL \mu
    + g_{RV(q)} \bar{e} \gamma^{\mu} \PR \mu
    \right) \bar{q} \gamma_{\mu} q \nonumber \\
    &&   \hspace*{2.4cm}
    +
    \left(
    g_{LA(q)} \bar{e} \gamma^{\mu} \PL \mu
    + g_{RA(q)} \bar{e} \gamma^{\mu} \PR \mu
    \right) \bar{q} \gamma_{\mu} \gamma_5 q \nonumber \\
    &&   \hspace*{2.4cm}
    + \ \frac{1}{2}
    \left(
    g_{LT(q)} \bar{e} \sigma^{\mu \nu} \PR \mu
    + g_{RT(q)} \bar{e} \sigma^{\mu \nu} \PL \mu
    \right) \bar{q} \sigma_{\mu \nu} q
	+ {\rm h.c.}
    \left. {\rule[-3mm]{0mm}{10mm}\ } \right]\,,\nonumber
\end{eqnarray}
where we have adapted the sign in front of the photon field to our
definition of the gauge covariant derivative.
In the MRSSM not all of the 4-fermion form factors are relevant. Like in the MSSM, the effective 4-fermion interaction is mainly generated by photon
penguin, $Z$-penguin, and box diagrams. These only give rise to the vector-like form factors
(up to terms suppressed by additional powers of Yukawa couplings). 

For an extensive discussion of the $Z$-penguins in the context of the MSSM, see Refs.~\cite{Arganda:2005ji,Krauss:2013gya,Abada:2014kba}.
Higgs-penguin diagrams have been discussed for the MSSM in Refs.\ \cite{Kitano:2003wn,Cirigliano:2009bz,Crivellin:2014cta}; they are negligible, except if the extra Higgs bosons are much lighter than the SUSY particles and if $\tan\beta$ is very large. Since we do not consider such a parameter scenario in the present paper, we neglect the Higgs penguins.

We write the generated
effective interaction Lagrangian excluding dipole contributions in the form of
Ref.\ \cite{Hisano:1995cp},
\begin{align}
  \label{eq:LHisano}
        {\cal{L}} ^{\text{Ref.\ \cite{Hisano:1995cp}}} _{\text{eff}} =&{} -e^2
        {\bar{e}}   \gamma _\alpha \left(\Acharg ^{{\TR}} \PR +
        \Acharg ^{{\TL}}  \PR \right)\mu\,
        \sum _{q=u,d} Q_q {\bar{q}} \gamma    ^\alpha q
        \nonumber\\ 
    &+\frac{g ^2 _Z}{m^2 _Z} \,
            {\bar{e}} \gamma ^\alpha
            \left(\AZ^{{\TR}} \PR + \AZ^{{\TL}} \PL\right)\mu \,
             \sum _{q=u,d} \frac{Z^\TR _{q} + Z^\TL_{q}}{2} {\bar{q}} \gamma _\alpha q
             \nonumber\\
    &+ e^2 \, {\bar{e}} \gamma ^{\alpha} ( \Abox ^{{\TR}}{} _q \PR +
             \Abox ^{{\TL}}{} _ q \PL ) \mu \,
             \sum _{q=u,d} {\bar{q}} \gamma _\alpha q \,,
\end{align}
with coefficients for the $Z$-penguin and the box diagrams.
Comparing this effective Lagrangian and the  definition of the
dipole coefficients in Eq.\ (\ref{GammallgammaFV}) with
Eq.\ (\ref{LeffKitano}) gives the relations
\begin{align}
  \ALKitano^* &= -\frac{\sqrt{2}\,e}{8\, G_\text{F}}   A_2^{\bar{e}\mu \TL}\,,
  \\
  \ARKitano^* &= -\frac{\sqrt{2}\,e}{8\, G_{\text{F}}}   A_2^{\bar{e}\mu \TR}\,,
  \\
  g _{LV(q)} &=
  \frac{\sqrt{2}}{G_{\text{F}}}
  \Big(
  e ^2 Q _q \Acharg ^{\TL}
  - \frac{g ^2 _Z}{m^2 _Z}\left(\frac{Z ^\TR _{q} + Z ^\TL _{q}}{2}\right)\AZ^{\TL}
  - e ^2 \Abox ^{\TL}{} _{q}
  \Big)\,,
  \\
  g _{{RV(q)}} &=
  \frac{\sqrt{2}}{G_{\text{F}}}
  \Big(
  e ^2 Q _q \Acharg ^{{\TR}}
  - \frac{g ^2 _Z}{m^2 _Z}\left(\frac{Z ^\TR_{q} + Z ^\TL _{q}}{2}\right)\AZ^{\TR}
  - e ^2 \Abox ^{\TR}{} _{q}
  \Big)\,.
\end{align}
With these relations we can define coefficients for protons and
neutrons
\begin{align}
  \tilde{g}^{(p)}_{LV,RV} = & 2 g_{LV,RV(u)} + g_{LV,RV(d)}\,, \\
  \tilde{g}^{(n)}_{LV,RV} = & g_{LV,RV(u)} + 2 g_{LV,RV(d)}\,,
\end{align}
and obtain the $\mu\to e$ conversion rate and branching
ratio  as \cite{Kitano:2002mt}
\begin{align}
  \omega_{\text{conv}}& = 2 G_{\text{F}}^2 \left| \ARKitano^*
  D  + \tilde{g}^{(p)}_{LV} V^{(p)} + \tilde{g}^{(n)}_{LV}
  V^{(n)}\right|^2 + L \leftrightarrow R \,,
  \\
  B_{\mu N\to e N}&=\frac{\omega_{\text{conv}}}{\omega_{\text{capt}}} \,,
\end{align}
where the constants $D$, $V^{(p,n)}$ correspond to overlap integrals
evaluated in Ref.\ \cite{Kitano:2002mt}. We will use numerical values specific for aluminum, where $D=0.0357$,
$V^{(p)}=0.0159$, $V^{(n)}=0.0169$ in units of $m_\mu^{5/2}$.
The capture rate for aluminium is
$\omega_{\text{capt}}=0.7054\times10^6/s=4.643\times10^{-19}$~GeV
\cite{Kitano:2002mt,Suzuki:1987jf}. 

The up-to-date experimental upper limit is obtained for the gold nucleus from
the SINDRUM experiment \cite{Bertl:2006up},
\begin{align}
  B_{\mu \text{Au}\to e \text{Au}}&<7\times10^{-13}\ (90\%\text{
    C.L.})\,.
  \label{SINDRUMlimit}
\end{align}
In the forthcoming few years, the limit will be improved substantially
by the COMET \cite{Cui:2009zz,Kuno:2013mha} and Mu2E
\cite{Abrams:2012er} experiments. Both of these experiments will
measure $\mu\to e$ conversion in an aluminium nucleus. The foreseen
limits are $7.2\times10^{-15}$ for COMET Phase 1  and better than
$10^{-16}$ for COMET Phase II and for Mu2e.

\begin{figure}
  \begin{center}
  \begin{subfigure}[c]{.23\textwidth}
    \begin{center}
      \includegraphics[scale=.45]{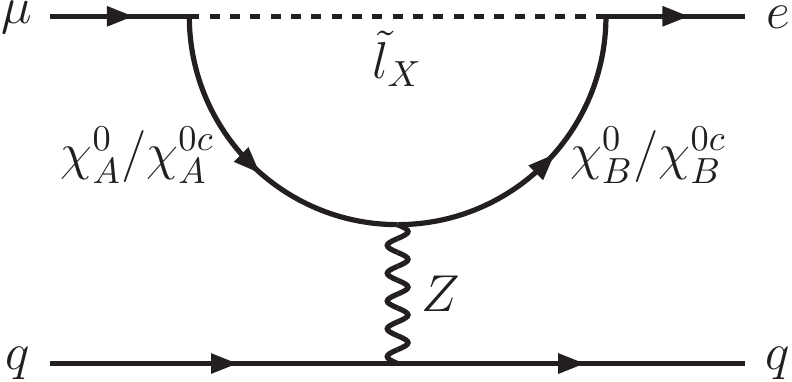}
    \end{center}
    \subcaption{Type 1 $\AZ ^{n\, \TL/\TR\, (1)}$}
    \label{fig:Zpenguin1}
  \end{subfigure}
  \,
   \begin{subfigure}[c]{.23\textwidth}
    \begin{center}
      \includegraphics[scale=.45]{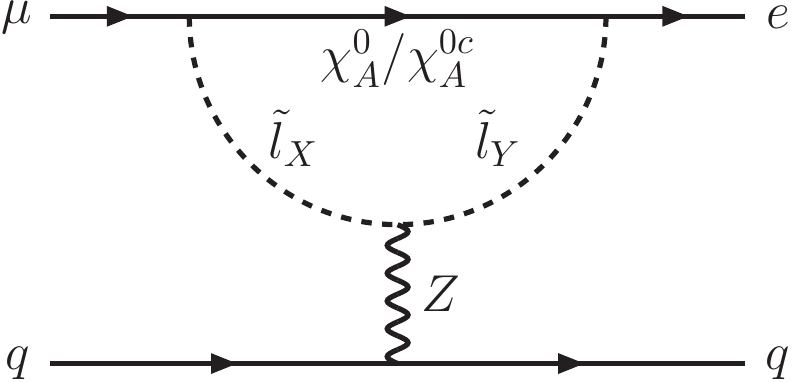}
    \end{center}
    \subcaption{Type 2 $\AZ ^{n\, \TL/\TR\,(2)}$}
    \label{fig:Zpenguin2}
   \end{subfigure}
   \,
 \begin{subfigure}[c]{.23\textwidth}
    \begin{center}
      \includegraphics[scale=.45]{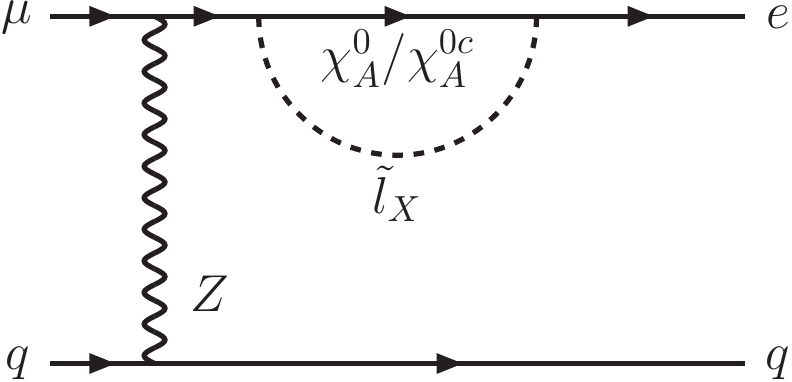}
    \end{center}
    \subcaption{Type 3 $\AZ ^{n\, \TL/\TR\, (3)}$}
    \label{fig:Zpenguin3}
 \end{subfigure}
 \,
 \begin{subfigure}[c]{.23\textwidth}
    \begin{center}
      \includegraphics[scale=.45]{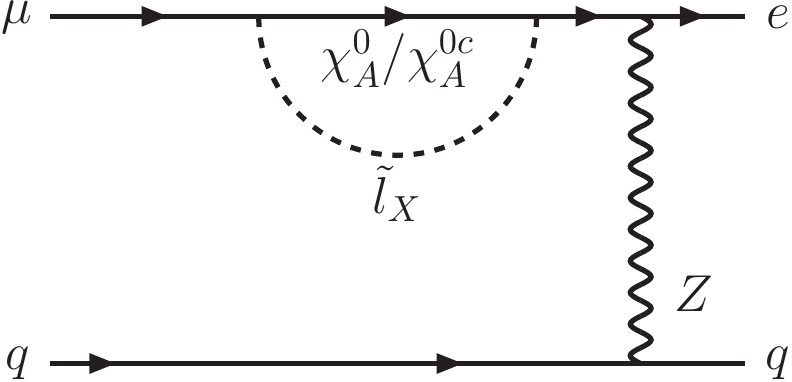}
    \end{center}
    \subcaption{Type 4 $\AZ ^{n\, \TL/\TR\, (4)}$}
    \label{fig:Zpenguin4}
 \end{subfigure}
 \end{center}
 \caption{The four types of $Z$-penguin diagrams corresponding to
   Eqs.\ (\ref{Zpenguin}). The chargino diagrams are analogous with the
   replacement of $\neut$ with
   $\charg$ and $\slep$ with $\tilde{\nu}$. }
 \label{fig:Zpenguin}
\end{figure}
We now turn to the actual one-loop results in the MRSSM.\footnote{%
  An implementation into FlexibleSUSY \cite{Athron:2014yba,Athron:2017fvs} is under
  development. The present paper uses a dedicated implementation into Mathematica.}
  The $Z$-penguin is given
by the diagrams in Fig.\ \ref{fig:Zpenguin}. The diagrams can be first
classified according to the exchanged particles: neutralinos and
right-handed sleptons with couplings $\cn$, antineutralinos and
left-handed sleptons with couplings $\co$, and $\chi$-charginos and
sneutrinos with couplings $\cc$. The diagrams can be further
classified into diagrams 
with $Z$ coupling to the neutralino/chargino (diagram type 1), to the
sfermion (type 2), and to the muon/electron (types 3, 4). Due to the
Ward-like identity corresponding to broken gauge invariance the
diagram types 2+3+4 exactly cancel in case of the (anti)neutralino
diagrams and partially cancel in case of the chargino diagrams. The
full contribution to the $Z$-penguin can thus be written as
\begin{subequations}
  \label{Zpenguin}
  \begin{align}
  \AZ^{{\TL}} =& \AZ^{n\TL (1)}+ \AZ ^{n\TL (2)}+\AZ ^{n\TL (3+4)}+\AZ
  ^{c\TL (1)}+\AZ ^{c \TL (2)}+\AZ ^{c \TL (3+4)} \,,
  \\
  \AZ^{{\TR}} =&\AZ^{n\TR (1)}+ \AZ ^{n\TR (2)}+\AZ ^{n\TR (3+4)}+\AZ
  ^{c\TR (1)}+\AZ ^{c \TR (2)}+\AZ ^{c \TR (3+4)} \,,
\end{align}
\end{subequations}
where the $Z$-penguin neutralino contributions  are
\begin{align}
  \AZ ^{n\TL (1)} =&
  \frac{1}{16 \pi ^2}
  \sum _{A, B, X} 
  \cn ^{\TL (l) \ast} _{1BX} \cn ^{\TL (l)} _{2AX}
  \Big\{
  -2 \fAZb (m ^2 _{\neut _{A}},m ^2 _{\neut _{B}},m ^2 _{{\slep _{X}}})
  \Big(
  \frac{-N^{2\ast}_{B3} N^{2}_{A3} + N^{2\ast} _{B4} N^{2} _{A4}}{2}
  \Big)
  \nonumber\\
  &+
  m_{\neut _{A}} m_{\neut _{B}}
  \fAZa (m ^2 _{\neut _{A}},m ^2 _{\neut _{B}},m ^2 _{\slep _{X}})
  \Big(
  \frac{-N ^{1} _{B3} N ^{1\ast} _{A3} + N ^{1} _{B4} N ^{1\ast} _{A4}}{2}
  \Big)
  \Big\}\nonumber\\
  &+
  \frac{1}{16 \pi ^2}
  \sum _{A, B, X}
  \co ^{\TL (l) \ast} _{1BX} \co ^{\TL (l)} _{2AX}
  \Big\{
  -2 \fAZb (m ^2 _{\neut _{A}},m ^2 _{\neut _{B}},m ^2 _{\slep _{X}})
  \Big(
  \frac{N^{1\ast}_{B3} N^{1}_{A3} - N^{1\ast} _{B4} N^{1} _{A4}}{2}
  \Big)\nonumber\\
  &+
  m_{\neut _{A}} m _{\neut _{B}}
  \fAZa (m ^2 _{\neut _{A}},m ^2 _{\neut _{B}},m ^2 _{\slep _{X}})
  \Big(
  \frac{N ^{2} _{B3} N ^{2\ast} _{A3} - N ^{2} _{B4} N ^{2\ast} _{A4}}{2}
  \Big)
  \Big\}
  \,,\\
  \AZ ^{n\TR (1)} =&
  \frac{1}{16 \pi ^2}
  \sum _{A, B, X}
  \cn ^{\TR (l) \ast} _{1BX} \cn ^{\TR (l)} _{2AX}
  \Big\{
  -2 \fAZb (m ^2 _{\neut _{A}},m ^2 _{\neut _{B}},m ^2 _{\slep _{X}})
  \Big(
  \frac{-N^{1}_{B3} N^{1\ast}_{A3} + N^{1} _{B4} N^{1\ast} _{A4}}{2}
  \Big)\nonumber\\
  &+
  m_{\neut _{A}} m _{\neut _{B}}
  \fAZa (m ^2 _{\neut _{A}},m ^2 _{\neut  _{B}},m ^2 _{\slep _{X}})
  \Big(
  \frac{-N ^{2\ast} _{B3} N ^{2} _{A3} + N ^{2\ast} _{B4} N ^{2} _{A4}}{2}
  \Big)
  \Big\}\nonumber\\
  &+
  \frac{1}{16 \pi ^2}
  \sum _{A, B, X}
  \co ^{\TR (l) \ast} _{1BX} \co ^{\TR (l)} _{2AX}
  \Big\{
  -2 \fAZb (m ^2 _{\neut _{A}},m ^2 _{\neut _{B}},m ^2 _{\slep _{X}})
  \Big(
  \frac{N^{2}_{B3} N^{2\ast}_{A3} - N^{2} _{B4} N^{2\ast} _{A4}}{2}
  \Big)\nonumber\\
  &+
  m_{\neut _{A}} m _{\neut _{B}}
  \fAZa (m ^2 _{\neut _{A}},m ^2 _{\neut _{B}},m ^2 _{\slep _{X}})
  \Big(
  \frac{N ^{1\ast} _{B3} N ^{1} _{A3} - N ^{1\ast} _{B4} N ^{1} _{A4}}{2}
  \Big)
  \Big\}\,,
\\
  \AZ ^{n\TL (2)} =&
  \frac{1}{16 \pi ^2}
  \sum _{A, X}
  \Big\{
  \big(
  \cn ^{\TL (l) \ast} _{1AY} \cn ^{\TL (l)} _{2AX}
  +
  \co ^{\TL (l) \ast} _{1AY} \co ^{\TL (l)} _{2AX}
  \big)
  \big(
    -2 \fAZb (m ^{2} _{\neut _{A}}, m ^2 _{\slep _{X}}, m ^{2} _{\slep _{Y}})
    \big)\nonumber\\
  &\times
  \Big[
    \sum _{g=1} ^{3} U ^{\slep} _{Y g} U ^{\slep \ast} _{X g} Z ^{\TL} _{l}
    +
    U ^{\slep} _{Y (g+3)} U ^{\slep \ast} _{X (g+3)} Z ^{\TR} _{l}
    \Big]
  \Big\}\,,\\
  \AZ ^{n\TR (2)} =&
  \frac{1}{16 \pi ^2}
  \sum _{A, X}
  \Big\{
  \big(
  \cn ^{\TR (l) \ast} _{1AY} \cn ^{\TR (l)} _{2AX}
  +
  \co ^{\TR (l) \ast} _{1AY} \co ^{\TR (l)} _{2AX}
  \big)
  \big(
    -2 \fAZb (m ^{2} _{\neut _{A}}, m ^2 _{\slep _{X}}, m ^{2} _{\slep _{Y}})
    \big]\nonumber\\
  &\times
  \Big[
    \sum _{g=1} ^{3} U ^{\slep} _{Y g} U ^{\slep \ast} _{X g} Z ^{\TL} _{l}
    +
    U ^{\slep} _{Y (g+3)} U ^{\slep \ast} _{X (g+3)} Z ^{\TR} _{l}
    \Big]
    \Big\}\,,
    \\
  \AZ ^{n\TL (3+4)} =&
  \frac{1}{16 \pi ^2}
  \sum _{A, X}
  \big(
  \cn ^{\TL (l) \ast} _{1AX} \cn ^{\TL (l)} _{2AX}
  +
  \co ^{\TL (l) \ast} _{1AX} \co ^{\TL (l)} _{2AX}
  \big)
  \fAZbf (m ^2 _{\neut _{A}}, m ^2 _{\slep _{X}})
  Z ^{\TL} _{l},\\
  \AZ ^{n\TR (3+4)} =&
  \frac{1}{16 \pi ^2}
  \sum _{A, X}
  \big(
  \cn ^{\TR (l) \ast} _{1AX} \cn ^{\TR (l)} _{2AX}
  +
  \co ^{\TR (l) \ast} _{1AX} \co ^{\TR (l)} _{2AX}
  \big)
  \fAZbf (m ^2 _{\neut _{A}}, m ^2 _{\slep _{X}})
  Z ^{\TR} _{l} \,,
\end{align}
where $Z ^{\TL} _l = -\frac{1}{2} + \sin^2\theta_W$ and $Z ^{\TR} _l
= \sin^2\theta_W$ as given in Eq.\ (\ref{ZLRcoefficients}).

The  $Z$-penguin chargino contributions are
given by
\begin{align}
  \AZ ^{c\TL (1)} =&
  \frac{1}{16 \pi ^2}
  \sum _{A, B, X}
  \cc ^{\TL (l) \ast} _{1BX} \cc ^{\TL (l)} _{2AX}
  \Big\{
  -2 \fAZb (m ^2 _{\charg _{A}}, m ^2 _{\charg _{B}}, m ^2 _{\sneu _{X}})
  \nonumber\\&\qquad\times
  \big\{-V ^{1\ast} _{B1} V ^{1} _{A1} -\frac{1}{2}V ^{1\ast} _{B2} V^{1} _{A2} + s ^2 _{\text{W}} \delta _{AB}\big\}\nonumber\\
  &+
  m_{\charg _{A}} m_{\charg _{B}}
  \fAZa (m ^2 _{\charg _{A}}, m ^2 _{\charg _{B}}, m ^2 _{\sneu _{X}})
  \big\{-U ^{1} _{B1} U ^{1\ast} _{A1} - \frac{1}{2} U ^{1} _{B2} U ^{1\ast} _{A2} + s ^2 _{\text{W}} \delta _{AB}\big\}
  \Big\} \,,\label{eq:F1cL}\\
  \AZ ^{c\TR (1)} =&
  \frac{1}{16 \pi ^2}
  \sum _{A, B, X}
  \cc ^{\TR (l) \ast} _{1BX} \cc ^{\TR (l)} _{2AX}
  \Big\{
  -2 \fAZb (m ^2 _{\charg _{A}}, m ^2 _{\charg _{B}}, m ^2 _{\sneu _{X}})
  \nonumber\\&\qquad\times
  \big\{-U ^{1} _{B1} U ^{1\ast} _{A1} -\frac{1}{2} U ^{1} _{B2} U^{1\ast} _{A2} + s ^2 _{\text{W}} \delta _{AB}\big\}\nonumber\\
  &+
  m_{\charg _{A}} m_{\charg _{B}}
  \fAZa (m ^2 _{\charg _{A}}, m ^2 _{\charg _{B}}, m ^2 _{\sneu _{X}})
  \big\{-V ^{1\ast} _{B1} V ^{1} _{A1} - \frac{1}{2} V ^{1\ast} _{B2} V ^{1} _{A2} + s ^2 _{\text{W}} \delta _{AB}\big\}
  \Big\} \,,\label{eq:F1cR}
  \\
  \AZ ^{c \TL (2)} =&
  \frac{1}{16 \pi ^2}
  \sum _{A, X, Y}
  \cc ^{\TL (l) \ast} _{1AY} \cc ^{\TL (l)} _{2AX}
  \big(
  -2 \fAZb (m ^2 _{\charg _{A}}, m ^2 _{\sneu _{X}}, m ^2 _{\sneu _{Y}})
  \big)\nonumber\\
  &\times
  \Big[
    \sum _{g=1} ^{3}
    U ^{\sneu} _{Y g} U ^{\sneu\ast} _{Xg} Z ^{\TL} _{\nu}
    +
    U ^{\sneu} _{Y (g+3)} U ^{\sneu\ast} _{X (g+3)} Z ^{\TR} _{\nu}
    \Big] \,,\\
  \AZ ^{c \TR (2)} =&
  \frac{1}{16 \pi ^2}
  \sum _{A, X, Y}
  \cc ^{\TR (l) \ast} _{1AY} \cc ^{\TR (l)} _{2AX}
  \big(
  -2 \fAZb (m ^2 _{\charg _{A}}, m ^2 _{\sneu _{X}}, m ^2 _{\sneu _{Y}})
  \big)\nonumber\\
  &\times
  \Big[
    \sum _{g=1} ^{3}
    U ^{\sneu} _{Y g} U ^{\sneu\ast} _{X g} Z ^{\TL} _{\nu}
    +
    U ^{\sneu} _{Y (g+3)} U ^{\sneu\ast} _{X (g+3)} Z ^{\TR} _{\nu}
    \Big] \,,
  \\
  \AZ ^{c \TL (3+4)} =&
  \frac{1}{16 \pi ^2}
  \sum _{A, X}
  \cc ^{\TL (l) \ast} _{1AX} \cc ^{\TL (l)} _{2AX}
  \big(
  \fAZbf (m ^2 _{\charg _{A}}, m ^2 _{\sneu _{X}}) Z ^{\TL} _{l}
  \big) \,,\\
  \AZ ^{c \TR (3+4)} =&
  \frac{1}{16 \pi ^2}
  \sum _{A, X}
  \cc ^{\TR (l) \ast} _{1AX} \cc ^{\TR (l)} _{2AX}
  \big(
  \fAZbf (m ^2 _{\charg _{A}}, m ^2 _{\sneu _{X}}) Z ^{\TR} _{l}
  \big) \,,
\end{align}
where again
$Z ^{\TL} _{l} = -\frac{1}{2} +\sin^2\theta_W$ and $Z ^{\TR} _{l}
=\sin^2\theta_W$, while
 $Z ^{\TL} _{\nu} = \frac{1}{2}$ and $Z ^{\TR} _{\nu} = 0$. 

The loop functions needed for the $Z$-penguin contributions are the ones
defined in Ref.\ \cite{Ilakovac:2012sh},
\begin{align}
  \fAZa (a,b,c) &= -\frac{1}{b-c} \Big(\frac{a\ln a - b \ln b}{a-b} - \frac{a\ln a - c \ln c}{a - c} \Big) \,,\\
  \fAZb (a,b,c) &=\frac{3}{8} - \frac{1}{4(b-c)} \Big( \frac{a ^2 \ln a - b ^2 \ln b}{a - b} - \frac{a ^2 \ln a - c^2 \ln c}{a - c} \Big)\,,\\
  \fAZbf (a,b) &= \frac{1}{2} - \frac{\ln b}{2} + \frac{ a^2 - b^2 + 2 a^2 (\ln b - \ln a )}{4(a-b)^2}\,.
\end{align}
The two loop functions $\fAZa (a,b,c)$ and $\fAZb (a,b,c)$ are totally
symmetric; the loop function $\fAZbf(a,b)$ is not symmetric.

\begin{figure}
  \begin{center}
    \begin{subfigure}{.23\textwidth}
      \begin{center}
        \includegraphics[scale=.45]{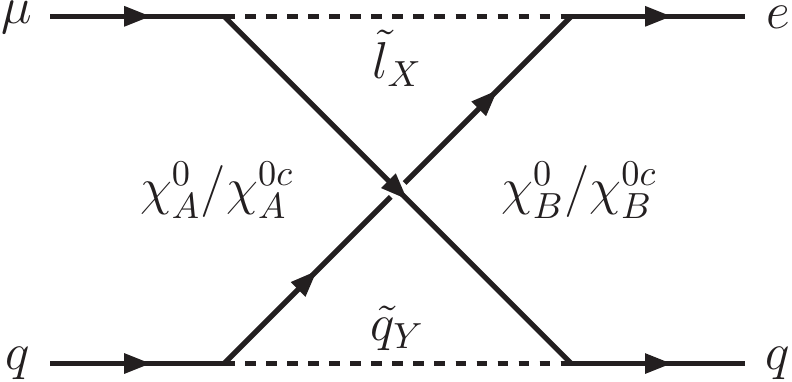}
      \end{center}
    \end{subfigure}
    \qquad
    \begin{subfigure}{.23\textwidth}
      \begin{center}
        \includegraphics[scale=.45]{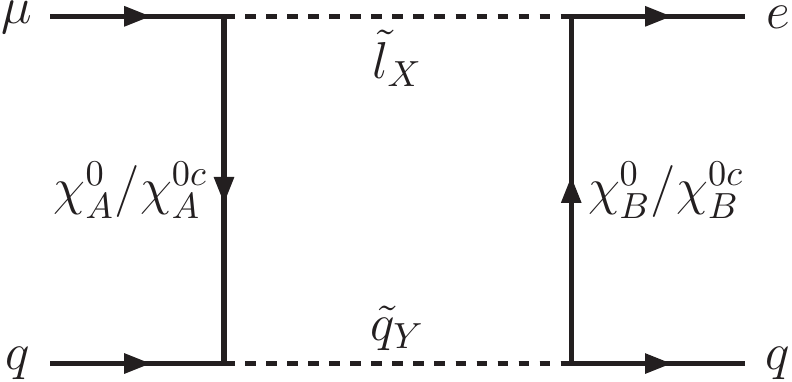}
      \end{center}
    \end{subfigure}
    \qquad
    \begin{subfigure}{.23\textwidth}
      \begin{center}
        \includegraphics[scale=.45]{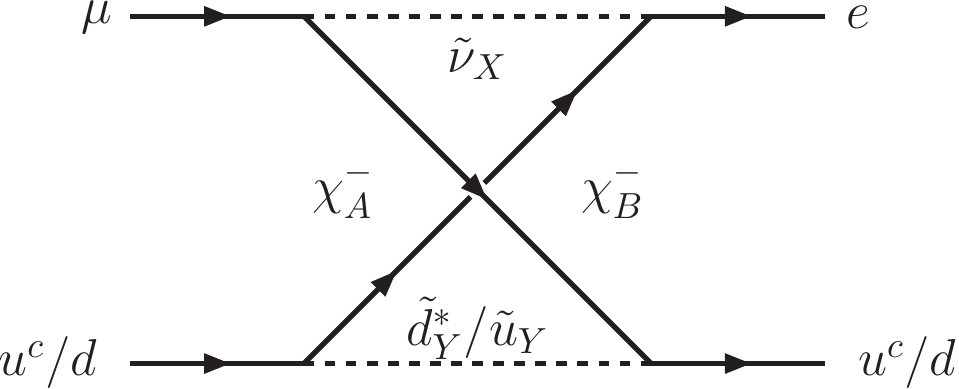}
      \end{center}
    \end{subfigure}
  \end{center}
  \caption{Box diagrams with neutralinos/antineutralinos and charginos. All indicated particles run in the direction of the arrows/from left to right.
    }
  \label{fig:boxdiagrams}
\end{figure}
The box diagrams are shown in Fig.\ \ref{fig:boxdiagrams}. The results
for the neutralino box diagrams are
\begin{align}
  {\Abox ^{n \TL}} _{q_g} &=
  \frac{1}{8 e ^2}
  \sum _{A, B, X, Y}
  \Big\{
  \fJfour (m_{\neut _{A}}, m_{\neut _{B}}, m _{\squark _{Y}}, m _{\slep _{X}})
  \nonumber\\
  &\times
  \Big(
  (\co ^{\TL (l) \ast} _{1BX} \co ^{\TL (l)} _{2AX}
  \co ^{\TL (q) \ast} _{g AY} \co ^{\TL (q)} _{g BY}
  +
  \cn ^{\TL (l) \ast} _{1BX} \cn ^{\TL (l)} _{2AX}
  \cn ^{\TL (q) \ast} _{g AY} \cn ^{\TL (q)} _{g BY})
  \nonumber\\
  &-
  (\co ^{\TL (l) \ast} _{1 BX} \co ^{\TL (l)} _{2 AX} 
  \cn ^{\TR (q) \ast} _{g AY} \cn ^{\TR (q)} _{g BY}
  +
  \cn ^{\TL (l) \ast} _{1 BX} \cn ^{\TL (l)} _{2 AX}
  \co ^{\TR (q) \ast} _{g AY} \co ^{\TR (q)} _{g BY})
  \Big)
  \Big\}
  \nonumber\\
  &-
  \frac{1}{4 e ^2}
  \sum _{A, B, X, Y}
  \Big\{
  m_{\neut _{A}} m_{\neut _{B}}
  \fIfour (m_{\neut _{A}}, m_{\neut _{B}}, m _{\squark _{Y}}, m _{\slep _{X}})
  \nonumber\\
  &\times
  \Big(
  (\co ^{\TL (l) \ast} _{1 BX} \co ^{\TL (l)} _{2 AX} 
  \co ^{\TR (q)} _{g AY} \co ^{\TR (q) \ast} _{g BY}
  +
  \cn ^{\TL (l) \ast} _{1 BX} \cn ^{\TL (l)} _{2 AX} 
  \cn ^{\TR (q) \ast} _{g AY} \cn ^{\TR (q)} _{g BY})
  \nonumber\\
  &-
  (\co ^{\TL (l) \ast} _{1 BX} \co ^{\TL (l)} _{2 AX} 
  \cn ^{\TL (q)} _{g AY} \cn ^{\TL (q) \ast} _{g BY}
  +
  \cn ^{\TL (l) \ast} _{1 BX} \cn ^{\TL (l)} _{2 AX} 
  \co ^{\TL (q)} _{g AY} \co ^{\TL (q) \ast} _{g BY})
  \Big)
  \Big\}\,,
\end{align}

\begin{align}
  {\Abox ^{n \TR}} _{q _g} &=
  \frac{1}{8 e ^2}
  \sum _{A, B, X, Y}
  \Big\{
  \fJfour (m_{\neut _{A}}, m_{\neut _{B}}, m _{\squark _{Y}}, m _{\slep _{X}})
  \nonumber\\
  &\times
  \Big(
  (\cn ^{\TR (l) \ast} _{1 BX} \cn ^{\TR (l)} _{2 AX}
  \cn ^{\TR (q) \ast} _{g AY} \cn ^{\TR (q)} _{g BY}
  +
  \co ^{\TR (l) \ast} _{1 BX} \co ^{\TR (l)} _{2 AX}
  \co ^{\TR (q) \ast} _{g AY} \co ^{\TR (q)} _{g BY})
  \nonumber\\
  &-
  (\cn ^{\TR (l) \ast} _{1 BX} \cn ^{\TR (l)} _{2 AX} 
  \co ^{\TL (q) \ast} _{g AY} \co ^{\TL (q)} _{g BY}
  +
  \co ^{\TR (l) \ast} _{1 BX} \co ^{\TR (l)} _{2 AX}
  \cn ^{\TL (q) \ast} _{g AY} \cn ^{\TL (q)} _{g BY})
  \Big)
  \Big\}
  \nonumber\\
  &-
  \frac{1}{4 e ^2}
  \sum _{A, B, X, Y}
  \Big\{
  m_{\neut _{A}} m_{\neut _{B}}
  \fIfour (m_{\neut _{A}}, m_{\neut _{B}}, m _{\squark _{Y}}, m _{\slep _{X}})
  \nonumber\\
  &\times
  \Big(
  (\co ^{\TR (l) \ast} _{1 BX} \co ^{\TR (l)} _{2 AX}
  \co ^{\TL (q)} _{g AY} \co ^{\TL (q) \ast} _{g BY}
  +
  \cn ^{\TR (l) \ast} _{1 BX} \cn ^{\TR (l)} _{2 AX} 
  \cn ^{\TL (q) \ast} _{g AY} \cn ^{\TL (q)} _{g BY})
  \nonumber\\
  &-
  (\co ^{\TR (l) \ast} _{1 BX} \co ^{\TR (l)} _{2 AX}
  \cn ^{\TR (q)} _{g AY} \cn ^{\TR (q) \ast} _{g BY}
  +
  \cn ^{\TR (l) \ast} _{1 BX} \cn ^{\TR (l)} _{2 AX}
  \co ^{\TR (q)} _{g AY} \co ^{\TR (q) \ast} _{g BY})
  \Big)
  \Big\}\,.
\end{align}
The chargino box diagram contributions are
\begin{align}
  {\Abox ^{c \TL}} _{u} &=
  \frac{1}{e ^2}
  \sum _{A, B, X, Y}
  \Big\{
  -\frac{1}{8}
  \fJfour (m_{\charg _{A}}, m_{\charg _{B}}, m _{\tilde{d} _{Y}}, m _{\slep _{X}})
  \cc ^{\TL (l)\ast} _{1 BX} \cc ^{\TL (l)}_{2 AX} 
  \cc ^{\TL (u) \ast} _{1 AY} \cc ^{\TL (u)} _{1 BY}
  \nonumber\\
  &+
  \frac{1}{4}
  m_{\charg _{A}} m_{\charg _{B}}
  \fIfour (m_{\charg _{A}}, m_{\charg _{B}}, m _{\tilde{d} _{Y}}, m _{\slep _{X}})
  \cc ^{\TL (l)\ast} _{1 BX} \cc ^{\TL (l)} _{2 AX}
  \cc ^{\TR (u) \ast} _{1 AY} \cc ^{\TR (u)} _{1 BY}
  \Big\}\,,
\end{align}

\begin{align}
 {\Abox ^{c \TR}} _{u} &=
 \frac{1}{e ^2}
 \sum _{A, B, X, Y}
  \Big\{
  -\frac{1}{8}
  \fJfour (m_{\charg _{A}}, m_{\charg _{B}}, m _{\tilde{d} _{Y}}, m _{\slep _{X}})
  \cc ^{\TR (l)\ast} _{1 BX} \cc ^{\TR (l)}_{2 AX}
  \cc ^{\TR  (u) \ast} _{1 AY} \cc ^{\TR (u)} _{1 BY}
  \nonumber\\
  &+
  \frac{1}{4}
  m_{\charg _{A}} m_{\charg _{B}}
  \fIfour (m_{\charg _{A}}, m_{\charg _{B}}, m _{\tilde{d} _{Y}}, m _{\slep _{X}})
  \cc ^{\TR (l)\ast} _{1 BX} \cc ^{\TR (l)} _{2 AX}
  \cc ^{\TL (u) \ast} _{1 AY} \cc ^{\TL (u)} _{1 BY}
  \Big\}\,,
\end{align}

\begin{align}
  {\Abox ^{c \TL}} _{d} &=
  \frac{1}{e ^2}
  \sum _{A, B, X, Y} 
  \Big\{
  \frac{1}{8}
  \fJfour (m_{\charg _{A}}, m_{\charg _{B}}, m _{\tilde{u} _{Y}}, m _{\slep _{X}})
  \cc ^{\TL (l)\ast} _{1 BX} \cc ^{\TL (l)}_{2 AX}
  \cc ^{\TL (d) \ast} _{1 AY} \cc ^{\TL (d)} _{1 BY}
  \nonumber\\
  &-
  \frac{1}{4}
  m_{\charg _{A}} m_{\charg _{B}}
  \fIfour (m_{\charg _{A}}, m_{\charg _{B}}, m _{\tilde{u} _{Y}}, m _{\slep _{X}})
  \cc ^{\TL (l)\ast} _{1 BX} \cc ^{\TL (l)} _{2 AX} 
  \cc ^{\TR (d) \ast} _{1 AY} \cc ^{\TR (d)} _{1 BY}
  \Big\}\,,
\end{align}

\begin{align}
  {\Abox ^{c \TR}} _{d} &=
  \frac{1}{e ^2}
  \sum _{A, B, X, Y}
  \Big\{
  \frac{1}{8}
  \fJfour (m_{\charg _{A}}, m_{\charg _{B}}, m _{\tilde{u} _{Y}}, m _{\slep _{X}})
  \cc ^{\TR (l)\ast} _{1 BX} \cc ^{\TR (l)}_{2 AX} 
  \cc ^{\TR (d) \ast} _{1 AY} \cc ^{\TR (d)} _{1 BY}
  \nonumber\\
  &-
  \frac{1}{4}
  m_{\charg _{A}} m_{\charg _{B}}
  \fIfour (m_{\charg _{A}}, m_{\charg _{B}}, m _{\tilde{u} _{Y}}, m _{\slep _{X}})
  \cc ^{\TR (l)\ast} _{1 BX} \cc ^{\TR (l)} _{2 AX} 
  \cc ^{\TL (d) \ast} _{1 AY} \cc ^{\TL (d)} _{1 BY}
  \Big\}\,.
\end{align}
Here the following loop functions appear:
\begin{align}
  \fIfour (a,b,c,d) &=
  \frac{1}{16 \pi ^2}
  \Big\{
  \frac{a^2 \ln\frac{a^2}{d^2}}{(a^2 - b^2)(a^2 - c^2)(d^2 - a^2)}
  -\frac{b^2 \ln\frac{b^2}{d^2}}{(a^2 - b^2)(b^2 - c^2)(d^2 - b^2)}
  \nonumber\\
  &+\frac{c^2 \ln\frac{c^2}{d^2}}{(a^2 - c^2)(b^2 - c^2)(d^2 - c^2)}
  \Big\}\,,
\end{align}

\begin{align}
  \fJfour (a,b,c,d) &=
  \frac{1}{16 \pi ^2}
  \Big\{
  \frac{a^4\frac{a^2}{d^2}}{(a^2 - b^2)(a^2 - c^2)(d^2 - a^2)}
  -\frac{b^4\frac{b^2}{d^2}}{(a^2 - b^2)(b^2 - c^2)(d^2 - b^2)}
  \nonumber\\
  &+\frac{c^4\frac{c^2}{d^2}}{(a^2 - c^2)(b^2 - c^2)(d^2 - c^2)}
  \Big\}\,.
\end{align}

\newpage
\section{Numerical results}

\label{sec:numerics}

\subsection{Relevant parameters and experimental constraints}
\label{sec:parameters}

We begin our discussion of the numerical results with a survey of the
relevant MRSSM parameters, an overview of characteristic regions of
parameter space, and a discussion of applicable experimental
constraints on the parameters. The three observables $a_\mu$, $\mu\to
e\gamma$ and $\mu\to e$ depend on an increasing number of
parameters. $a_\mu$ depends only (up to small effects due to mixing)
on the masses of the gauginos and down-type Higgsinos and thus on the
parameters
\begin{align}
M_B^D,\, M_W^D,\, \mu_d,\, \lambda_d,\, \Lambda_d
\end{align}
as well as on the slepton mass parameters\footnote{we generally
abbreviate 
$m_{\tilde{l},ij}\equiv\sqrt{(m_{\tilde{l}}^2)_{ij}}$.}
\begin{align}
m_{\tilde{l},22},\,
 m_{\tilde{e},22}.
\end{align}
The dependence on the up-type Higgsino mass $\mu_u$ and on $\tan\beta$
is very weak, in contrast to the MSSM.

The observable $\mu\to e\gamma$ only depends on photon dipole
operators and has thus a similar parameter dependence as $a_\mu$, but
it involves slepton mass and mixing parameters of the first and second
generation.\footnote{We assume mixing with the third generation to be
absent, but this does not change the results in a substantial way.}
We use the common dimensionless LFV parameters
\begin{align}
  \delta^{\TL}_{12}&\equiv
  \frac{(m_{\tilde{l}} ^2)_{12}}{m_{\tilde{l},11}\,m_{\tilde{l},22}}
  &\delta^{\TR}_{12}&\equiv
  \frac{(m_{\tilde{e}} ^2)_{12}}{m_{\tilde{e},11}\,m_{\tilde{e},22}} \,.
  \label{Defdeltas}
\end{align}
Furthermore, we keep the ratio of the slepton masses around order 1 and set always the selectron masses to $1.5$ times the corresponding smuon 
masses,
\begin{align}
m_{\tilde{l},11} &=1.5\,m_{\tilde{l},22}\,,
&
m_{\tilde{e},11} &=1.5\,m_{\tilde{e},22}\,.
\label{selectronmasses}
\end{align}
This is not a significant restriction. We have explicitly checked that
the phenomenological results presented below remain essentially the
same if the factor $1.5$ is changed to a factor $1$ or anything of the
order 1. Indeed the LFV observables mainly
depend on the dimensionless parameters $\delta_{12}^{\TL,\TR}$, unless there is
a significant hierarchy between the different slepton masses, in which
case the observables are simply suppressed by the heavier mass scale.

The muon-to-electron conversion $\mu\to e$ depends on additional types
of diagrams and thus on additional parameters. The $Z$-penguin diagrams
have a significant dependence also on the up-type Higgsino parameters
and $\tan\beta$
\begin{align}
\mu_u,\, \lambda_u,\, \Lambda_u,\, \tan\beta,
\end{align}
and the box diagrams depend on the squark masses. For simplicity we
choose a common squark mass scale without squark flavour mixing,
\begin{align}
(m_{\tilde{q}}^2)_{ij} = (m_{\tilde{u}}^2)_{ij} = (m_{\tilde{d}}^2)_{ij}\equiv \delta_{ij} m_{\tilde{q}}^2\,.
\end{align}

The parameter space can be best explored by investigating various different
parameter scenarios, corresponding to 
distinct patterns of hierarchies between light/heavy SUSY
particles. In this way we can isolate several
parameter influences,  separate leading and subleading terms and
obtain a complete understanding of the parameter 
dependence. In the MSSM, corresponding parameter scenarios have been
defined e.g.\ in Refs.\ \cite{Moroi:1995yh,Cho:2011rk,DSreview,Fargnoli:2013zia} for
studies of $a_\mu$ and in Refs.\ \cite{Kersten:2014xaa,CalibbiShadmi} for
LFV observables.

All Feynman diagrams for the considered observables
involve the exchange of at least one neutralino/chargino
and one slepton/sneutrino, and all diagrams have a generic
$1/\msusy^2$ mass suppression. Hence we need at least two light SUSY
particles in order to have non-negligible results: at least one
neutralino/chargino and at least one slepton/sneutrino. In addition,
at least one light neutralino/chargino must be gaugino-like since
otherwise all contributions are suppressed by additional powers of
Yukawa couplings.

Hence we end up with seven distinct parameter scenarios with
characteristic mass hierarchies, which we denote
as BL, BR, WL, BHL, BHR, WHL, equal-mass. To be concrete, we define the following
patterns:
\begin{subequations}
\label{masspatterns}
\begin{align}
  \text{BL: }&&M_B^D=&\msusy,& m_{\tilde{l},{22}}=&(1\ldots1.5)\times\msusy,&
  \lambda_d,\,\delta^{\TL}_{12}=&\text{free},\\
  \text{BR: }&&M_B^D=&\msusy,& m_{\tilde{e},{22}}=&(1\ldots1.5)\times\msusy,&
  \lambda_d,\,\delta^{\TR}_{12}=&\text{free},\\
    \text{WL: }&&M_W^D=&\msusy,& m_{\tilde{l},{22}}=&(1\ldots1.5)\times\msusy,&
  \Lambda_d,\,\delta^{\TL}_{12}=&\text{free},\\
  \text{BHL: }&&M_B^D=\mu_d=&\msusy,& m_{\tilde{l},{22}}=&(1\ldots1.5)\times\msusy,&
  \lambda_d,\,\delta^{\TL}_{12}=&\text{free},\\
  \text{BHR: }&&M_B^D=\mu_d=&\msusy,& m_{\tilde{e},{22}}=&(1\ldots1.5)\times\msusy,&
  \lambda_d,\,\delta^{\TR}_{12}=&\text{free},\\
    \text{WHL: }&&M_W^D=\mu_d=&\msusy,& m_{\tilde{l},{22}}=&(1\ldots1.5)\times\msusy,&
    \Lambda_d,\,\delta^{\TL}_{12}=&\text{free},
    \end{align}
{as well as}
\begin{align}
\text{equal-mass:
    }&&M_B^D=M_W^D=\mu_d=&\msusy,&
    m_{\tilde{l},{22}}=m_{\tilde{e},{22}}=&(1\ldots1.5)\times\msusy,
    \nonumber\\&&\lambda_d,\,\Lambda_d,\,\delta^{\TL}_{12},\,\delta^{\TR}_{12}=&\text{free}.
\end{align}
\end{subequations}
In each scenario of Eq.\ (\ref{masspatterns}), all other masses not
listed in the corresponding equation  are set
to very high values (in 
practice we choose $50$~TeV). Unless noted otherwise, the other, not
listed 
$\lambda_i,\Lambda_i$ and $\delta^i_{12}$ are 
set to zero, and $\tan\beta$ is set to $\tan\beta=40$; a standard
value for the non-vanishing flavour
mixings  $\delta^{\TL,\TR}_{12}=10^{-4}$ is chosen. We also always set the
selectron masses as given in Eq.\ (\ref{selectronmasses}). 

Thus each scenario is then characterized by one common mass scale
$\msusy$.  The BL, BR, WL
regions contain only two light masses (one gaugino and one smuon). The parameter dependence in these scenario is
particularly simple as mainly gauge couplings enter. The three
further scenarios BHL, BHR, WHL with a light
Higgsino show a more interesting parameter
dependence; enhancements driven by $\lambda_d$ and $\Lambda_d$ become
possible. The equal-mass case contains 
interferences between different contributions.

Now we turn to constraints on the relevant parameters from existing
data. The SUSY masses are constrained by direct collider searches at LEP, Tevatron and
LHC. There is a multitude of analyses of
collider searches for different assumptions on the SUSY
spectrum. For our purposes, the most conservative and robust bounds
are the most important. They correspond to assuming small mass splittings
between SUSY particles, i.e.\ compressed SUSY
spectra such as the ones in Eq.\ (\ref{masspatterns}). As we will see,
these are the spectra which maximize the contributions to the
observables studied in the present paper. The bounds on the
relevant particle masses under the assumption of small mass splittings
are \cite{PDG}
\begin{align}
m_{\chi^\pm_1}&>92\text{ GeV}
&
m_{\tilde{\mu}_\TR}&>94\text{ GeV} \,.
\label{masslimits}
\end{align}
Since these limits originate from kinematics, they apply equally to
the MSSM and the MRSSM. For a recast of more model-dependent chargino limits to
the MRSSM see Ref.\ \cite{Diessner:2015iln}; this reference also
confirms that the limits become as weak as in Eq.\ (\ref{masslimits})
in case of compressed spectra.

The LHC limits from chargino and neutralino
searches do not apply for our case of very compressed chargino
and neutralino masses. 
However, limits from slepton searches are
relevant. Up-to-date limits from
Refs.\ \cite{PDG,Aaboud:2017leg,ATLAS-CONF-2019-008} show that certain
mass splittings $\Delta m(\tilde{l}_1,\chi^0_1)$ are allowed: e.g.\
for $m_{\chi^0_1}$ around $100$ GeV, mass 
splittings between below $1$ GeV and between $20 \ldots 60$ GeV are
allowed, while other mass splittings are excluded under
certain assumptions: mass degeneracy of all
sleptons of the first and second 
generation and $100\%$ branching
ratio of slepton decays into $\chi^0_1$, which are not fulfilled in our scenarios. As a result, our choice of
varying the slepton masses in the range $(1\ldots1.5)\times\msusy$ in
the scenarios (\ref{masspatterns}) is
subject to these exclusion limits, and we expect that a fraction of
data points in our plots might be excluded by LHC limits. However,  
the most interesting parameter points of each scenario are at the
lower borders of the mass ranges: they involve
particularly compressed spectra and are thus not excluded, and as we
will see later they 
provide the maximum results for the observables. Hence we do not
use these LHC slepton mass limits to constrain our forthcoming plots.

In the following we will only use the mass limits of Eq.\
(\ref{masslimits}) since these are the only ones relevant
for our goal of delineating the ranges of possible values of the
considered observables. But we note that under certain conditions,
e.g.\ if the lightest chargino is wino-like 
and if $m_{\chi^\pm_1}\gg m_{\tilde{l}_1}\gg m_{\chi^0_1}$, the limits
are far stronger.

Apart from the masses, the parameters $\lambda_{d,u}$ and
$\Lambda_{d,u}$ are important. As mentioned above, these parameters
are Yukawa-like superpotential parameters. Therefore, we always impose
the perturbativity constraint
\begin{align}
\label{lambdaconstraints1}
|\lambda_{u,d}|,\,|\Lambda_{u,d}|\le4\,.
\end{align}
Stronger limits can be obtained from phenomenology. In particular,
these parameters enter the MRSSM predictions for the lightest Higgs
boson mass and for  electroweak precision observables, particularly
via  the $T$-parameter \cite{Diessner:2014ksa,Bertuzzo:2014bwa}.
In particular, the $T$-parameter gets contributions
$\propto \Lambda_u^4 v_u^4+\Lambda_d^4 v_d^4$ and similarly for
$\lambda_{u,d}$ in the appropriate limit. Precise limits on the
$\Lambda$'s from electroweak precision observables, however, would
depend on the detailed spectrum in the stop/sbottom and other sectors,
which is not relevant for the considered observables. For this reason
we only use the approximate bounds
\begin{align}
\label{lambdaconstraints2}
|\Lambda_u|,|\lambda_u|\le2\,,
\end{align}
which guarantee that the $T$-parameter contributions from this sector
do not significantly overshoot the experimental limits. This limit
applies for $\tan\beta\gg1$. For $\tan\beta\lesssim1$, a similar limit
would apply to $\Lambda_d$ and $\lambda_d$, but we will not use such
small values of $\tan\beta$ in the present paper.

\subsection{Analysis of $a_\mu$ in the MRSSM}

\label{sec:amuresults}

In this subsection we present a detailed
analysis of the muon magnetic moment $a_\mu$. The contributing MRSSM
diagrams are shown in Fig.\ \ref{fig:muegDiagrams} and the analytical
results are given in Eqs.\ (\ref{eq:amuanalytical})--(\ref{eq:amuanalyticalcharg}). 
All Feynman diagrams involve the exchange of one neutralino/chargino
and one slepton/sneutrino, and all diagrams have a generic
$1/\msusy^2$ mass suppression. Hence we need at least two light SUSY
particles in order to have non-negligible $a_\mu$: at least one
neutralino/chargino and at least one slepton/sneutrino. At least one
light neutralino/chargino must be gaugino-like since otherwise all
contributions are suppressed by additional powers of Yukawa couplings.

As described above, it is instructive to distinguish several
distinct patterns of light/heavy SUSY particles, see e.g.\
Refs.\ \cite{Moroi:1995yh,Cho:2011rk,DSreview,Fargnoli:2013zia,Kersten:2014xaa,CalibbiShadmi}
for corresponding discussions in the MSSM.

The first three patterns in Eq.\ (\ref{masspatterns}) involve only two
light SUSY masses:
\begin{align}
  \text{BL: }&\text{light }M_B^D,\ m_{\tilde{l}}\,,\\
  \text{BR: }&\text{light }M_B^D,\ m_{\tilde{e}}\,,\\
  \text{WL: }&\text{light }M_W^D,\ m_{\tilde{l}}\,.
\end{align}
In these cases $a_\mu$ is essentially given by the $\fAdipac$- and
$\fAdipan$-terms in Eqs.\ (\ref{amuMRSSM}) and simply proportional to $g_{1,2}^2/\msusy^2$,
where $\msusy$ is the scale of the two light masses. $a_\mu$ in these
cases can be expected to be very small.

The other patterns involve three or more light SUSY masses:
\begin{align}
  \text{BHL: }&\text{light }M_B^D,\ \mu_d,\ m_{\tilde{l}}\,,\\
  \text{BHR: }&\text{light }M_B^D,\ \mu_d,\ m_{\tilde{e}}\,,\\
  \text{WHL: }&\text{light }M_W^D,\ \mu_d,\ m_{\tilde{l}}\,,\\
  \text{equal-mass: }&\text{light }M_B^D,\ M_W^D,\ \mu_d,\ m_{\tilde{l}},\ m_{\tilde{e}}\,.
\end{align}
In the cases with light Higgsinos, $a_\mu$ can be enhanced by
additional sources of chirality flips, similarly to the MSSM
\cite{Moroi:1995yh,DSreview}. However, the origin of the enhancement is quite
different from the MSSM. In the MSSM, a transition from
$d$-Higgsino to $u$-Higgsino is possible, governed by the MSSM
$\mu$-term. This leads to the well-known $\tan\beta$-enhancement of
$a_\mu$ in the MSSM. A well-established way to understand the
$\tan\beta$-enhancement is provided by mass-insertion diagrams
involving insertions of the $\mu$-parameter and Majorana gaugino
masses. Recently an extensive study has confirmed the high accuracy of
the mass-insertion method \cite{Crivellin:2018mqz}.

\begin{figure}
  \begin{center}
    \begin{subfigure}[c]{.23\textwidth}
      \begin{center}
        \includegraphics[scale=.45]{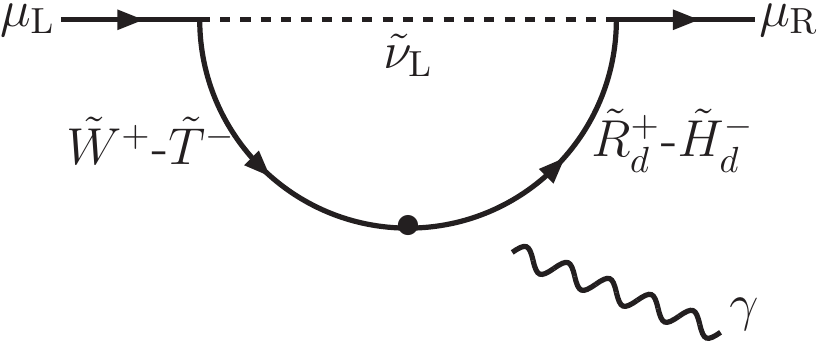}
      \end{center}
      \subcaption{$a _\mu ^{\text{MI}\text{ (c)}}$}
    \end{subfigure}
    \,
    \begin{subfigure}[c]{.23\textwidth}
      \begin{center}
        \includegraphics[scale=.45]{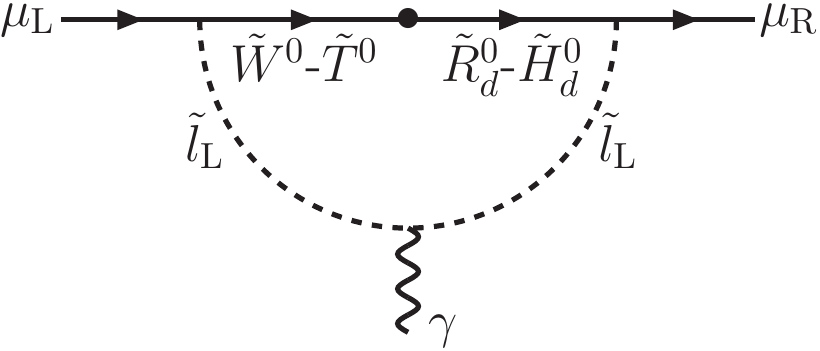}
      \end{center}
      \subcaption{$a _\mu ^{\text{MI}\text{ WHL}}$}
    \end{subfigure}
    \,
    \begin{subfigure}[c]{.23\textwidth}
      \begin{center}
        \includegraphics[scale=.45]{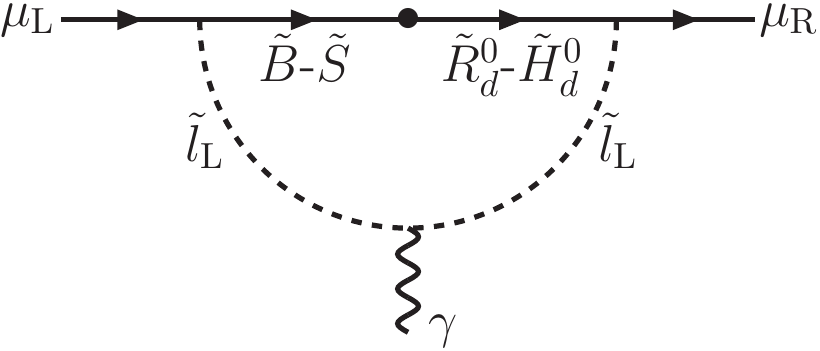}
      \end{center}
      \subcaption{$a _\mu ^{\text{MI}\text{ BHL}}$}
    \end{subfigure}
    \,
    \begin{subfigure}[c]{.23\textwidth}
      \begin{center}
        \includegraphics[scale=.45]{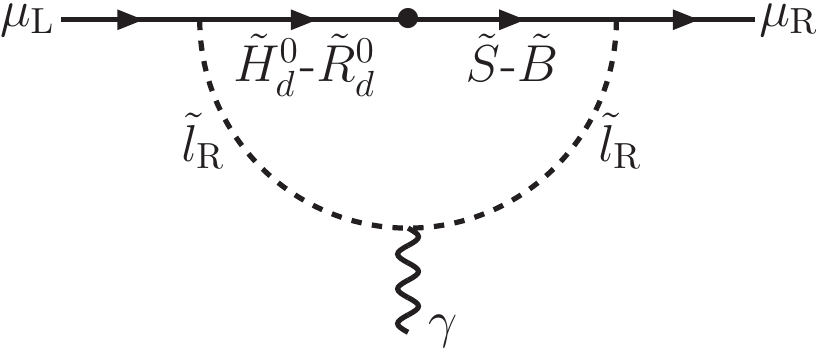}
      \end{center}
      \subcaption{$a _\mu ^{\text{MI}\text{ BHR}}$}
    \end{subfigure}
  \end{center}
  \caption{  \label{fig:gm2MIdiagrams}
Mass-insertion diagrams corresponding to Eq.\ (\ref{amuMI}). In these
    diagrams the charginos and neutralinos have definite compositions
    as bino--singlino, wino--triplino, Higgsino--R-Higgsino states,
    and the off-diagonal entries of the mass matrices (\ref{eq:cha1-massmatrix}),
    (\ref{eq:neut-massmatrix}) are inserted as vertices. In diagram
    (a), the photon can couple to all charged lines.}
\end{figure}

In contrast, the $\mu$-term and Majorana gaugino masses do not exist in
the MRSSM and consequently 
$a_\mu$ is not enhanced by $\tan\beta$. Instead, however, $a_\mu$ can
be enhanced by mass-insertion diagrams which involve a transition from
$d$-(R)Higgsino to singlino/triplino to bino/wino via the Yukawa-like
parameters $\lambda_d$ and $\Lambda_d$. Similarly to the MSSM, one can
approximate this effect using mass-insertion diagrams with one
insertion of the $\lambda_d$/$\Lambda_d$-entries in the
chargino/neutralino mass matrices. The results are the simple formulas
\begin{subequations}
\label{amuMI}
\begin{align}
  a _\mu ^{\text{MI}\, (c)}
  &=
  \frac{1}{8 \pi ^2}
  g _2
  \Lambda _d
  m _{\mu} ^{2}
  \frac{(g_2 v _T + M _W ^D) \mu ^{\text{eff},-} _d}{m _{\tilde{\nu}} ^4}
  \ffa\Big(\frac{(g_2 v _T + M _{W} ^{D}) ^2}{m _{\tilde{\nu}} ^2}, \frac{(\mu _d ^{\text{eff,}-}) ^2}{m _{\tilde{\nu}} ^2}
  \Big)\,,
  \\
  a _\mu ^{\text{MI}\text{ WHL}}
  &=
-  \frac{1}{16 \pi ^2}
  g _2
  \Lambda _d
  m _{\mu} ^{2}
  \frac{M_W ^D \mu _d ^{\text{eff},+}}{m _{\tilde{\mu} _\TL} ^4}
  \ffb\Big(\frac{(M _{W} ^{D}) ^2}{m _{\tilde{\mu} _\TL} ^2}, \frac{(\mu _d ^{\text{eff,}+}) ^2}{m _{\tilde{\mu} _\TL} ^2} \Big)\,,
  \\
  a _\mu ^{\text{MI}\text{ BHL}}
  &=  -\frac{1}{16 \pi ^2}
  g _1
\sqrt{2}  \lambda _d
  m _\mu ^2
  \frac{M _B ^D \mu _d ^{\text{eff},+}}{m _{\tilde{\mu} _\TL} ^4}
  \ffb\Big(\frac{(M _{B} ^{D}) ^2}{m _{\tilde{\mu}_\TL} ^2},\frac{(\mu _d ^{\text{eff,}+}) ^2}{m _{\tilde{\mu}_\TL} ^2} \Big)\,, 
  \\
  a _\mu ^{\text{MI}\text{ BHR}}
  &=
  \frac{1}{8 \pi ^2}
  g _1
\sqrt{2}  \lambda _d
  m _{\mu} ^{2}
  \frac{M_B ^D \mu _d ^{\text{eff},+}}{m _{\tilde{\mu} _\TR} ^4}
  \ffb\Big(\frac{(M _{B} ^{D}) ^2}{m _{\tilde{\mu} _\TR}
  ^2}, \frac{(\mu _d ^{\text{eff,}+}) ^2}{m _{\tilde{\mu} _\TR}
  ^2} \Big)\,,
\end{align}
\end{subequations}
which can be directly compared to the corresponding formulas in the
MSSM \cite{Moroi:1995yh}, quoted e.g.\ in Refs.\ \cite{Cho:2011rk,Fargnoli:2013zia} in
the present form.
The loop functions appearing here are defined as
\begin{subequations}
\begin{align}
\ffa(x, y) &= -\frac{G_{3}(x) - G_{3}(y)}{x - y}\,, \\
\ffb(x, y) &= -\frac{G_{4}(x) - G_{4}(y)}{x - y}\,,
\end{align}
\end{subequations}
with
\begin{subequations}
\begin{align}
G_{3} (x) &= \frac{1}{2(x-1)^{3}}\Big[ (x-1)(x-3) + 2\log x\Big]\,,\\
G_{4} (x) &= \frac{1}{2(x-1)^{3}}\Big[ (x-1)(x+1) - 2 x \log x \Big] \,,
\end{align}
\end{subequations}
with the normalization $\ffa(1,1)=1/4$, $\ffb(1,1)=1/12$.

The comparison to the MSSM immediately shows that the MRSSM results
for $a_\mu$ will be rather small. The  suppression can be roughly expressed as
\begin{align}
  a_\mu^{\text{MRSSM}}\sim
  a_\mu^{\text{MSSM}} \times\left(\frac{\Lambda_d}{g_2\,\tan\beta},\,
  \frac{\lambda_d}{g_1\,\tan\beta}\right)\,.
\end{align}
We recall that the $\lambda_d,\Lambda_d$-parameters are Yukawa-like
parameters; values of around unity are similar to the top-Yukawa
coupling, and we restrict them by Eqs.\
(\ref{lambdaconstraints1},\ref{lambdaconstraints2}) in 
view of perturbativity. Then the MRSSM
results for $a_\mu$ will be far
smaller than corresponding MSSM results for $\tan\beta=50$.

\begin{figure}[t]
  \centerline{\includegraphics[width=.45\textwidth]{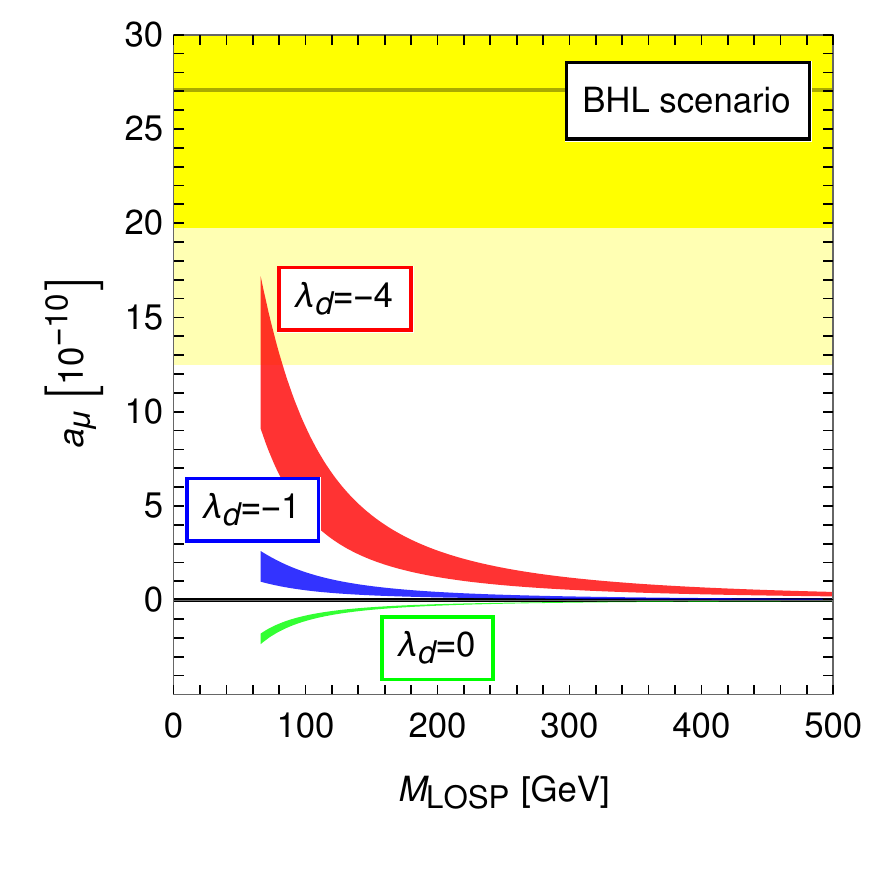}
  \qquad
  \includegraphics[width=.45\textwidth]{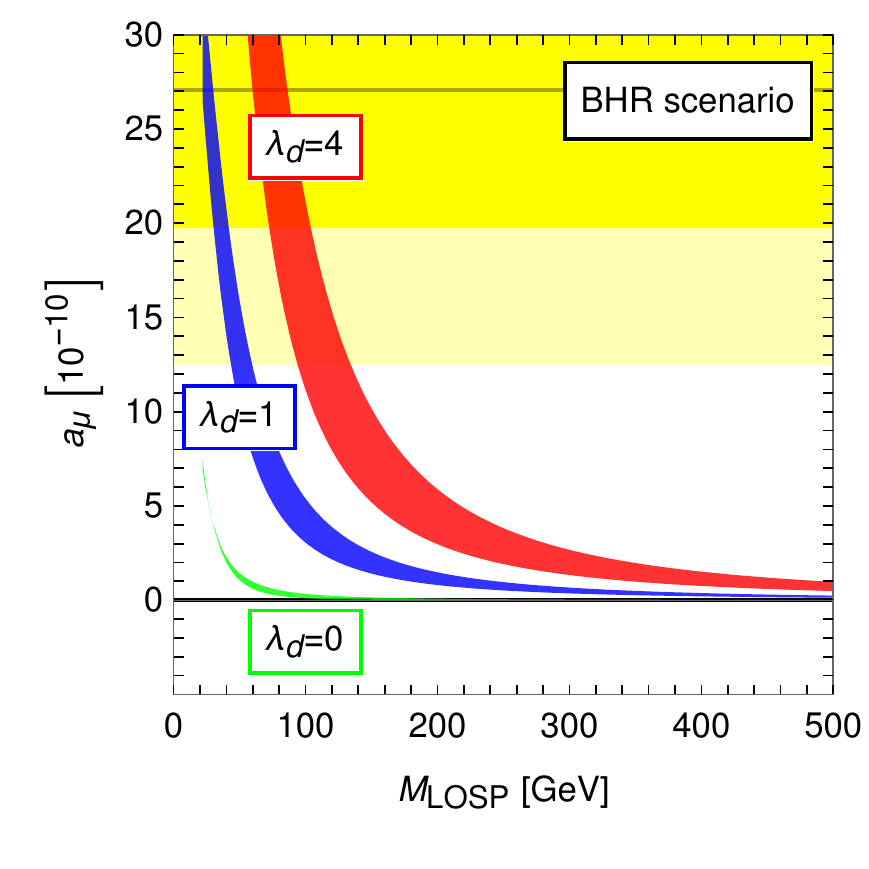}}
  \caption{\label{fig:BHLBHRamu}$a_\mu$ in the BHL and BHR
    scenarios. In each plot, the colours of the bands correspond to
    different values of $\lambda_d$ as indicated; the
    width of each band corresponds to a variation of the slepton
    masses by a factor $1.5$: the borders of each band correspond to
    the choices $m_{\tilde{l},\tilde{e},{22}}=\msusy$ or
    $=1.5\msusy$. The dark and light yellow horizontal bands
    correspond to the $1\sigma$ and $2\sigma$ bands given by Eq.\ (\ref{eq:Deltaamu}).}
\end{figure}
\begin{figure}[t]
         \centerline{\includegraphics[width=.45\textwidth]{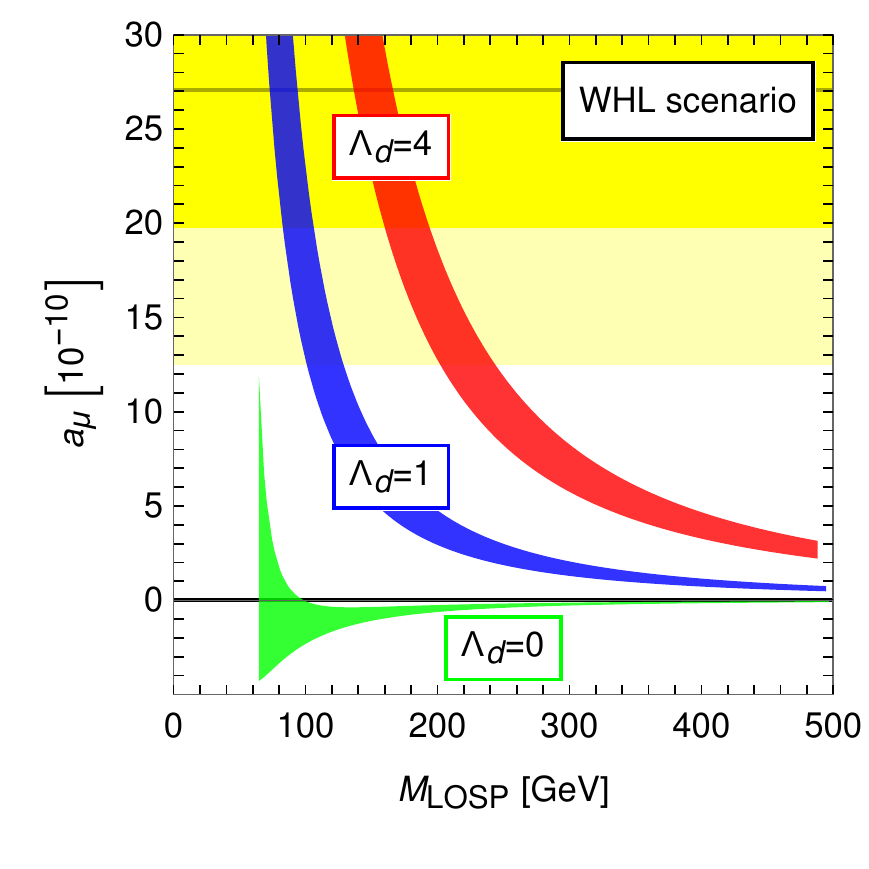}
         \qquad
         \includegraphics[width=.45\textwidth]{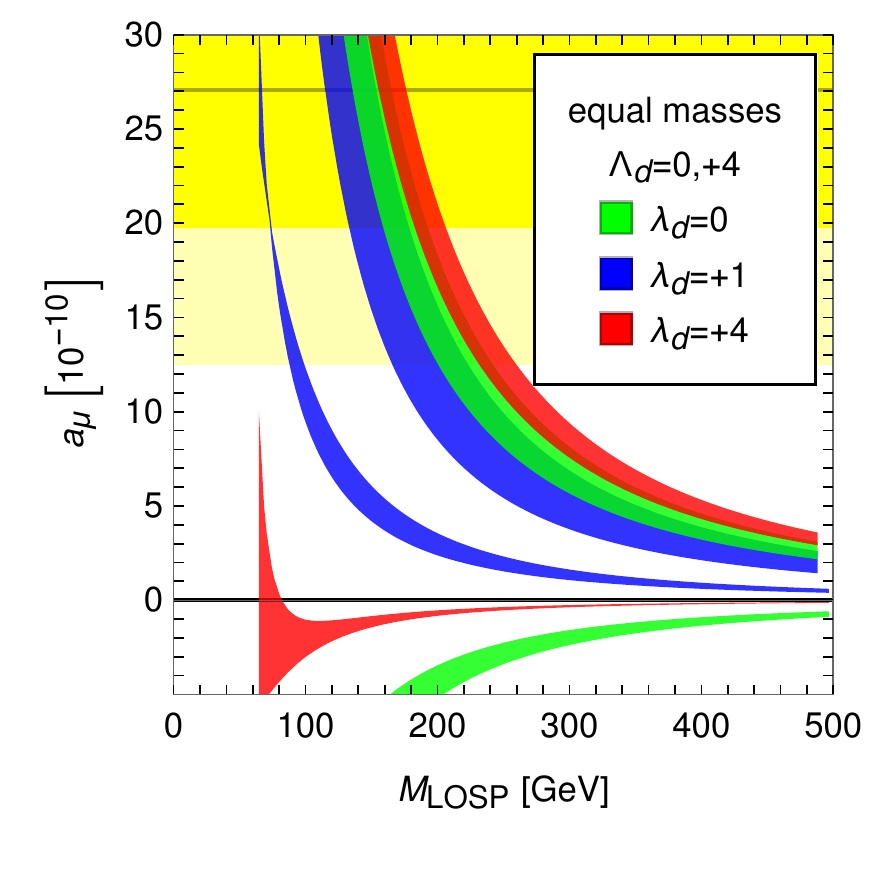}}
        \caption{\label{fig:WHLandallMS} As
          Fig.\ \ref{fig:BHLBHRamu}, but for the WHL and equal-mass
          scenarios; in the left plot, the values of $\Lambda_d$ are
        indicated; in the right plot, the topmost three bands correspond to
          $\Lambda_d=+4$, the lower three bands to
          $\Lambda_d=0$.  }
  \end{figure}

Figure \ref{fig:BHLBHRamu} and Figure\ \ref{fig:WHLandallMS} (left)
show the results for $a_\mu$  in 
the BHL, BHR, WHL scenarios. The results are shown as 
functions of the ``lightest observable SUSY mass'' $\mLOSP$, defined as
the minimum of the electrically charged SUSY particle masses. The
$\lambda_d,\Lambda_d$ parameters are varied, and their signs are
chosen such that $a_\mu$ is positive.

As expected from the result of the mass-insertion diagrams above,
$a_\mu$ is essentially proportional to $\lambda_d$ or $\Lambda_d$, as
appropriate. And as expected the results are
significantly smaller than the corresponding MSSM results for
large $\tan\beta$ due to the absence of a $\tan\beta$ enhancement. 
The largest contribution can be obtained from the WHL
scenario because of the larger SU(2) gauge coupling. In the blue
bands, $\Lambda_d=1$ (corresponding to a value similar to the
top-Yukawa coupling). Here
the current $a_\mu$ deviation can be explained for $\mLOSP$ around 100
GeV at the $2\sigma$ level; for $\Lambda_d=4$ (red bands), the current
deviation 
can be explained  up to $\mLOSP$ around 200~GeV. We do not consider
larger values of $\Lambda_d$ because of perturbativity.

The BHR contribution is slightly smaller than the WHL contribution,
and the BHL contribution is again smaller, because of the
smaller hypercharge. In the BHR case $a_\mu$ is just large enough for
a $1\sigma$ explanation of the current deviation if $\mLOSP$ is around
100~GeV and $\lambda_d=4$. In the BHL case this is impossible, and
$a_\mu$ reaches at most around $10\times10^{-10}$.

The remaining contribution for vanishing $\lambda$'s, $\lambda_d=\Lambda_d=0$ is
tiny; its magnitude is always below $2\times10^{-10}$ as long as
$\mLOSP>100$~GeV.

Figure \ref{fig:WHLandallMS}(right) shows $a_\mu$ in the
equal-mass scenario. We see that the value of $\Lambda_d$ is far more
important than the value of $\lambda_d$. The maximum is reached if
both $\Lambda_d$ and $\lambda_d$ are large, i.e.\ if the WHL and BHR
contributions add up constructively. For
$\Lambda_d=\lambda_d=4$, the current $a_\mu$ deviation can be
explained at the $1\sigma$ ($2\sigma$) level for $\mLOSP$ slightly
higher than 200~GeV (250~GeV). 

\begin{figure}[t]
\centerline{\includegraphics[width=.5\textwidth]{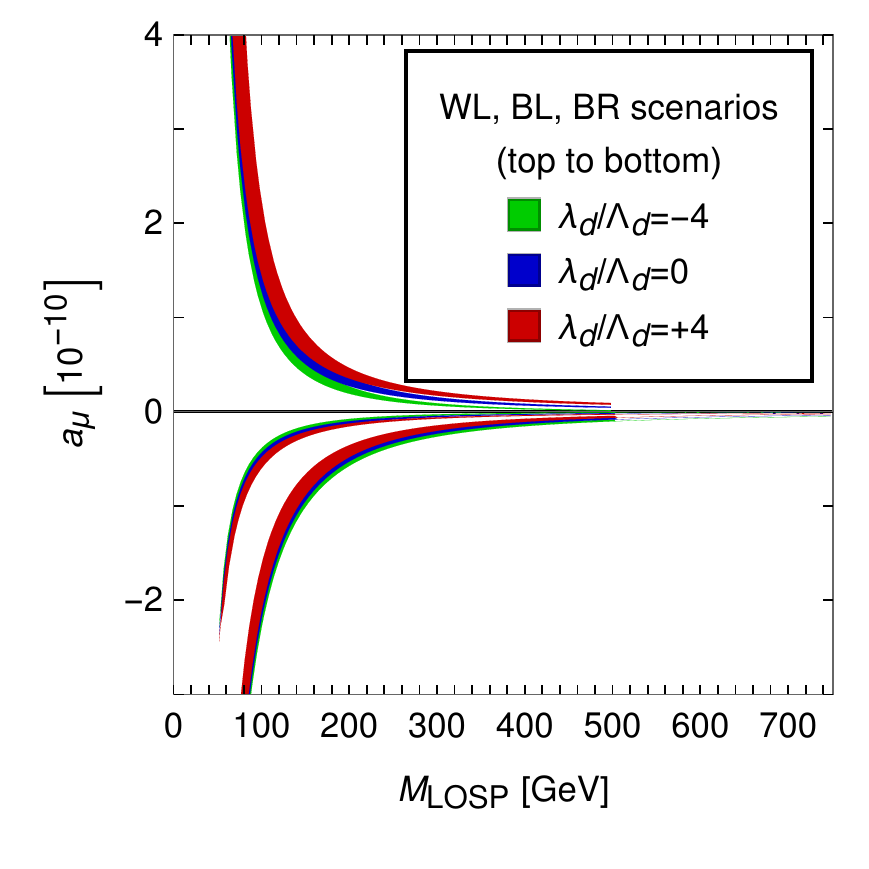}}
        \caption{\label{fig:BLBRWLamu}           $a_\mu$ in the WL (top three bands), BL
          (middle three bands),  BR (bottom three bands) scenarios. The
          three bands correspond to three different choices of
          $\lambda_d$, $\Lambda_d$, as appropriate; the width of the
          bands is defined as in Fig.\ \ref{fig:BHLBHRamu}.}
          \end{figure}

Figure \ref{fig:BLBRWLamu} shows $a_\mu$ in the BL, BR,
WL scenarios. In these scenarios no enhanced chirality-flips are
possible, and the $\lambda_d,\Lambda_d$-parameters play a minor role
via the chargino/neutralino mass matrices. Overall the values of
$a_\mu$ are tiny and almost entirely negligible.

The width of the bands in
Figs.\ \ref{fig:BHLBHRamu},\ref{fig:WHLandallMS},\ref{fig:BLBRWLamu}
corresponds to the variation of the slepton masses by a factor $1.5$,
see the definition of the scenarios in Eq.\ (\ref{masspatterns}). In
all cases, the maximum results for $a_\mu$ are obtained for the case
of fully degenerate spectra, while mass splittings tend to decrease
$a_\mu$. As 
discussed in sec.\ \ref{sec:parameters}, some parameter
points with certain non-vanishing mass splittings might be
excluded by LHC data; thus we see here that these exclusions cannot not
affect the upper 
borders of the bands and the overall maximum contributions for
$a_\mu$. Similar
comments will apply to the plots in the forthcoming subsections.

\subsection{Analysis of $\mu\to e \gamma$ in the MRSSM}

Now we turn to $\mu\to e \gamma$ in the MRSSM. Clearly there is a
strong similarity between $\mu\to e\gamma$ and $a_\mu$. Both are given
by dipole amplitudes; the difference is the existence of a
flavour transition in $\mu\to e\gamma$. The dependence on the LFV
parameters $\delta_{12}^{\TL,\TR}$ is very simple. Neglecting terms
suppressed by the electron mass, the amplitude $  A_2 ^{\bar{e}\mu \TL}$
is proportional to $\delta_{12}^\TL$ and  $  A_2 ^{\bar{e}\mu \TR}$
is proportional to $\delta_{12}^\TR$ to a very good approximation. Hence
we may write  $  A_2 ^{\bar{e}\mu \TL}\approx A_{2\text{ red}}
^{\bar{e}\mu \TL}\times\delta_{12}^\TL$  with a ``reduced'' amplitude $A_{2\text{ red}}
^{\bar{e}\mu \TL}$, and similarly for the
right-handed amplitude. Schematically we then have
\begin{align}
\label{mu2egbehaviour}
B_{\mu\to e\gamma} \propto
|  A_{2\text{ red}}
^{\bar{e}\mu \TL}|^2 \times|\delta_{12}^\TL|^2
+
|  A_{2\text{ red}}
^{\bar{e}\mu \TR}|^2 \times|\delta_{12}^\TR|^2 \,,
\end{align}
while $a_\mu$ can be expressed as
\begin{align}
a_\mu\propto
  A_{2}
^{\bar{\mu}\mu \TL}
+
  A_{2}
^{\bar{\mu}\mu \TR} \,,
\end{align}
with an obvious extension of the notation introduced in
Sec.\ \ref{sec:observables}.
These two equations specify the dependence on the $\delta$'s and the
relation between the two observables.

As a result, the analysis of the dependence on SUSY masses and
$\Lambda_d$, $\lambda_d$ of the previous subsection on $a_\mu$
carries over to
$\mu\to e \gamma$, and there is a strong correlation with $a_\mu$.

\begin{figure}[t]
\centerline{
\includegraphics[width=.45\textwidth]{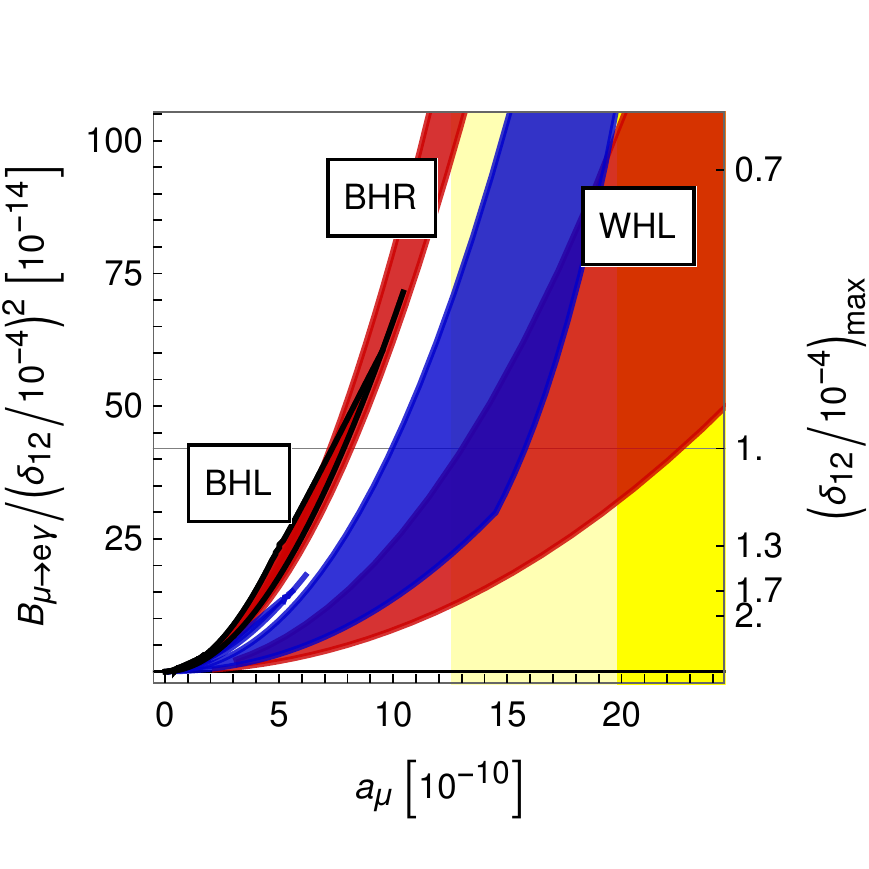}
\qquad \includegraphics[width=.45\textwidth]{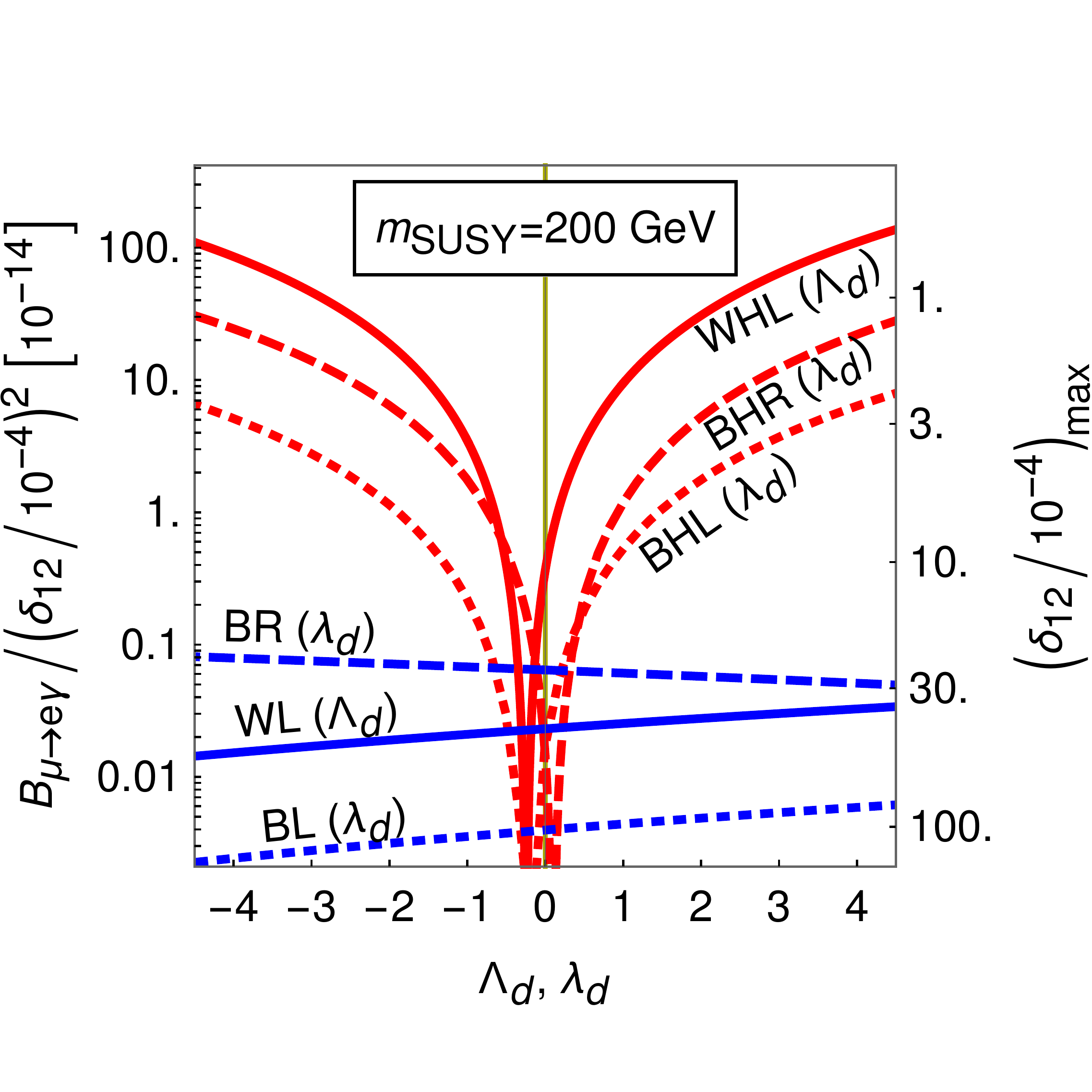}}
        \caption{\label{fig:mu2egLambdaVs}Left: Correlation between
        $a_\mu$ and $B_{\mu\to e\gamma}$ in the WHL, BHL, BHR
        scenarios, for $|\lambda_d|,|\Lambda_d|=1$ (blue) and $=4$
        (red). The signs for the $\Lambda_d,\lambda_d$ are chosen as
        in Figs.\ \ref{fig:BHLBHRamu}, \ref{fig:WHLandallMS}(left). The axis on the left shows
        $B_{\mu\to e\gamma} $ 
for the fixed value of the appropriate $\delta_{12}=10^{-4}$, the axis on
        the right shows the maximum $\delta_{12}$ allowed by the MEG limit
        (\ref{eq:mu2eglimit}),
        see text for details. The very small blue region corresponds
        to BHR ($\lambda_d=1$);
        the black edgy contour corresponds to the BHL region with $|\lambda _d|=4$;
        the BHL region with $|\lambda_d|=1$ is
        invisibly small.
        Right: Detailed dependence of $B_{\mu\to e\gamma}$ on
        $\Lambda_d,\lambda_d$ in the WHL, BHL, BHR, WL, BL, BR
        scenarios. In each scenario only one of the
        $\Lambda_d,\lambda_d$ and only one of the $\delta_{12}^{\TL}$,
        $\delta_{12}^{\TR}$ is nonzero, see text for details. The axes
        are as for the left plot.}
  \end{figure}

Figure \ref{fig:mu2egLambdaVs}(left) shows this correlation and displays
$\mu\to e\gamma$ as a function of $a_\mu$ for the three scenarios BHL,
BHR, WHL (see Eq.\ (\ref{masspatterns})) and for different choices for
$\Lambda_d$, $\lambda_d$ as indicated. In the BHL and WHL scenarios
only $\delta_{12}^\TL$ is nonzero while $\delta_{12}^\TR=0$, see Eq.\
(\ref{masspatterns}). In the BHR scenario only
$\delta_{12}^\TR$ is nonzero and $\delta_{12}^\TL=0$.
We see that in each scenario, $\mu\to e\gamma$ is essentially
proportional to $|a_\mu|^2$ as expected. The proportionality
coefficient depends on the case --- for fixed $a_\mu$, the BHL and BHR
scenarios give slightly larger $\mu\to e\gamma$. In each scenario the
correlation furthermore depends on $\Lambda_d$ or $\lambda_d$ (as
appropriate), and on the mass ratio between the smuon and the other
light masses. The borders of the regions are defined by taking the
respective smuon mass as either $\msusy$ or $1.5\times\msusy$ in Eq.\
(\ref{masspatterns}).

The result for $\mu\to e\gamma$ can be interpreted in two ways, as
indicated by the axis labels on the left and right border of the plot.
On the left border we indicate the value of $ B_{\mu\to e\gamma} $
for the fixed value of the appropriate $\delta_{12}=10^{-4}$. In all
scenarios 
$\mu\to e\gamma$ then varies in the range up to around $10^{-12}$ in the
considered range for $a_\mu$. On the other hand, in view of Eq.\
(\ref{mu2egbehaviour}) this allows to obtain $ B_{\mu\to e\gamma} $
for any other value of the appropriate $\delta_{12}$. Conversely, it
also allows to determine the value of the $\delta_{12}$, for which 
$ B_{\mu\to e\gamma} $ is equal to the MEG limit
(\ref{eq:mu2eglimit}). This is shown on the axis on the right border
of the plot. This axis corresponds to the maximum $\delta_{12}$'s allowed
for the corresponding points in the plot. We see that they are in the
range $(0.5\ldots2)\times10^{-4}$ if $a_\mu$ is in the $2\sigma$-band
around its experimental value.

In the plot we do not show the scenarios BL, BR, WL and
$\Lambda_d,\lambda_d=0$ since these cases lead to
tiny $a_\mu$. However the correlation would be of a similar kind. We
also do not show the equal-mass scenario since there destructive
interference between different amplitudes can happen, as will be
studied below. Overall, Fig.\ \ref{fig:mu2egLambdaVs}(left) is very similar
to the corresponding MSSM result shown in
Ref.\ \cite{Kersten:2014xaa}.

Figure \ref{fig:mu2egLambdaVs}(right) shows the more detailed
dependence of $\mu\to e\gamma$ in six scenarios, including the BL,
BR, WL scenarios. In each case, only one of the two $\delta_{12}$'s is
nonzero, as appropriate, see (\ref{masspatterns}): for BHR and BR,
only $\delta_{12}^{\TR}$ is 
nonzero, in the other cases only $\delta_{12}^{\TL}$. The plot then
shows $ B_{\mu\to e\gamma} $ as a function of either
$\Lambda_d$ (for WHL and WL) or $\lambda_d$ (all other cases).
The axis on the right border of the plot shows the respective maximum
allowed value of $\delta_{12}$ for the respective point in the plot,
computed as for Fig.\ \ref{fig:mu2egLambdaVs}(left).
All results in Fig.\ \ref{fig:mu2egLambdaVs}(right) are shown for the
fixed value of $\msusy=200$ GeV. The behaviour for other values of
$\msusy$ would be very similar; the branching ratio simply scales as
$1/\msusy^4$, and the maximum $\delta_{12}$'s scale as $\msusy^2$.

The results shown in the plot are similar to the corresponding
results for $a_\mu$: The scenarios without Higgsinos WL, BL, BR give
very small contributions, which depend on $\Lambda_d,\lambda_d$ only
in a minor way via the chargino/neutralino masses. The branching ratio
reaches the MEG limit in these scenarios for values of the
$\delta_{12}$'s 
around $(30\ldots100)\times10^{-4}$. Due to the scaling with
$\msusy^2$, this equivalently implies that e.g.\ values of
$\delta_{12}$ around $10\%$ are allowed if $\msusy$ is in the few TeV
range. 

The contributions in the scenarios with light Higgsino are
significantly enhanced by large $\Lambda_d$ and $\lambda_d$;
the amplitudes
are enhanced linearly, while the branching ratios are enhanced quadratically. Like for
$a_\mu$, the contributions can be largest in the WHL scenario,
followed by the BHR and BHL scenarios. For 
$\Lambda_d$, $\lambda_d$ bigger than around unity, the contributions in
these scenarios reach the MEG limit for around $\delta_{12}
\sim10^{-4}\ldots10^{-3}$.

The scenarios with light Higgsino show an interesting behaviour at
  small $\Lambda_d,\lambda_d$. If $\Lambda_d,\lambda_d=0$, only
  contributions governed by gauge couplings remain, and the branching
  ratio becomes similarly small as in the WL, BL, BR scenarios. For
  certain small but nonzero values of $\Lambda_d,\lambda_d$ the
  amplitudes pass through zero and the branching ratio vanishes.

  \begin{figure}[t]
    \centerline{
    \begin{tabular}{c}
    \includegraphics[width=.45\textwidth]{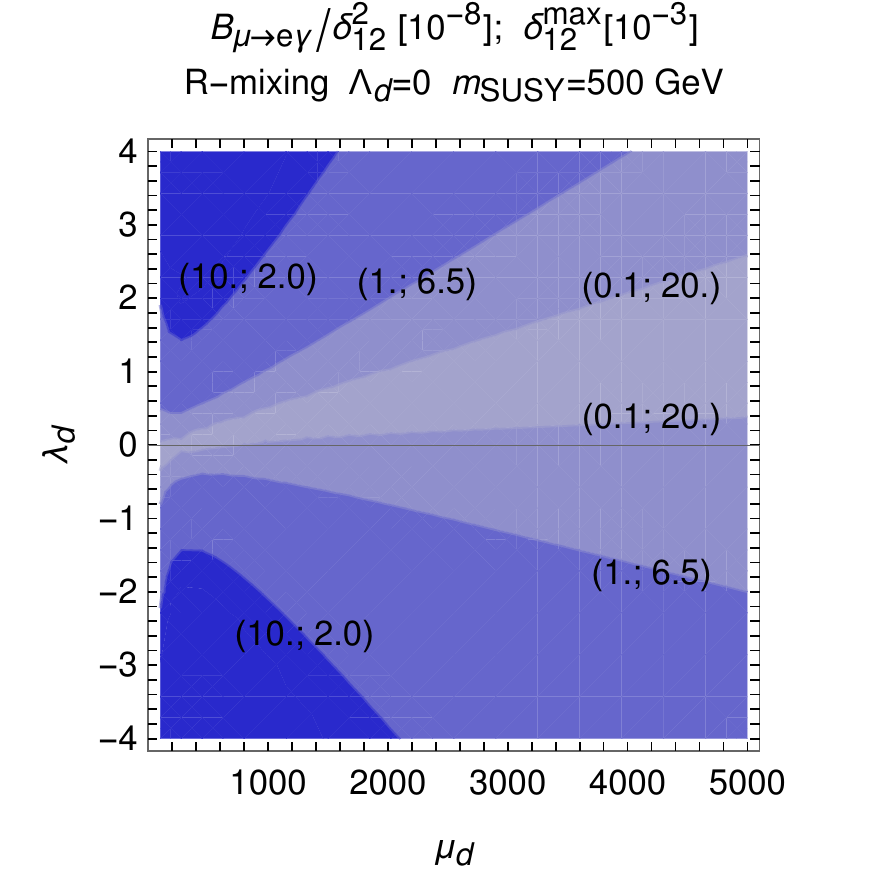}\\(a)
    \end{tabular}
    \qquad
    \begin{tabular}{c}
    \includegraphics[width=.45\textwidth]{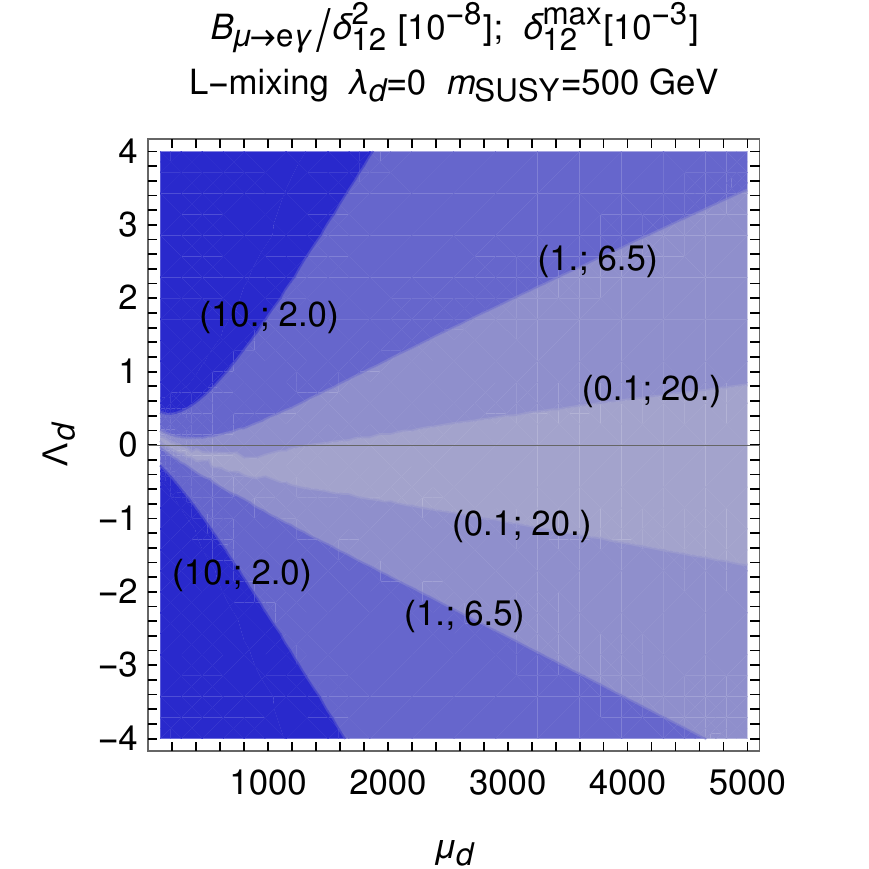}\\(b)
    \end{tabular}
    }
    \vspace{.5em}
    \centerline{
    \begin{tabular}{c}
    \includegraphics[width=.45\textwidth]{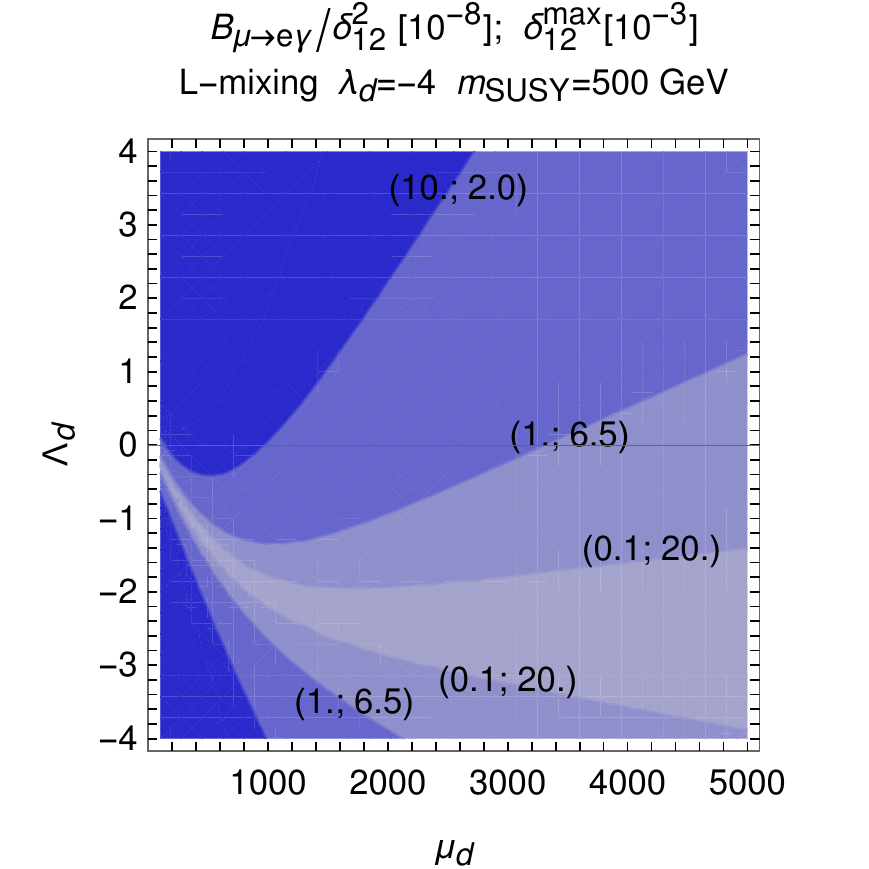}\\(c)
    \end{tabular}
    \qquad
    \begin{tabular}{c}
    \includegraphics[width=.45\textwidth]{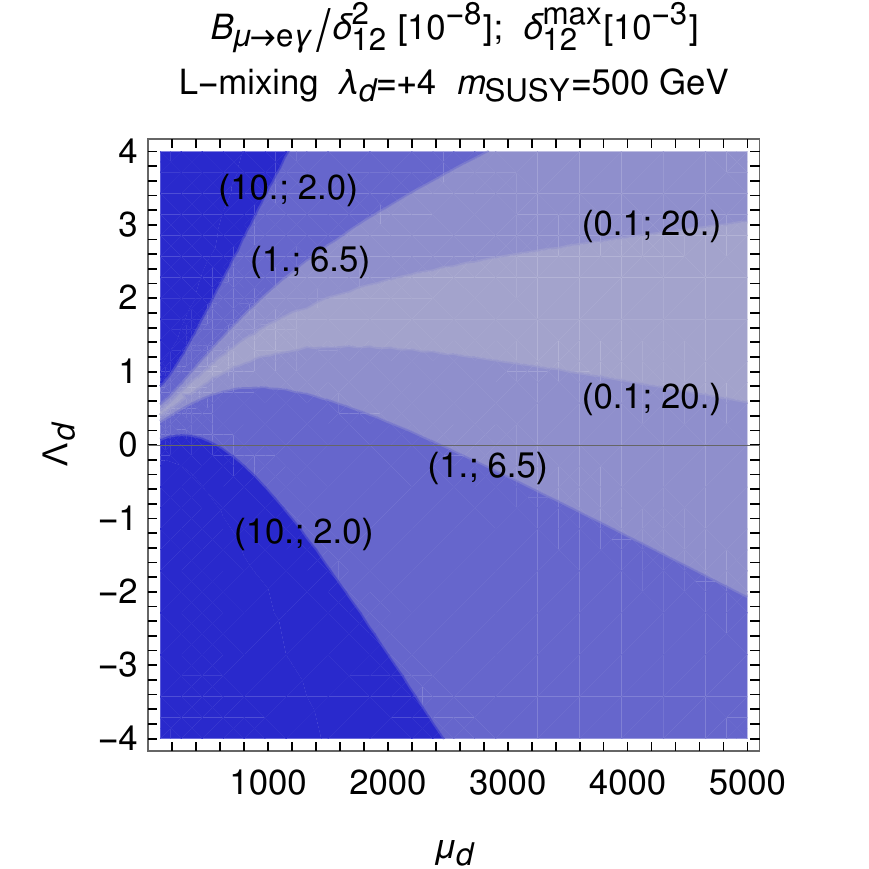}\\(d)
    \end{tabular}
    }
        \caption{\label{fig:mu2egContours}$\mu\to e\gamma$ in the
    equal-mass scenario, with $\mu_d$ kept as a free parameter. In
    plot (a) only $\delta_{12}^{\TR}$ is nonzero, while
    $\delta_{12}^{\TL}=0$; opposite in plots (b,c,d). The values of
    $\Lambda_d,\lambda_d$ are indicated in the plots. The contours
    correspond on the one hand to $B_{\mu\to e\gamma}/(\delta_{12})^2$
    and on the other hand to the maximum $\delta_{12}$ allowed by the
    MEG limit. See Eqs.\
    (\ref{interferenceapprox1},\ref{interferenceapprox2}) for an
    approximation of the behaviour.}
  \end{figure}

Figure \ref{fig:mu2egContours} shows the interference between
contributions with and without Higgsinos. We begin with describing
plot \ref{fig:mu2egContours}(a). Here only right-handed mixing
$\delta_{12}^{\TR}$ is nonzero, while $\delta_{12}^{\TL}=0$. We
consider the equal-mass scenario of Eq.\ (\ref{masspatterns}) with
$\msusy=500$ GeV, with the exception of the Higgsino mass $\mu_d$,
which is kept as a variable. The contour plot then shows $\mu\to
e\gamma$ as a function of $\mu_d$ and $\lambda_d$ for
$\Lambda_d=0$. Again, the contours are interpreted in two ways. On the
one hand they indicate $B_{\mu\to e\gamma}/(\delta_{12}^{\TR})^2$, on
the other hand they allow to read off the maximum $\delta_{12}^{\TR}$
allowed by the MEG limit.

The behaviour of Fig.\ \ref{fig:mu2egContours}(a) arises from
interference of BHR-type and BR-type contributions. For small $\mu_d$
and large $\lambda_d$, the BHR-type contributions dominate and show
the expected $\lambda_d$-enhancement. The corresponding parts of the
amplitude behave as $\lambda_d/\mu_d$. The BR-type
contributions are approximately independent of $\lambda_d$ and $\mu_d$.
 In fact, the behaviour in plot
(a) can be
well approximated by the simple fit
\begin{align}
\label{interferenceapprox1}
B_{\mu\to e\gamma}&\approx
\left(\frac{1500\text{ GeV}}{\mu_d}\lambda_d -0.4\right)^2
\times { 10^{-8}} \left(\delta_{12}^{\TR}\right)^2 \,,
\end{align}
at large $\mu_d$, exhibiting the two types of contributions.
This also allows to understand the triangular region with very small
$B_{\mu\to e\gamma}$ in which the $\lambda_d$-enhanced terms are
cancelled by the BR-type contributions.

Figures \ref{fig:mu2egContours}(b,c,d) are similar but for nonzero
left-handed
mixing $\delta_{12}^{\TL}$ and as a function of $\Lambda_d$
for different choices of $\lambda_d$. In plot (b) the behaviour is
similar to the one in plot (a) but it arises essentially from a
combination of WHL- and WL-type contributions; in plots (c,d) also the
BHL- and BL-type contributions matter and shift the contours according
to the choice of $\lambda_d$. The behaviour in these plots
at large $\mu_d$ can be
approximated by
\begin{align}
\label{interferenceapprox2}
B_{\mu\to e\gamma}&\approx
\left(\frac{1200\text{ GeV}}{\mu_d}
\left(\Lambda_d-\frac{\lambda_d}{2}\right)
+ 0.11\right)^2\times 10^{-8} \left(\delta_{12}^{\TL}\right)^2\,.
\end{align}

\subsection{Analysis of $\mu\to e$ conversion in the MRSSM}

The next observable we investigate is $\mu\to e$ conversion. This
observable has a much more complicated parameter dependence than the
previous observables. It depends on four types of form factors
$\Acharg$, $\Adip$, $\AZ$, $\Abox$, corresponding to the charge
radius, the dipole, the $Z$-penguin and box diagrams. The previous
observables only depended on dipole form factors.

The behaviour of the dipole form factor has been discussed
in the the previous subsections in detail, so here we focus
particularly on the 
additional parameter dependence arising from the new form factors. Not
all parameters lead to a nontrivial dependence. All form factors are
linear in the generation mixing parameters $\delta_{12}^{\TL,\TR}$ to
a very good 
approximation. Therefore we fix the $\delta_{12}$'s to the standard values
of Sec.\ \ref{sec:parameters},
i.e.\ to $10^{-4}$ or to zero, depending on the scenario. Furthermore,
all form factors are proportional to $1/\msusy^2$, where $\msusy$ is a
representative mass scale. Hence we also fix the overall mass scale to
$\msusy=500$ GeV for our discussion.

A very useful quantity to study is the ratio between the two branching
ratios for $\mu\to e$ and $\mu\to e\gamma$,
\begin{align}
\Rmue\equiv\frac{{B_{\mu \text{Al}\to e \text{Al}}}}{B_{\mu \to
e\gamma}}
\,.
\end{align}
In this ratio, the generic dependence on the $\delta$'s and the masses
drops out. Knowledge of $\Rmue$ also tells us the maximum possible
$\mu\to e$ conversion rate given the MEG limit on $\mu\to e\gamma$ for
any given parameter scenario. Interestingly, if $\mu\to e$ is 
dominated by the dipole form factors, then all model dependence, i.e.\
the full form factors $\Adip$ drop out in the ratio $\Rmue$ and there
is a perfect correlation. The correlation only depends on the element
used in the experiment;
for Aluminum the prediction for dipole dominance is \cite{Kitano:2002mt}
\begin{align}
\Rmuedipole&=0.0026\,.
\label{Rmuedipole}
\end{align}
Deviations of the actual result for $\Rmue$ from this prediction then
highlight the impact of the additional form factors $\Acharg$, $\AZ$,
$\Abox$.

\begin{figure}[t]
  \centerline{\includegraphics[width=.33\textwidth]{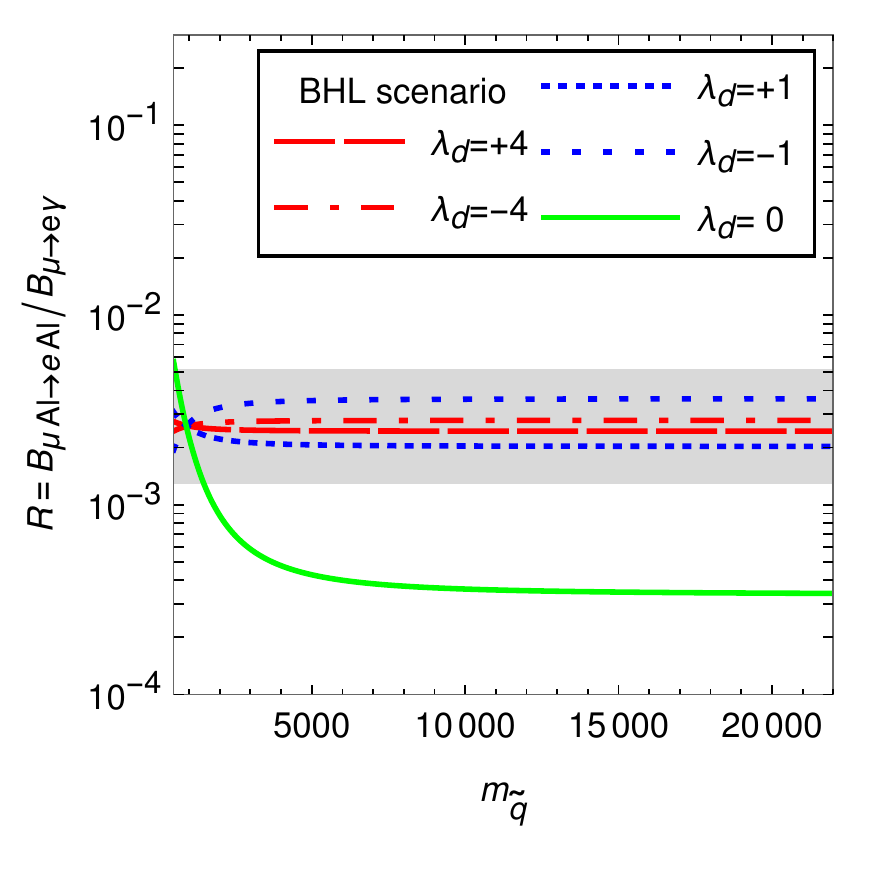}\includegraphics[width=.33\textwidth]{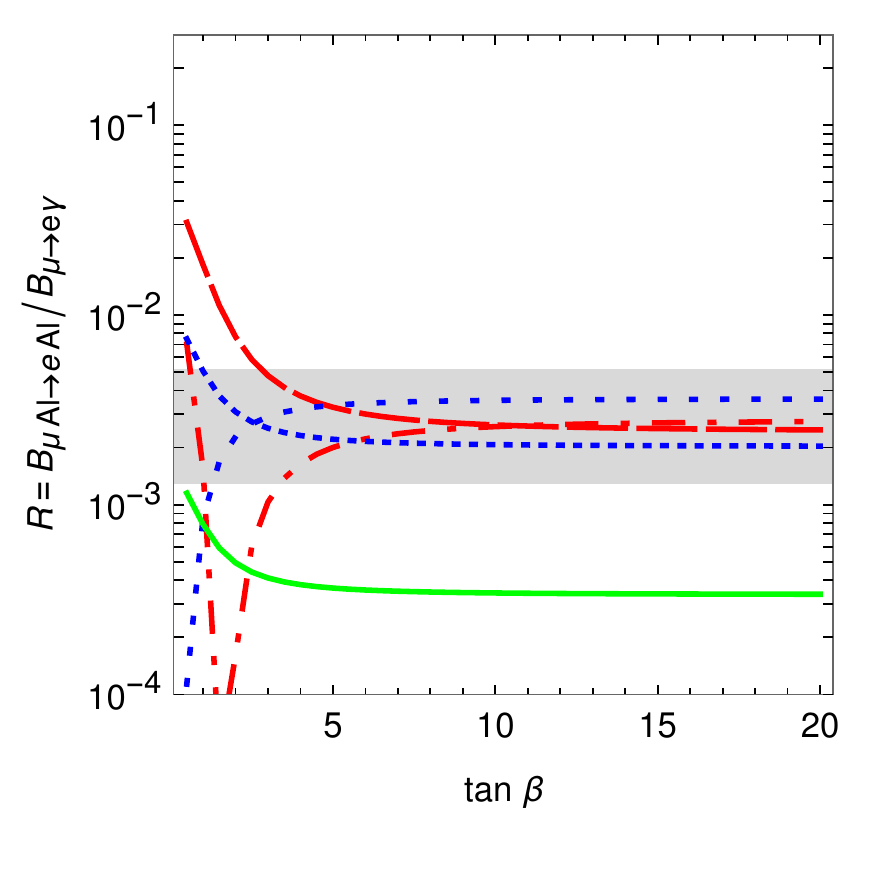}\includegraphics[width=.33\textwidth]{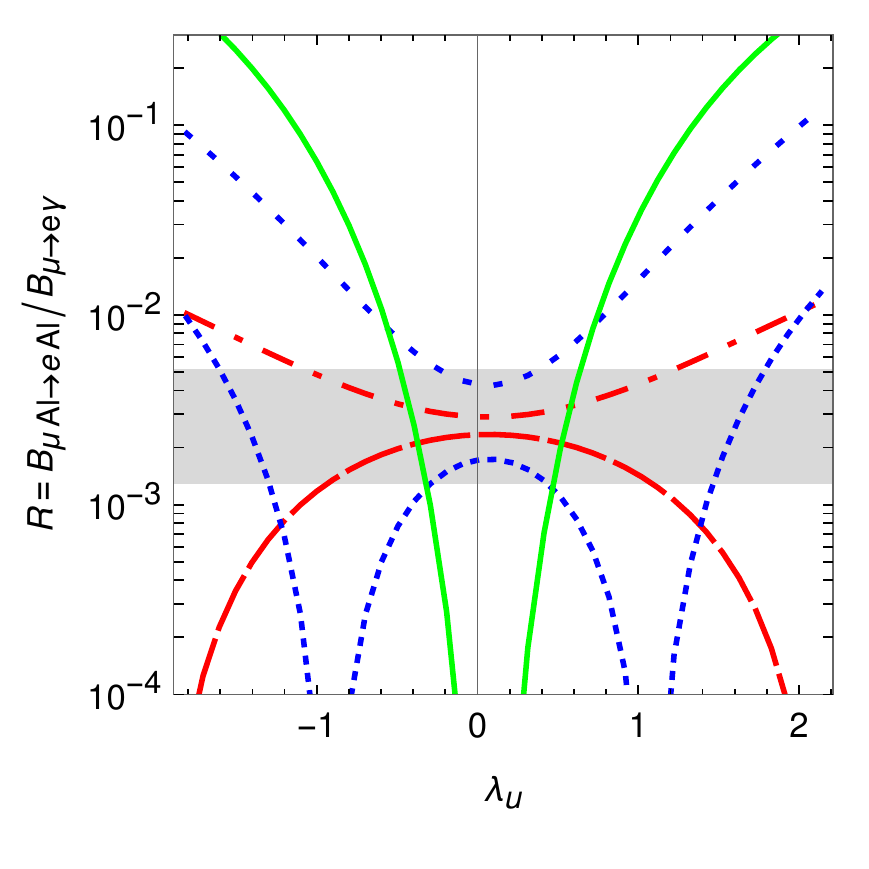}}
  \centerline{\includegraphics[width=.33\textwidth]{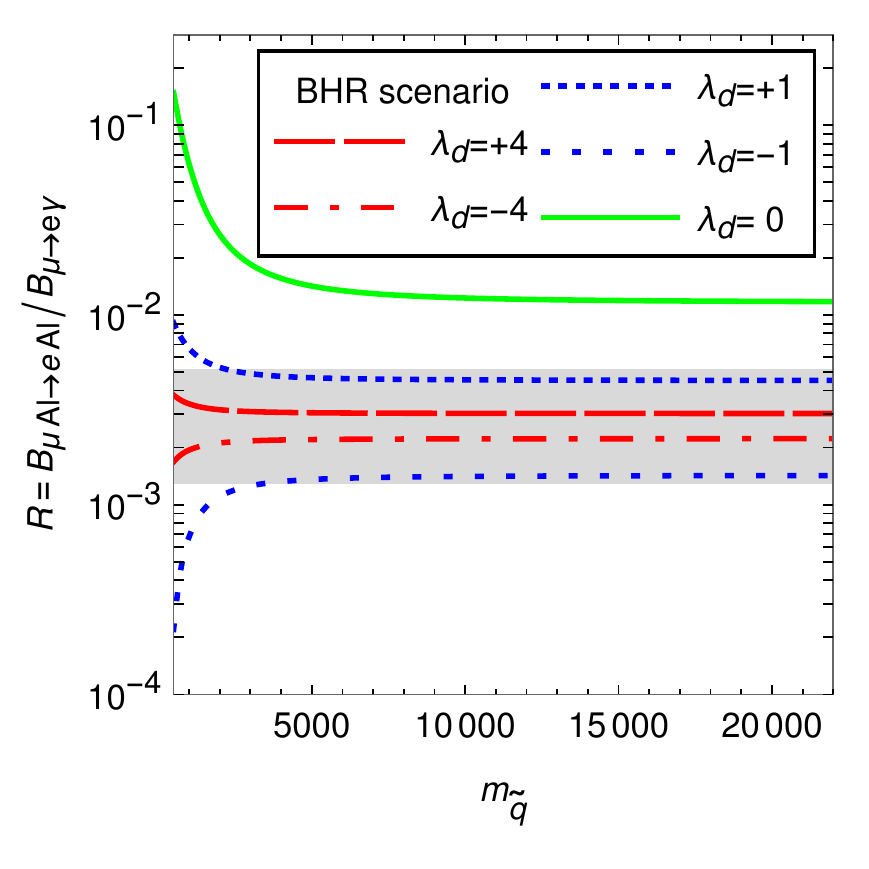}\includegraphics[width=.33\textwidth]{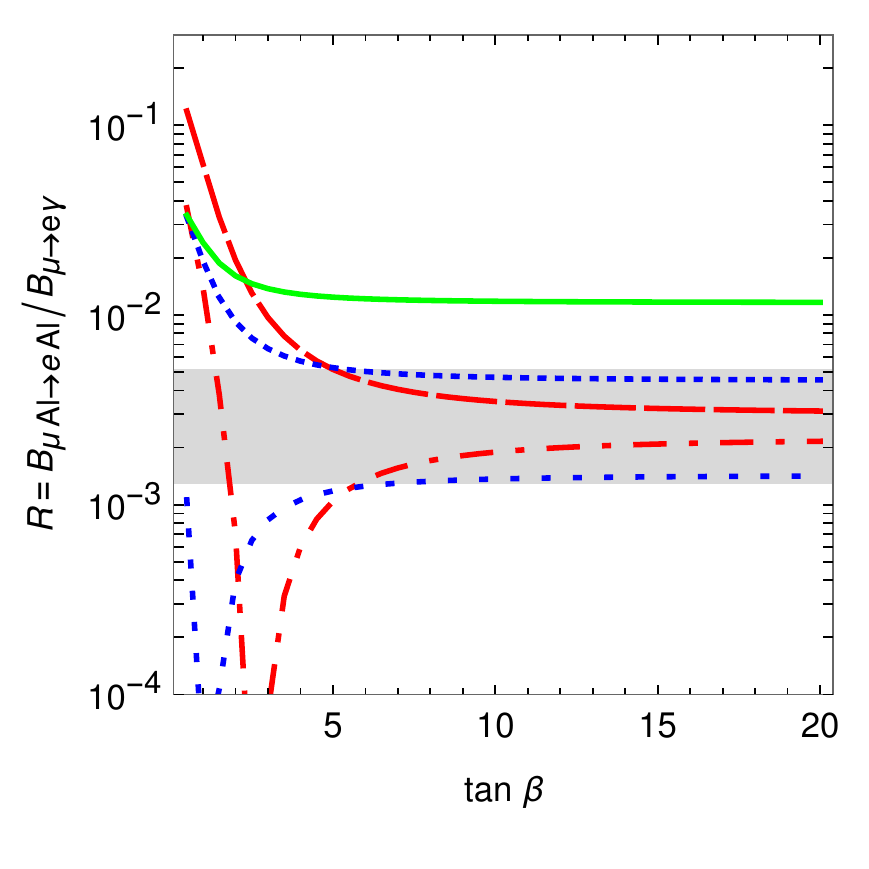}\includegraphics[width=.33\textwidth]{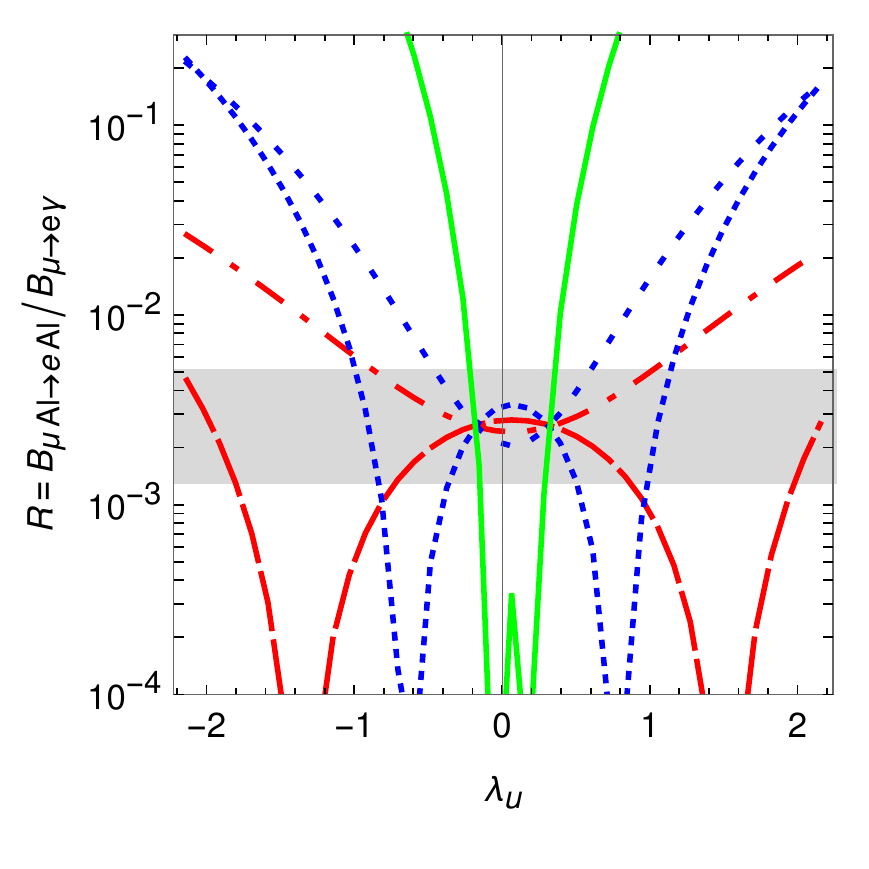}}
  \centerline{\includegraphics[width=.33\textwidth]{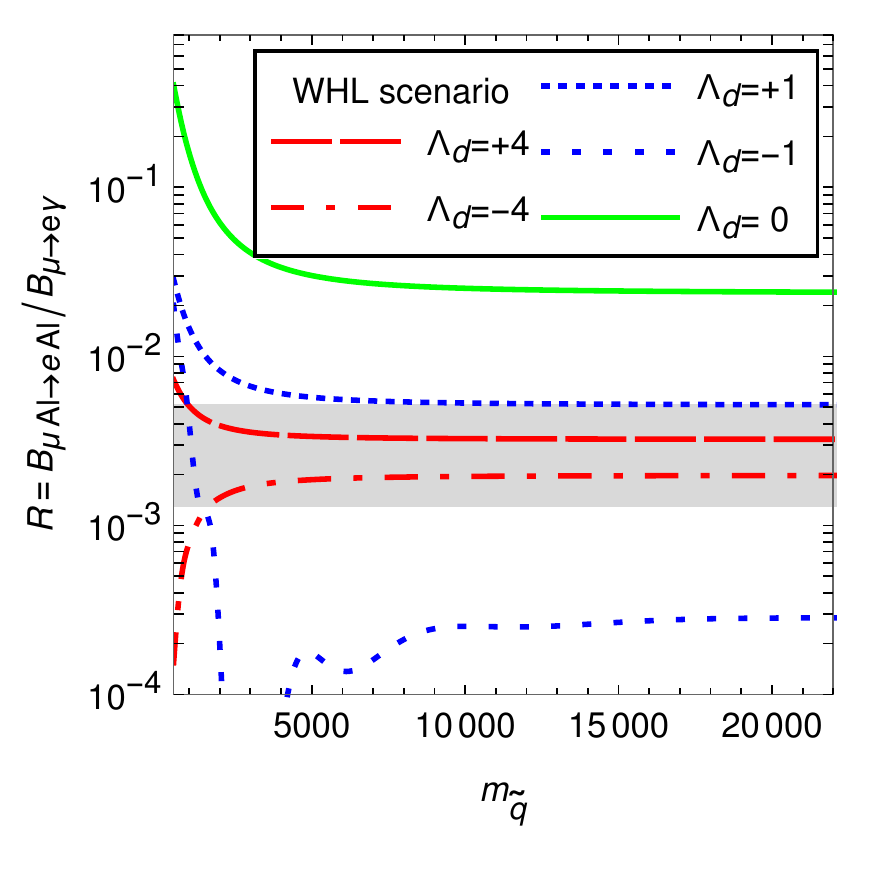}\includegraphics[width=.33\textwidth]{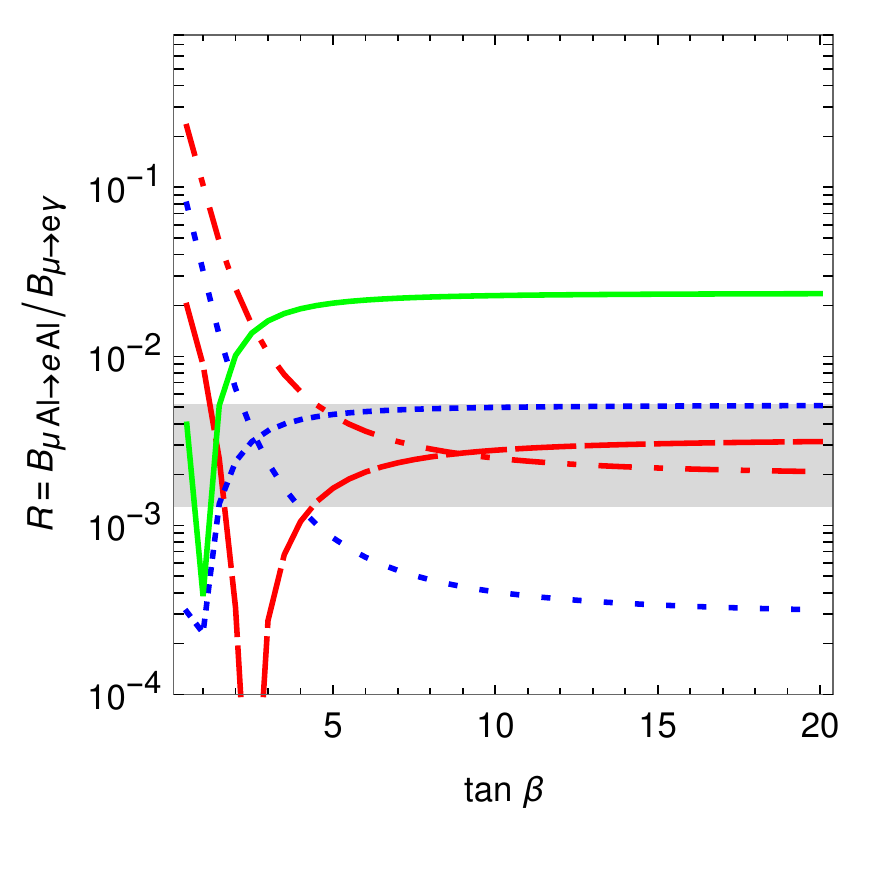}\includegraphics[width=.33\textwidth]{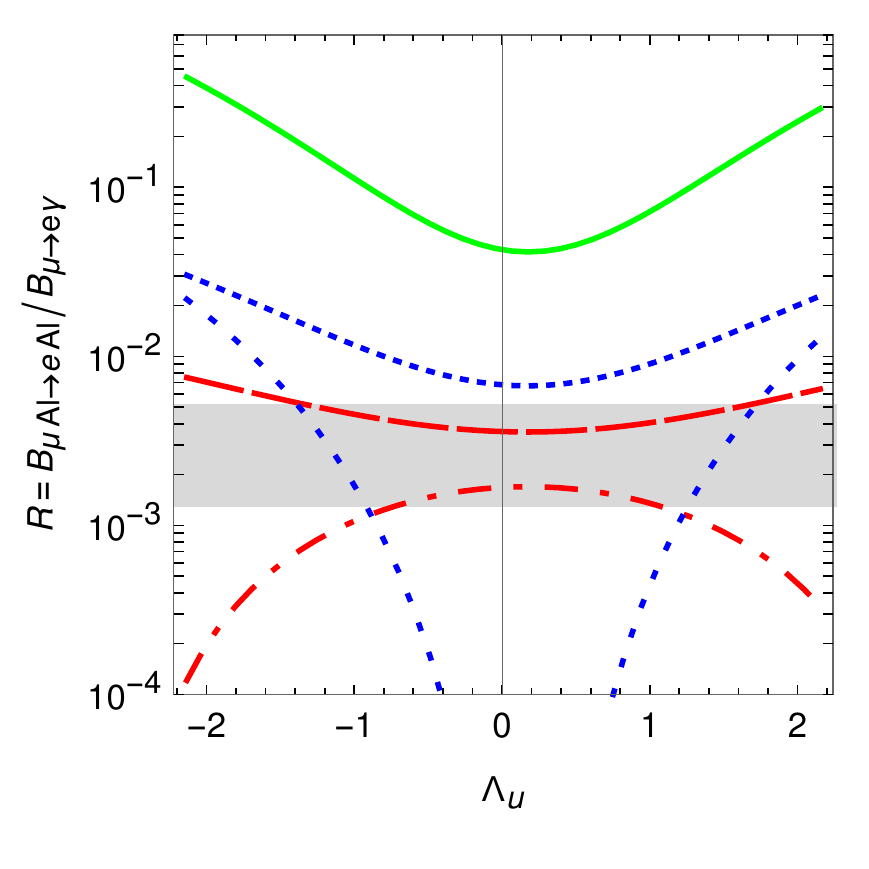}}
    \caption{\label{fig:mu2ePart1}
  The ratio $\Rmue$ of the $\mu\to e$ and $\mu\to e\gamma$ branching
  ratios as function of the squark mass, $\tan\beta$, and of
  $\lambda_u$ or $\Lambda_u$, as appropriate. The plots in the first,
  second, and third rows correspond to the BHL, BHR, WHL scenarios,
  respectively; the legends correspond to all plots in each row. In
  each plot all parameters are fixed to the standard 
  values for each scenario with $\msusy=500$ GeV except in the $\lambda_u$, $\Lambda_u$
  plots, where $\mu_u=\msusy$. The gray band indicates the
expectation corresponding to dipole dominance, Eq.\
(\ref{Rmuedipole}), allowing a fluctuation by a factor 2 up or
down.} 
\vspace{6em}\end{figure}

Before describing detailed plots we provide an overview of the
behaviour of the four form factors.
\begin{itemize}
\item $\Acharg$: $\Acharg$ is dominated by diagrams with exchange of
gaugino-like charginos/neutralinos. There is only a mild parameter
dependence and no significant enhancement by light or heavy Higgsino
masses, by large or small $\lambda$'s, by large or small
$\tan\beta$. $\Acharg$ is largest in scenarios with light wino and
slightly smaller if only the bino mass is light.
\item $\Adip$: As discussed for $a_\mu$ and $\mu\to e\gamma$, the
dipole form factor is essentially linearly enhanced by $\lambda_d$,
$\Lambda_d$ if $\mu_d$ is light. The remaining terms are small and of
a similar size as $\Acharg$. Hence if the dipole is {\em not}
enhanced, there can be significant constructive or destructive
interference between $\Acharg$ and $\Adip$ within $\mu\to e$
conversion.
\item $\AZ$: The $Z$-penguin contributions are smaller than the
$\Acharg$ and $\Adip$ contributions in a large parameter region. The
Feynman diagrammatic reason is that the $Z$ boson only couples to
Higgsino-like charginos/neutralinos (this is obvious from the
neutralino--$Z$ Feynman rule; for the charginos there is a cancellation
between diagram types 1,2,3,4 in Fig.\ \ref{fig:Zpenguin}, see also
the MSSM case \cite{Hisano:1995cp}). On the other hand, for the same reason
the $Z$-penguin can be strongly enhanced proportional to
\begin{subequations}
\label{Zpenguinenhancements}
\begin{align}
&\propto v_d^2\lambda_d^2\,, &&\propto v_d^2\Lambda_d^2 \,,\\
&\propto v_u^2\lambda_u^2\,, &&\propto v_u^2\Lambda_u^2 \,,
\end{align}
\end{subequations}
corresponding to two insertions of gaugino--Higgsino mixing terms. The
enhancements proportional to $v_d^2$ become important
for small $\tan\beta$, leading to a $1/\tan^2\beta$ enhancement. The
enhancements proportional to $v_u^2$ become important if the up-type
Higgsino $\mu_u$ is light. A similar but smaller enhancement governed
by gauge couplings instead of $\lambda$'s has been discussed in
Ref.\ \cite{Fok:2010vk}.
\item $\Abox$: The box diagrams are negligible for large squark
masses; for small $m_{\tilde{q}}$ they reach similar values as
$\Acharg$ and have a similarly mild dependence on all other parameters.
\end{itemize}

\begin{figure}[t]
  \centerline{\includegraphics[width=.33\textwidth]{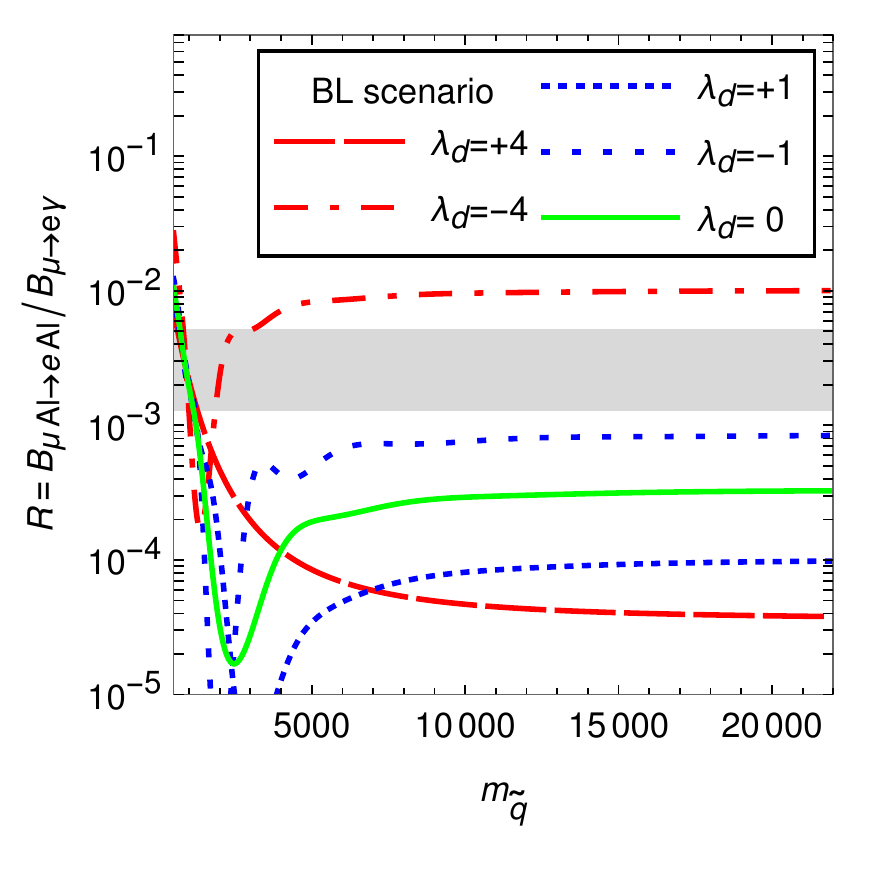}\includegraphics[width=.33\textwidth]{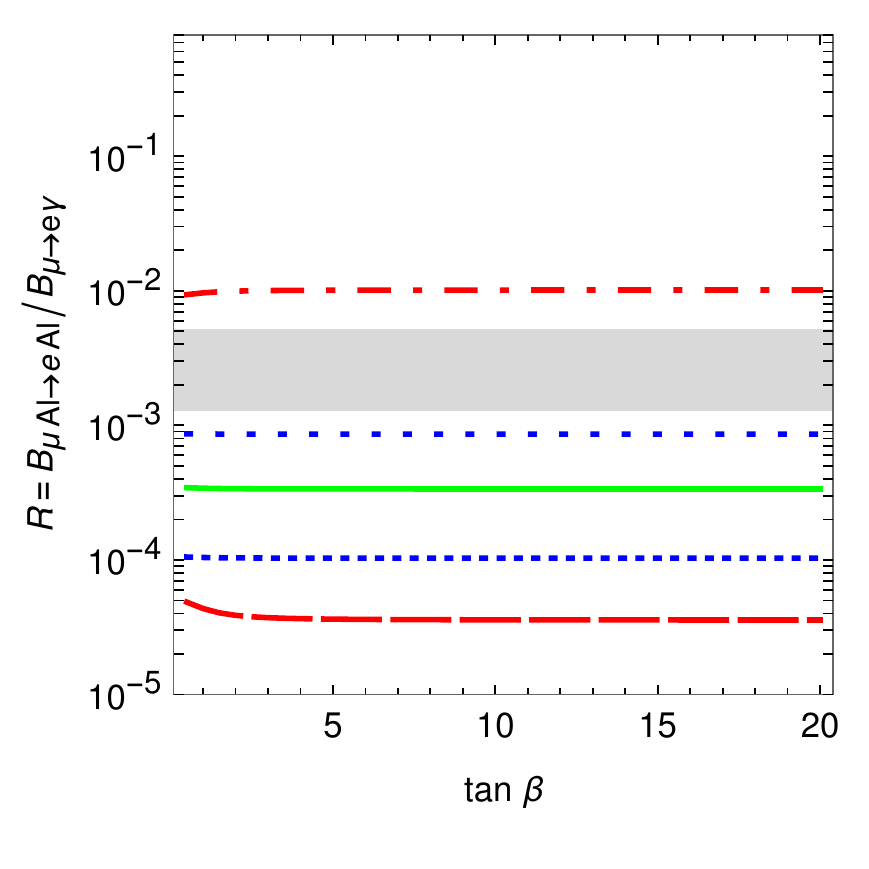}\includegraphics[width=.33\textwidth]{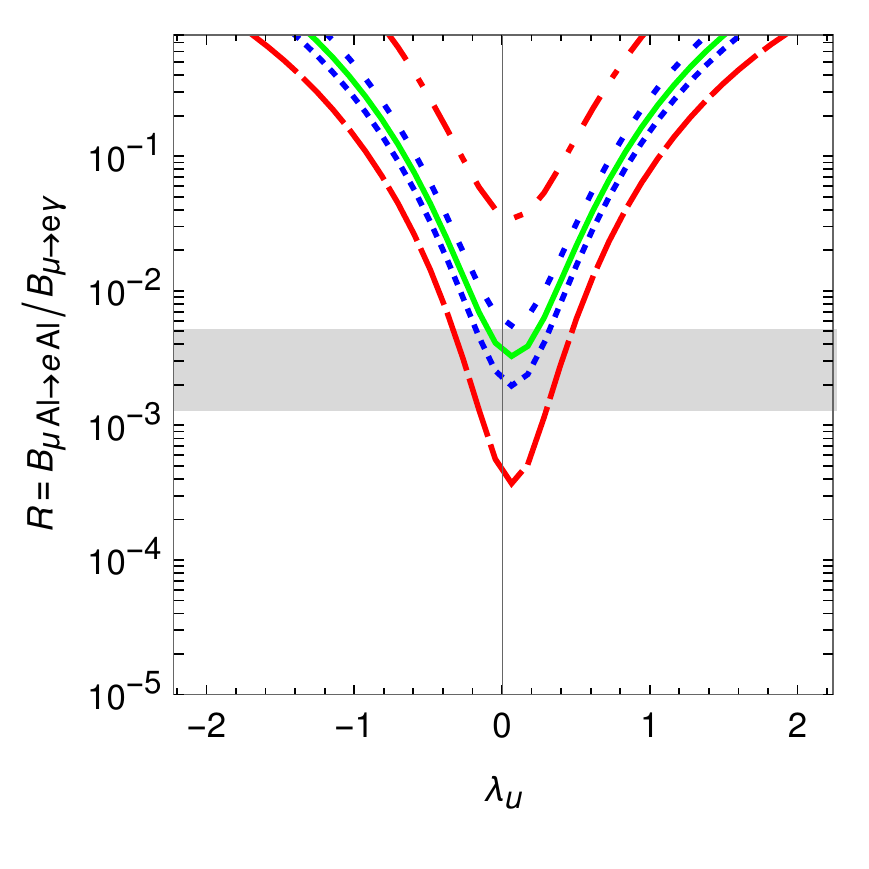}}
  \centerline{\includegraphics[width=.33\textwidth]{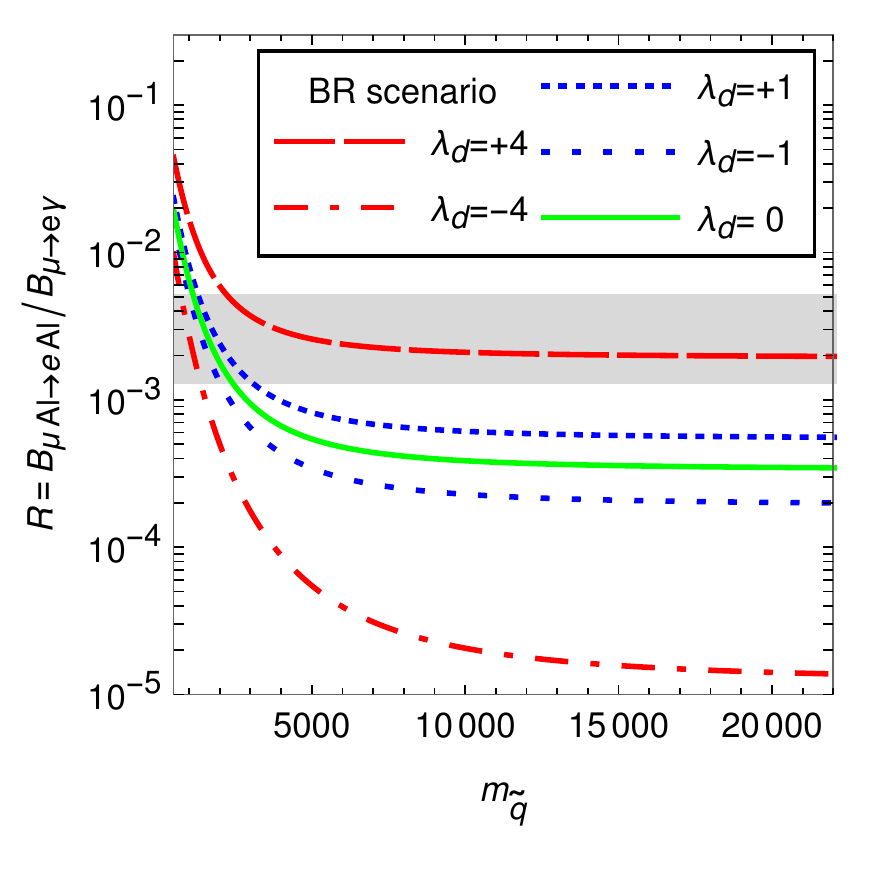}\includegraphics[width=.33\textwidth]{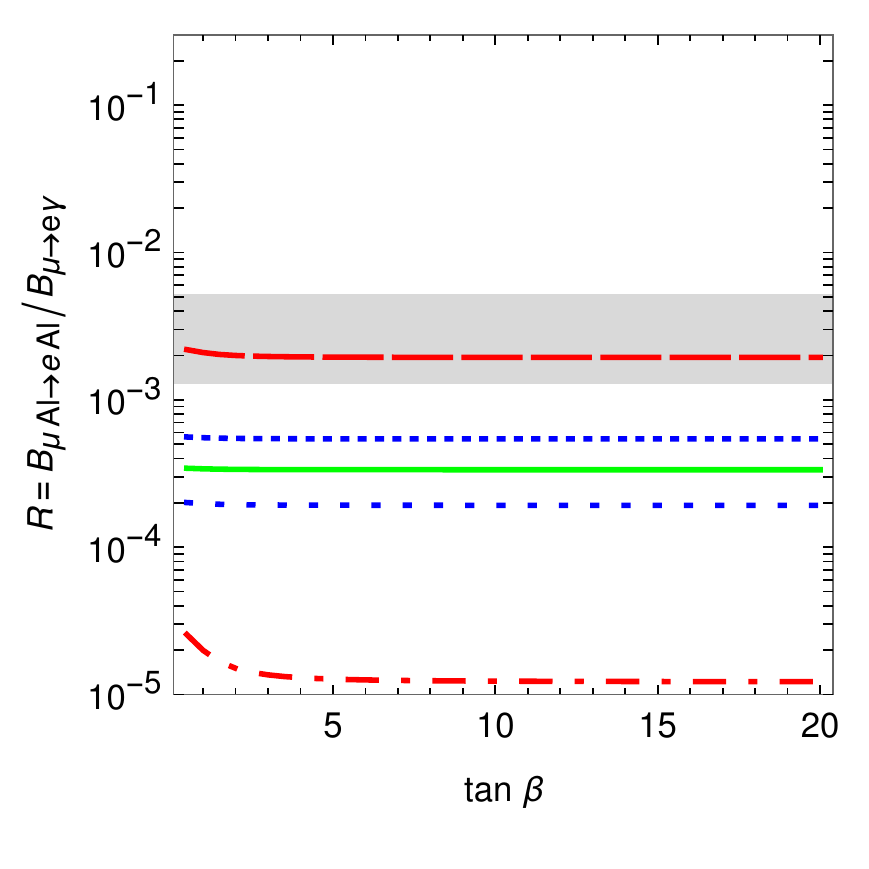}\includegraphics[width=.33\textwidth]{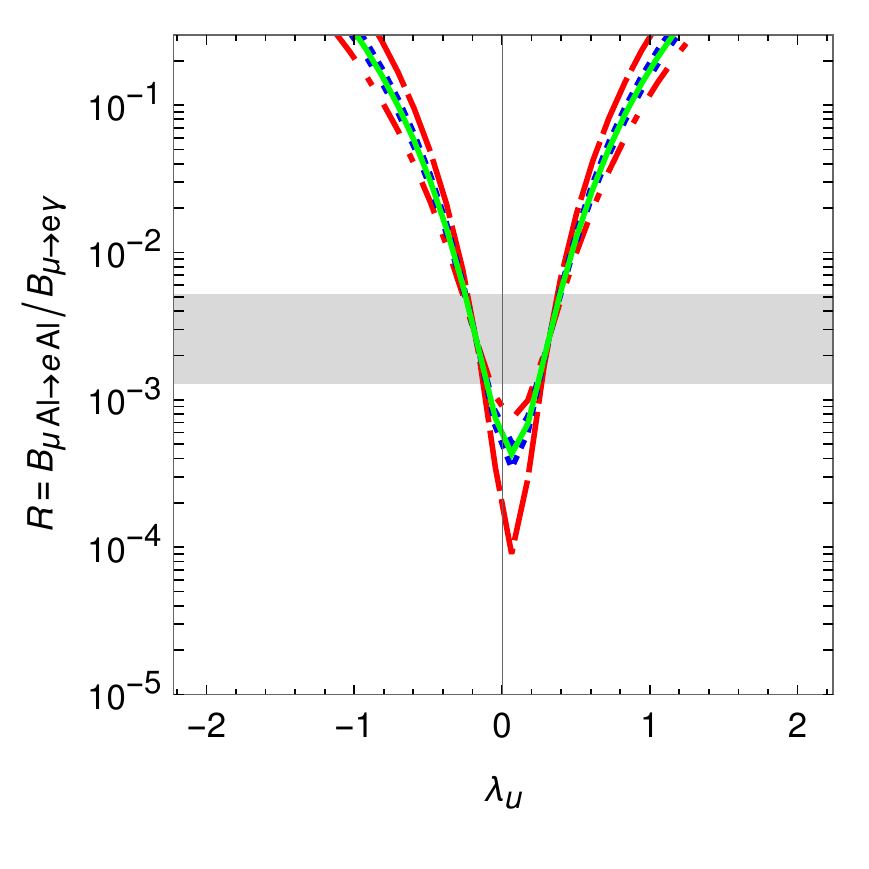}}
  \centerline{\includegraphics[width=.33\textwidth]{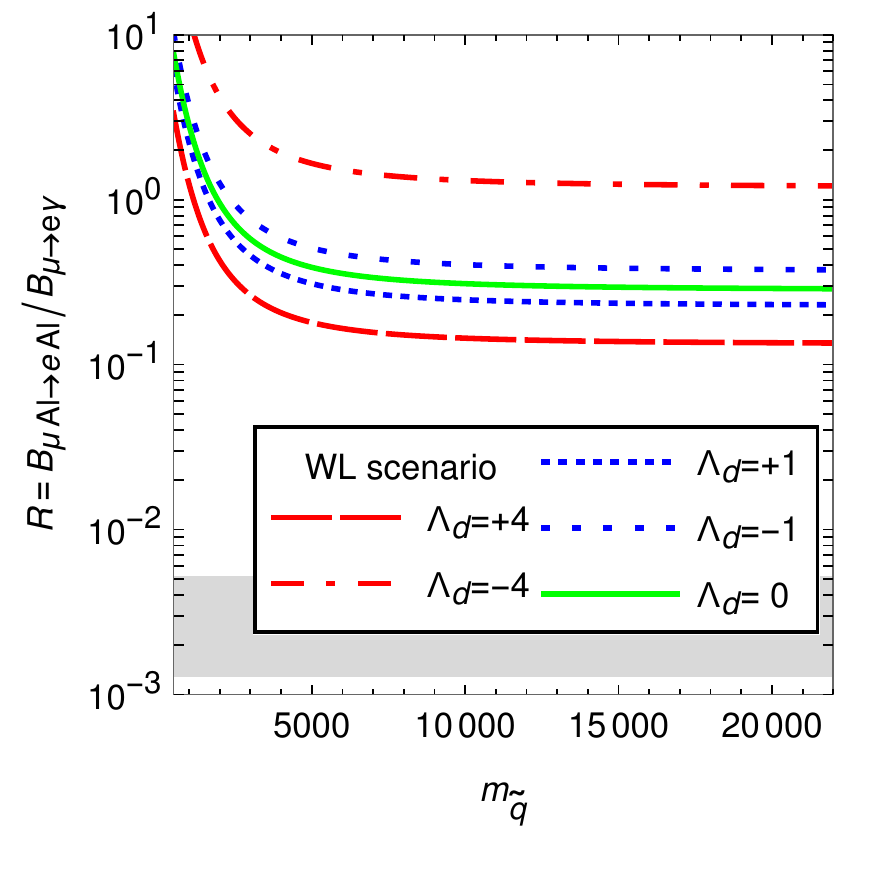}\includegraphics[width=.33\textwidth]{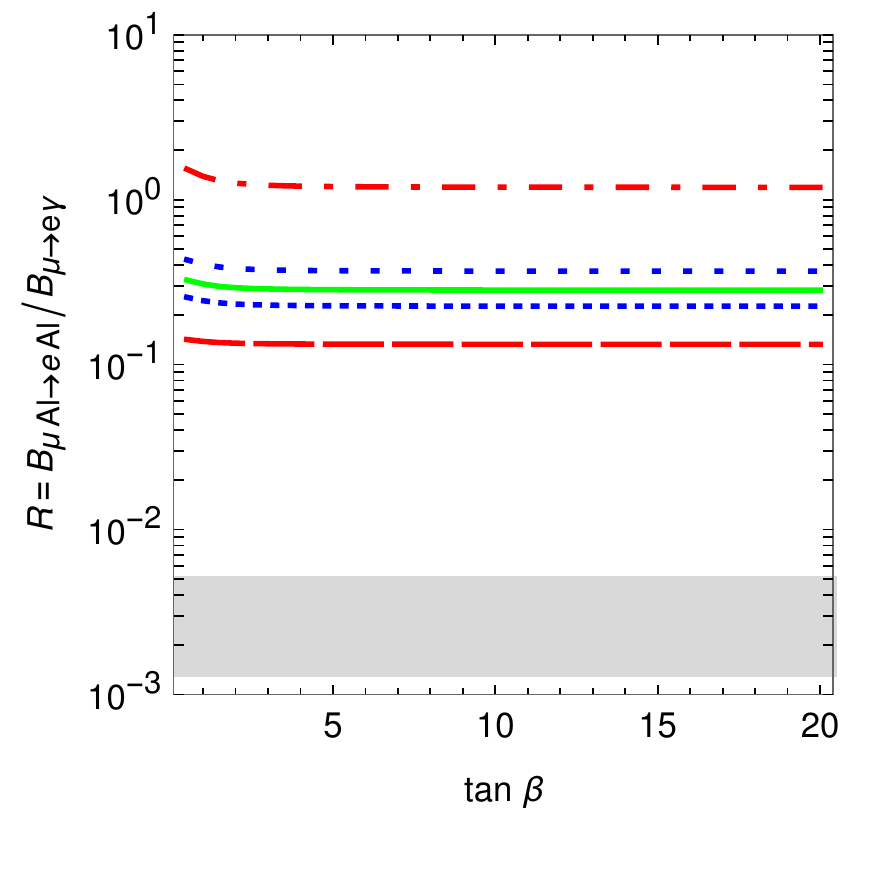}\includegraphics[width=.33\textwidth]{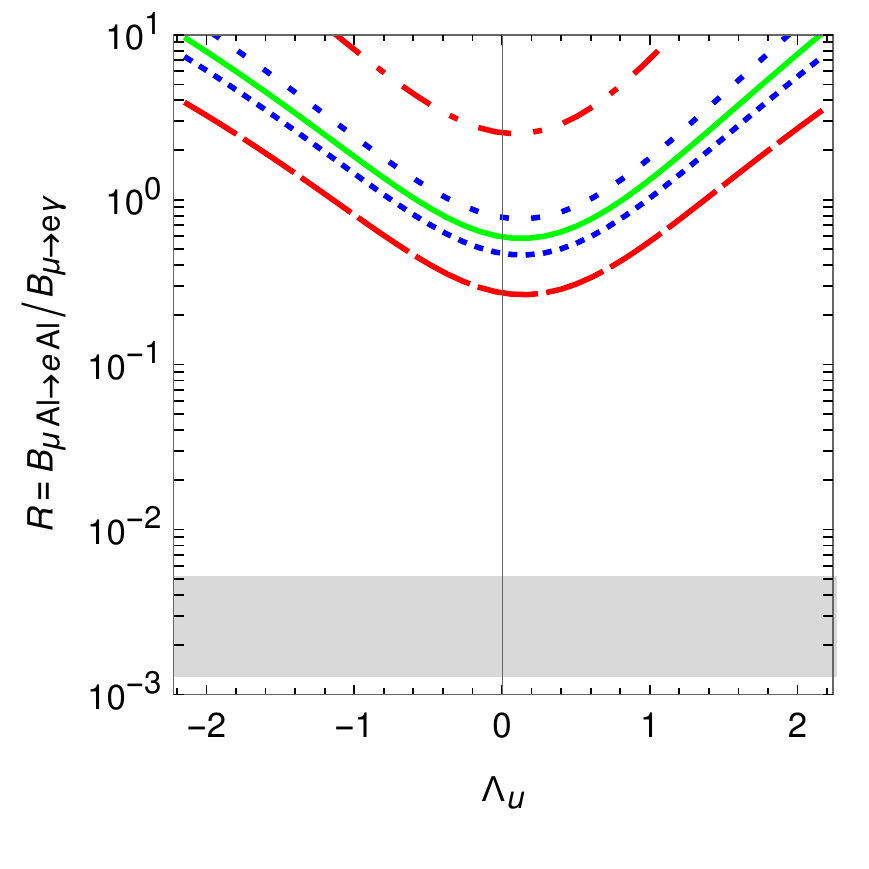}}
  \caption{\label{fig:mu2ePart2}
  As Fig.\ \ref{fig:mu2ePart1}, but for the scenarios BL, BR, WL without light $\mu_d$.}
\end{figure}

Figures \ref{fig:mu2ePart1} and \ref{fig:mu2ePart2} show the behaviour
of the ratio of branching ratios $\Rmue$ as a function of all relevant
parameters in the six scenarios BHL, BHR, WHL, BL, BR, WL. Each row in
the Figures corresponds to one of the scenarios. For each
scenario, $\msusy=500$ GeV fixed and $\lambda_d$ or $\Lambda_d$ is set
to the five values $\pm4$, $\pm1$, 0. The first plot in each row shows
$\Rmue$ as a
function of the squark mass 
$m_{\tilde{q}}$, the second plot as a function of $\tan\beta$, and the
third  plot
as  a function of $\lambda_u$ or $\Lambda_u$ (as appropriate) while
$\mu_u=\msusy$. All other parameters are set to the standard values
explained in Sec.\ \ref{sec:parameters}. The gray band indicates the
expectation corresponding to dipole dominance, Eq.\
(\ref{Rmuedipole}), allowing an up or down fluctuation by a factor 2.

The plots allow to easily read off under which conditions dipole
dominance holds and under which conditions $\mu\to e$ can be
enhanced relative to $\mu\to e\gamma$. The cases with enhanced $\mu\to
e$ are very interesting in view of the forthcoming COMET and Mu2e
experiments since they allow signals in those experiments without
violating the MEG limit on $\mu\to e\gamma$.

For the discussion we first focus on the regions of large
$m_{\tilde{q}}$ in the squark mass plots in the first column and large 
$\tan\beta$ in the 
$\tan\beta$ plots in the second column. These regions show the
``baseline behaviour'' 
resulting only from the form factors $\Acharg$ and $\Adip$, while the
$Z$-penguin and box diagrams are negligible. The results for large
$|\lambda_d|$, $|\Lambda_d|$ in the BHL, BHR, WHL scenarios are in the
gray region: this corresponds to the expected dipole dominance for
light Higgsino mass $\mu_d$ and large $|\lambda_d|$, $|\Lambda_d|$
resulting from the mass insertion diagrams discussed in
Sec.\ \ref{sec:amuresults}.

If $\lambda_d,\Lambda_d=0$ in the BHL, BHR, WHL scenarios, the
form factors $\Acharg$ and $\Adip$ are of a similar size. The same is
true in the BL, BR, WL scenarios, independently of $\lambda_d$,
$\Lambda_d$. Hence in all these cases we get strong deviations from
dipole dominance $\Rmuedipole$. The actual ratio ranges from
$\Rmue\sim10^{-5}$ in the BR scenario (due to an accidental
cancellation between $\Acharg$ and $\Adip$ which happens around
$\lambda_d\approx-4$) up to $\Rmue\sim1$ in the WL scenario (where
$\Acharg$ is a few times larger than $\Adip$).

Next we focus on the dependence on $m_{\tilde{q}}$, which arises only
from the
box diagrams. For large squark masses they
are negligible and we obtain the baseline behaviour discussed before;
for smaller squark masses below around $5\times\msusy$ they become
relevant. Of course, their impact is particularly pronounced in cases
where the dipole $\Adip$ is small, i.e.\ for small
$\lambda_d,\Lambda_d$ and/or in the BL, BR, WL scenarios. In these
cases the box diagrams can increase $\mu\to e$ by a factor of a few.

Finally we describe the influence of the $Z$-penguin contributions. They
are enhanced by two powers of the gaugino--Higgsino mixing, see Eq.\ 
(\ref{Zpenguinenhancements}). The enhancements governed by $v_d^2$ and
$\lambda_d^2,\Lambda_d^2$ are visible in the $\tan\beta$ plots at
small $\tan\beta$, where these terms become large. For $\tan\beta\le5$,
this effect can lead to dramatic enhancements in the scenarios with
light Higgsino mass $\mu_d$.

The enhancements governed by $v_u^2$ and $\lambda_u^2,\Lambda_u^2$ can
be seen in the $\lambda_u$, $\Lambda_u$ plots in the third column.
For small $\lambda_u$,
$\Lambda_u$, the results are similar to the baseline behaviour
discussed above (the slight differences are due to $\mu_u=\msusy$
instead of $\mu_u\gg\msusy$). For larger values of $\lambda_u$,
$\Lambda_u$ the $Z$-penguin dominates, leading to very strong
enhancements as well as to zeroes in $\mu\to e$ due to cancellations
between the different form factors.

The largest overall values of the ratio of branching ratios can reach
more than \mbox{$\Rmue>10$} in the scenarios with small dipoles (i.e.\ for
small $\lambda_d,\Lambda_d$ and/or the scenarios BL, BR, WL with heavy
$\mu_d$). In scenarios with large dipole, e.g.\ in the WHL scenario
with $\Lambda_d=1$, $\Rmue$ can still be 10 times larger than the
value $\Rmuedipole$.

\begin{figure}[t]
   \centerline{\includegraphics[width=.95\textwidth]{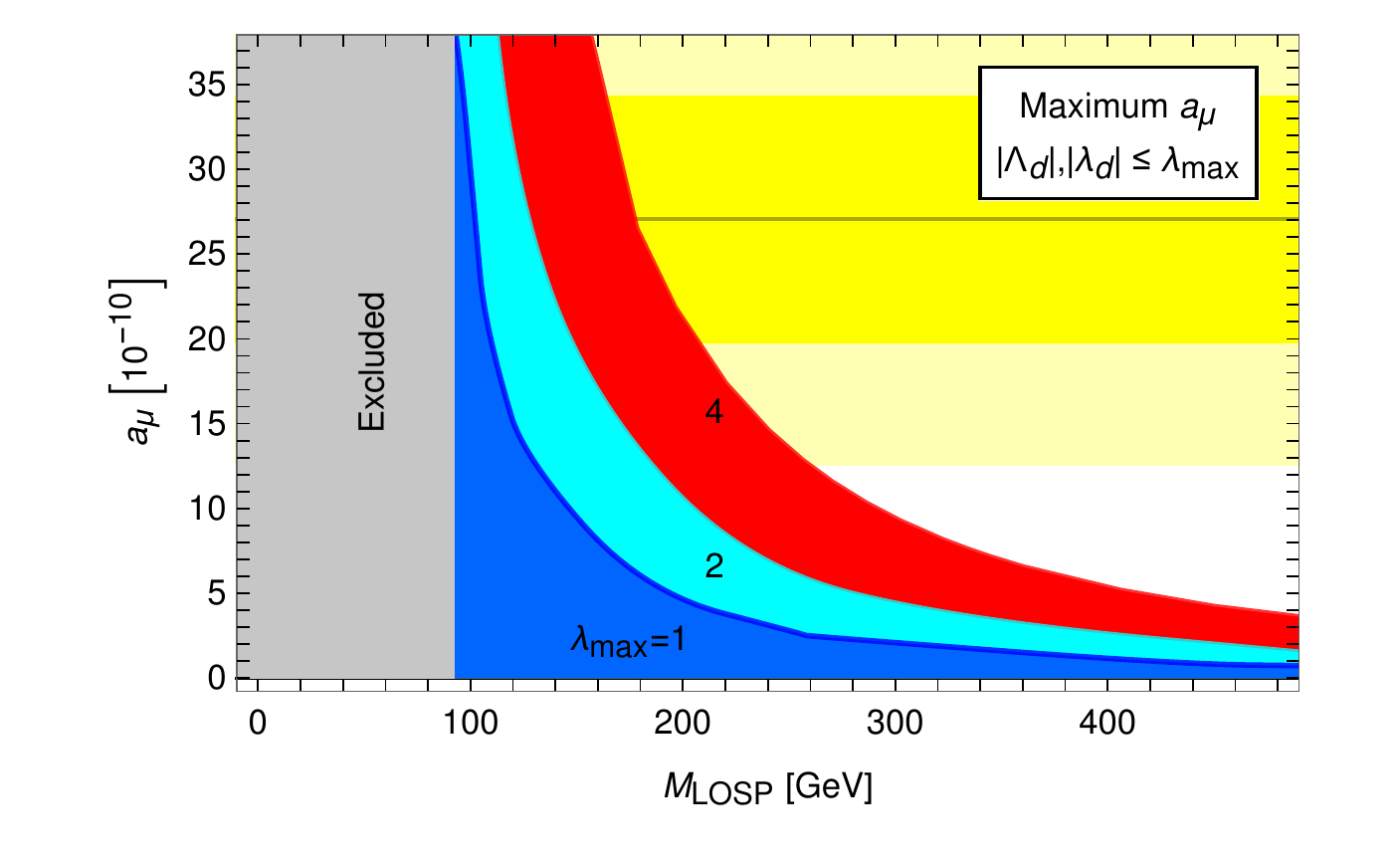}}
        \caption{\label{fig:maxamu} Maximum possible results for
    $a_\mu$ in the MRSSM obtained from a parameter scan applying
    the constraints of Sec.\ \ref{sec:parameters} and maximum values
    for $\Lambda_d$, $\lambda_d$ as indicated in the plot. The dark
    and light yellow horizontal bands 
    correspond to the $1\sigma$ and $2\sigma$ bands given by Eq.\
    (\ref{eq:Deltaamu}). The vertical band corresponds to the
    exclusion limit (\ref{masslimits}).}
  \end{figure}

\subsection{Summary plots based on scans}

The previous subsections have analyzed the detailed parameter
dependences of all three observables $a_\mu$, $\mu\to e\gamma$ and
$\mu\to e$ conversion in the MRSSM. In the present subsection we will
show several plots based on parameter scans. These plots summarize the
generic behaviour and show maximum possible results and the
correlations between the observables.

Figure \ref{fig:maxamu} shows the maximum possible results for $a_\mu$
in the MRSSM, as a function of the LOSP mass, i.e.\ the 
lightest electrically charged SUSY particle mass. It is based on a
scan in parameter space where the ratios between the masses are varied
and various upper limits on $\Lambda_d,\lambda_d$ are imposed. As
expected from section \ref{sec:amuresults} and
Fig.\ \ref{fig:WHLandallMS} the maximum $a_\mu$ is
obtained in scenarios where the WHL- and BHR-like contributions add up
constructively and the corresponding masses are all similar. Again the
plot shows that very small masses are required to explain the current
$a_\mu$ deviation. For $|\Lambda_d|<4$, a $1\sigma$ explanation
requires $m_{\text{LOSP}}<200$~GeV, and for $|\Lambda_d|<2$, it requires
$m_{\text{LOSP}}<150$~GeV.

  \begin{figure}[t]
    \centerline{\includegraphics[width=.6\textwidth]{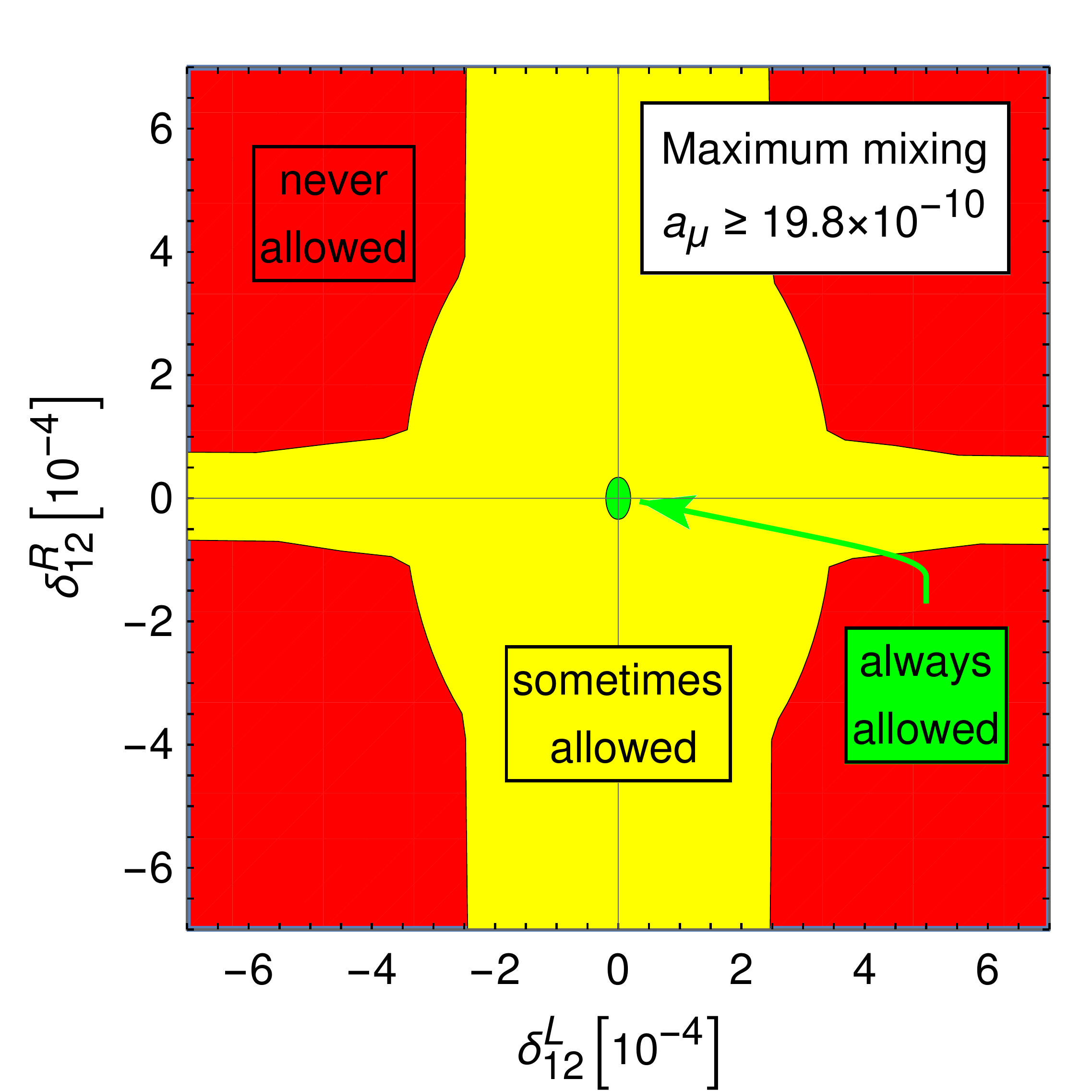}}
        \caption{\label{fig:mu2egMaxdelta} Scan over parameter choices
    for which $a_\mu$ agrees with the deviation (\ref{eq:Deltaamu}),
    displaying regions for $\delta^{\TL}_{12}$ and $\delta^{\TR}_{12}$
    allowed by the MEG limit on $\mu\to e\gamma$ (\ref{eq:mu2eglimit}). The
    small, inner green region is allowed by all parameter choices
    which explain $a_\mu$, the cross-shaped large yellow region is
    allowed by some parameter choices, and the outer red region is not
    allowed by any parameter choice.}
  \end{figure}

Figure \ref{fig:mu2egMaxdelta} focuses on the correlation between
$a_\mu$ and $\mu\to e\gamma$. It derives limits on the
flavour-violating parameters $\delta^{\TL,\TR}_{12}$, valid under the
condition that the $a_\mu$ deviation is fully explained by the
MRSSM. The logic behind this plot is as follows. For each parameter
choice with a certain value of $a_\mu$, the prediction for $\mu\to
e\gamma$ is essentially fixed, since both observables are governed by
dipole form factors, see Fig.\ \ref{fig:mu2egLambdaVs}. The only
remaining free parameters\footnote{We 
always keep the choice of the selectron masses (\ref{selectronmasses})
fixed.}
are the $\delta^{\TL,\TR}_{12}$, which enter as in Eq.\
(\ref{mu2egbehaviour}). As a result, for each parameter choice which
explains the $a_\mu$ deviation, there is a certain ellipse-shaped
region in the
$\delta^{\TL}_{12}$--$\delta^{\TR}_{12}$-space allowed by the MEG limit on $\mu\to e\gamma$.

Fig.\ \ref{fig:mu2egMaxdelta} shows the results of a scan over all
parameter choices for which the current $a_\mu$ deviation is
explained, and for which the parameter constraints of
Sec.\ \ref{sec:parameters} are met. The values of the $\delta$'s in
the small green inner contour are allowed by all parameter choices
(i.e.\ $\mu\to e\gamma$ is always below the MEG limit). This inner
contour arises from the intersection of all ellipses and has itself
approximately the shape of an ellipse.
On the other hand, the values of the $\delta$'s in the large
cross-like yellow region are allowed by some parameter choices and
forbidden by others; this region corresponds to the union of all
ellipses. The cross-like shape arises because for certain parameter
choices the ellipses degenerate to large rectangles: For the WHL-like
case shown in Fig.\ \ref{fig:mu2egLambdaVs}, $\mu\to e\gamma$ only
depends on $\delta^{\TL}_{12}$ and hence there is an upper limit on
$\delta^{\TL}_{12}$ but $\delta^{\TR}_{12}$ can be arbitrarily large;
similarly BHR-like parameter choices lead to unconstrained
$\delta^{\TL}_{12}$. Numerically, values of the $\delta$'s below
around $10^{-5}$ are always allowed. On the other hand,
choices where both
$\delta_{12}$'s are significantly above around $10^{-4}$ are always
forbidden.

  \begin{figure}[t]
    \centerline{\includegraphics[width=.95\textwidth]{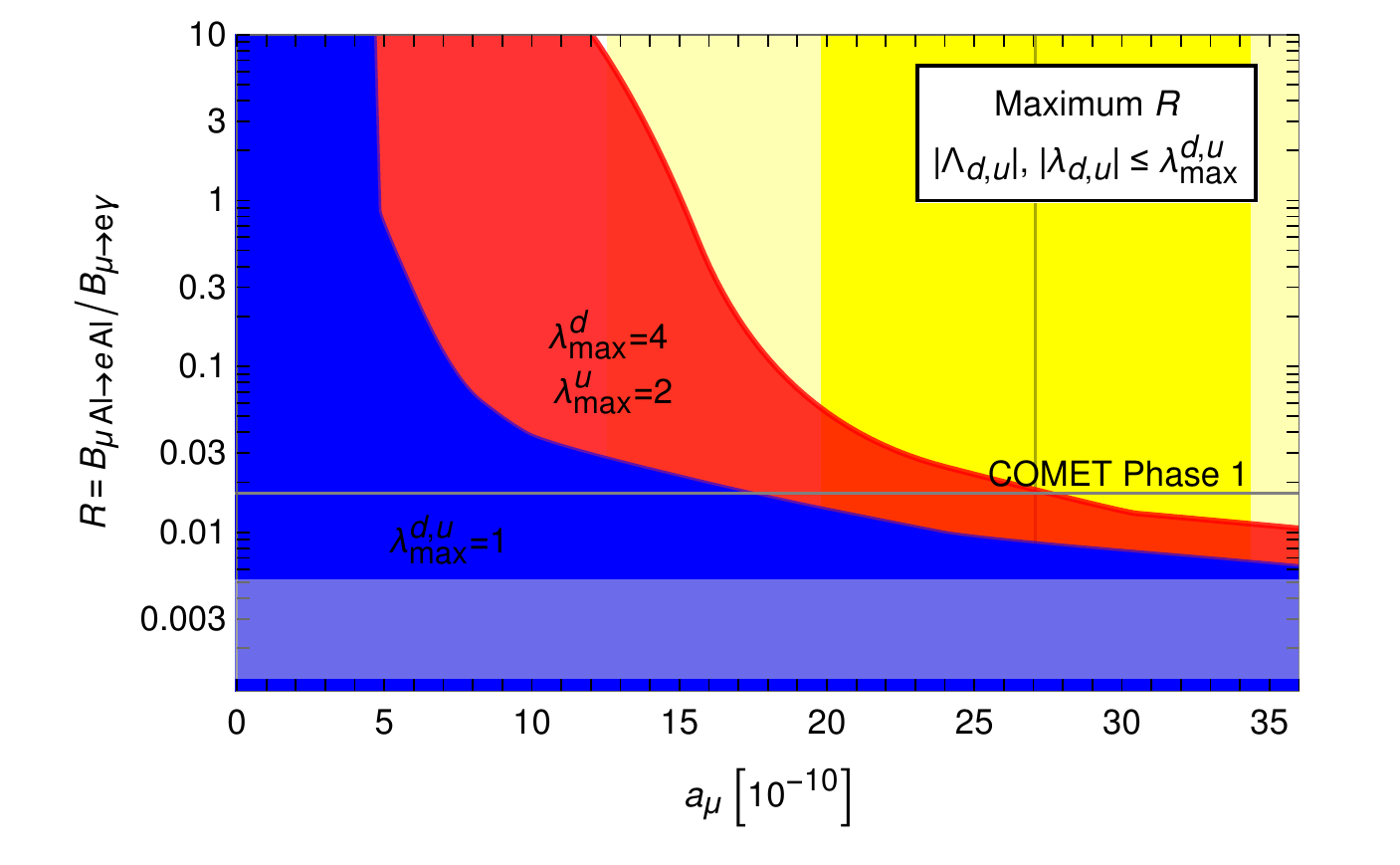}}
        \caption{\label{fig:mu2ecorrelation} The correlation between $\mu\to e$
    conversion and $\mu\to e\gamma$, as a function of $a_\mu$;
    equivalently, the maximum value for $\mu\to e$ conversion for a
    given value of $\mu\to e\gamma$. The
    possible ranges have been obtained from a parameter scan
    respecting the constraints of Sec.\ \ref{sec:parameters}. In
    particular, in the outer red region, the $\lambda$'s take the
    maximum values allowed in Sec.\ \ref{sec:parameters}; in the blue
    inner region the $\lambda$'s are constrained to be less than
    unity. The vertical yellow band is defined as in Fig.\ \ref{fig:BHLBHRamu};
    the horizontal light gray band indicates the
expectation corresponding to dipole dominance as in
    Fig.\ \ref{fig:mu2ePart1}.
    The thin horizontal line indicates the minimum value of $\Rmue$ for
    which COMET Phase 1 is sensitive to $\mu\to e$ conversion, given
    the current MEG limit on $\mu\to e\gamma$.}
  \end{figure}

Figure \ref{fig:mu2ecorrelation} focuses on the correlation of all
three observables $a_\mu$, $\mu\to e\gamma$ and $\mu\to e$
conversion. It is based on the following expectation. If $a_\mu$ is
large, the dipole form factor must be large, and in dipole-dominated
cases $\mu\to e\gamma$ and $\mu\to e$ are strongly correlated, see
Eq.\ (\ref{Rmuedipole}). If $a_\mu$ is small, the dipole form factor
can be small and $\mu\to e\gamma$ and $\mu\to e$ become
uncorrelated. Therefore Fig.\ \ref{fig:mu2ecorrelation} shows the
ratio $\Rmue={B_{\mu \text{Al}\to e \text{Al}}}/{B_{\mu \to
e\gamma}}$ as a function of $a_\mu$ in a parameter scan fulfilling the
constraints of Sec.\ \ref{sec:parameters}. The scan is further
constrained by $\delta^{\TL}_{12}=\delta^{\TR}_{12}$.

The result has the expected behaviour. If $a_\mu\gtrsim30\times10^{-10}$, the ratio $\Rmue$
is within the expectation of dipole dominance up to a factor
5, and even up to a factor 2 if all $\lambda_i$, $\Lambda_i$ are
constrained to be below unity.
Combining this upper limit on $\Rmue$ with the MEG limit on $\mu\to
e\gamma$ shows that in this parameter region the $\mu\to e$ conversion
rate is below the reach of COMET Phase 1. In the plot, this can be
seen with the help of the thin horizontal line, which corresponds to
$\Rmue=0.017$, the ratio of the COMET Phase 1 sensitivity and the MEG
limit (\ref{eq:mu2eglimit}).

If $a_\mu>12\times10^{-10}$, just within the $2\sigma$ region around
the observed deviation, the ratio $\Rmue$ can be 10 times larger even
for moderate $\lambda_i$, $\Lambda_i$. This is interesting in view of the
forthcoming COMET Phase 1 measurement of $\mu\to e$ conversion: a
positive signal at COMET Phase 1 is possible while $\mu\to e\gamma$
remains below the current MEG limit.

For lower
$a_\mu$ and/or larger values of the $\lambda_i$ and $\Lambda_i$,
$\Rmue$ can be even larger. The parameter choices which 
maximize $\Rmue$ are choices where $\Lambda_u$ and $\lambda_u$ take values
at the border of the allowed region and where all masses except the
Higgsino masses are very similar. In such parameter regions the MRSSM
prediction for $\mu\to e$ conversion can be easily in reach of COMET Phase
1 even if $\mu\to e\gamma$ is orders of magnitude below the current
MEG limit.

As mentioned above, Fig.\ \ref{fig:mu2ecorrelation} uses the constraint
$\delta^{\TL}_{12}=\delta^{\TR}_{12}$ (the actual value   drops out
in the ratio $\Rmue$ and is not
important).  If this constraint is dropped and $\delta^{\TL}_{12}=0$
or $\delta^{\TR}_{12}=0$ are allowed, the ratio $\Rmue$ becomes
unconstrained. E.g.\ we could choose a WHL-like mass pattern with large
$a_\mu$ and tune the masses and $\lambda_d$ such that the right-handed
dipole amplitude vanishes. If we then set $\delta^{\TL}_{12}=0$ but
$\delta^{\TR}_{12}\ne0$, all flavour-violating dipole amplitudes
vanish and $\mu\to e\gamma$ is impossible, while $\mu\to e$ conversion
is still possible due to the other form factors. Accordingly $\Rmue$
can be arbitrarily large independently of $a_\mu$ if one of the
$\delta$'s is allowed to vanish.

\section{Conclusions}

The MRSSM provides an attractive alternative realization of SUSY with
promising phenomenological properties. 
In the present paper we have considered the MRSSM predictions for
$a_\mu$ and the lepton-flavour violating observables $\mu\to e\gamma$
and $\mu\to e$. We presented analytic one-loop results, useful
compact approximations and a detailed numerical analysis.

A striking difference to the familiar MSSM case is the absence of
$\tan\beta$ enhancements in all dipole amplitudes. The reason is that
the $\tan\beta$ enhancement in the MSSM originates from insertions of
the MSSM $\mu$-parameter and Majorana gaugino masses. Both are
forbidden in the MRSSM by R-symmetry. The absence of $\tan\beta$
enhancements alters the phenomenology significantly.

In spite of this we have found that dipole amplitudes can be enhanced
in the MRSSM by MRSSM-specific superpotential parameters $\Lambda_d$,
$\lambda_d$. The mechanism is similar to the $\tan\beta$ enhancements
in the MSSM but its numerical impact is restricted by constraints on
the superpotential parameters from electroweak precision observables
and perturbativity.

The analysis of $a_\mu$ has shown that it is very hard to explain the
currently observed deviation in the MRSSM. An explanation is possible
only in particular corners of the MRSSM parameter space: several SUSY
masses, among 
them at least one smuon, one gaugino and 
one Higgsino, should be around 200~GeV or below and the values of $\Lambda_d$
and/or $\lambda_d$ should be at least as large as the top Yukawa
coupling, preferably larger. Such parameter choices are viable since
compressed light spectra are not excluded by LHC data.

The required
large values of $\Lambda_d$, $\lambda_d$ are intriguing. Similar
large values of $\Lambda_u$, $\lambda_u$ have been found helpful in
explaining the measured value of the Higgs boson mass
\cite{Diessner:2014ksa}; on the other hand very large values of these
parameters are constrained by electroweak precision data and are
difficult to reconcile with embedding the MRSSM into an $N=2$ SUSY
theory.

The decay $\mu\to e\gamma$ is strongly
correlated to $a_\mu$ if $a_\mu$ is large. As a result we could derive
limits on the flavour-violating parameters $\delta^{\TL,\TR}_{12}$
valid under the assumption that the MRSSM explains the current $a_\mu$
deviation. As shown in the traffic-light-like colours of
Fig.\ \ref{fig:mu2egMaxdelta} values for the $\delta$'s below around
$10^{-5}$ are generally allowed and higher values can be allowed,
depending on the choice of parameters. It is however also of interest
to discuss $\mu\to e\gamma$ in scenarios with small $a_\mu$ --- future
$a_\mu$ measurements could be closer to the SM prediction or non-MRSSM
new physics could explain the deviation. In such scenarios larger SUSY
masses and small $\Lambda_d$, $\lambda_d$ are possible and larger
$\delta$'s are allowed. Combining Figures
\ref{fig:mu2egLambdaVs}(right) and 
\ref{fig:mu2egContours} with the known mass scaling allows to conclude
that $\delta_{12}$'s
around $10\%$ become possible for SUSY masses in the few TeV range.

Our reason to consider particularly  $\mu\to
e$ conversion as another lepton flavour violating observable was
threefold.  The forthcoming COMET and Mu2e experiments promise to
improve the sensitivity to this process by orders of magnitude; an
earlier study in Ref.\ \cite{Fok:2010vk} already revealed that this
process can provide limits on the MRSSM, and we expected
characteristic differences between the MRSSM and MSSM predictions for
this process. In the MSSM, the process is typically dominated by
dipole amplitudes and strongly correlated with $\mu\to e\gamma$, see
Eq.\ (\ref{Rmuedipole}).

Indeed we found strong deviations from dipole dominance. There are two
main sources for these deviations. If the dipole amplitudes are small,
the charge radius form factors become relatively important and can
dominate strongly. And even if the dipole amplitudes are large, the
$Z$-penguin contributions can also be large --- they are enhanced
$\propto\Lambda_i^2 v_i^2$ ($i=u,d$). In the fully general case, where
mixing in the left-handed and right-handed slepton sectors is
independent, there is no correlation
 between $\mu\to
e$  and $\mu\to e\gamma$. Due to possible cancellations either of
these observables could be zero while the other is large.

We have then studied the (non-)correlation for the  specific
condition $\delta^{\TL}_{12}=\delta^{\TR}_{12}$. The result is
Fig.\ \ref{fig:mu2ecorrelation}, which shows the ratio between $\mu\to
e$  and $\mu\to e\gamma$ as a function of $a_\mu$. It shows that if
$a_\mu$ is as large as the current deviation, the correlation
between $\mu\to e$ and $\mu\to e\gamma$ is rather strong,
though not as
strong as in case of dipole dominance,
Eq.\ (\ref{Rmuedipole}). In this case, the current MEG limit on
$\mu\to e\gamma$ implies an upper limit on the possible MRSSM
prediction to $B_{\mu \text{Al}\to e \text{Al}}$ of a few times
$10^{-15}$, just touching the reach of COMET Phase I but well in reach of
COMET Phase II and the Mu2e experiment. 

On the other hand, if $a_\mu$ is not quite as large, the correlation
between $\mu\to e$ and $\mu\to e\gamma$ becomes weaker. For $a_\mu$
contributions below $20\times10^{-10}$,
Fig.\ \ref{fig:mu2ecorrelation} together with the MEG limit allows
$B_{\mu \text{Al}\to e \text{Al}}$ well in reach of COMET Phase I. Turning
the argument around, if COMET Phase I finds a signal for $\mu\to e$
conversion and if the MRSSM is realized in the scenario of
Fig.\ \ref{fig:mu2ecorrelation}, the MRSSM cannot explain the current
$a_\mu$ deviation at the $1\sigma$ level.

The present paper has focused on a detailed and comprehensive survey,
but we have restricted ourselves to three observables, and
all our results have been obtained at leading nonvanishing
order. We leave the study of further observables such as $\mu\to eee$
and the inclusion of higher-order corrections such as the ones
considered in
Refs.\ \cite{Davidson:2016edt,Davidson:2016utf,Crivellin:2017rmk} for
later work.

\acknowledgments
We thank Uladzimir Khasianevich for careful reading of the
manuscript and Seungwon Baek and Jae-hyeon Park for discussions.
This research was supported by the German Research Foundation (DFG)
under grant numbers STO 876/4-1, STO 876/6-1, STO 876/7-1, and the
Polish National Science Centre through the HARMONIA project under
contract UMO-2015/18/M/ST2/00518 (2016-2019).
\appendix
\section{MRSSM Feynman rules}

Here we provide the values of the coupling coefficients introduced in
section \ref{sec:Feynmanrules} and the resulting Feynman rules.

\subsection*{Lepton--sleptons--neutralinos/charginos}
\vspace{2cm}
\includegraphics[scale=1.]{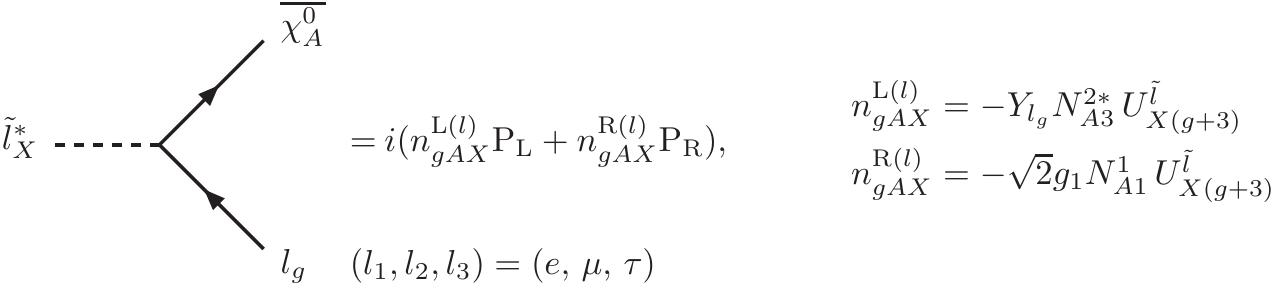}
\vspace{1.5cm}\\
\includegraphics[scale=1.]{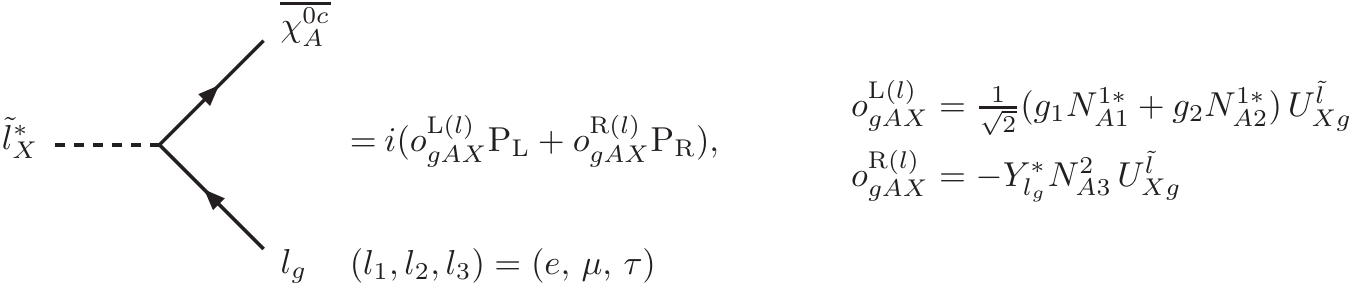}
\vspace{1.5cm}\\
\includegraphics[scale=1.]{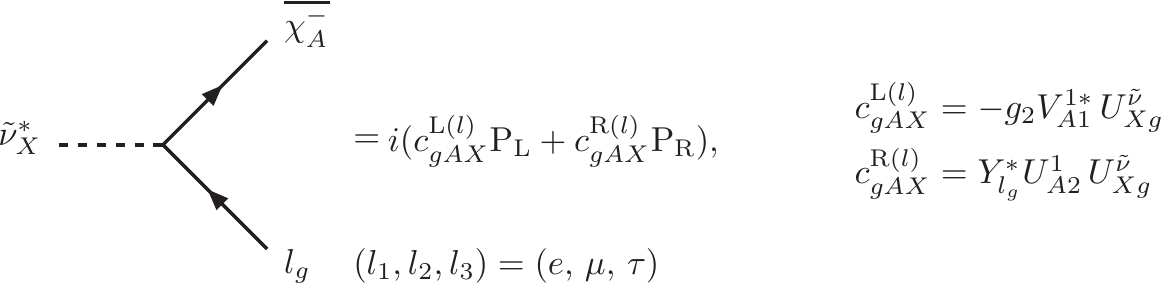}
\vspace{1.5cm}\\
\includegraphics[scale=1.]{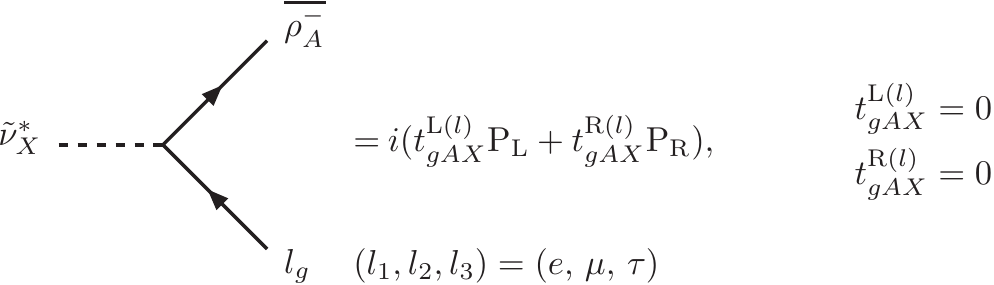}

\subsection*{Up--quarks--squarks--neutralinos/charginos}
\vspace{2cm}
\includegraphics[scale=1.]{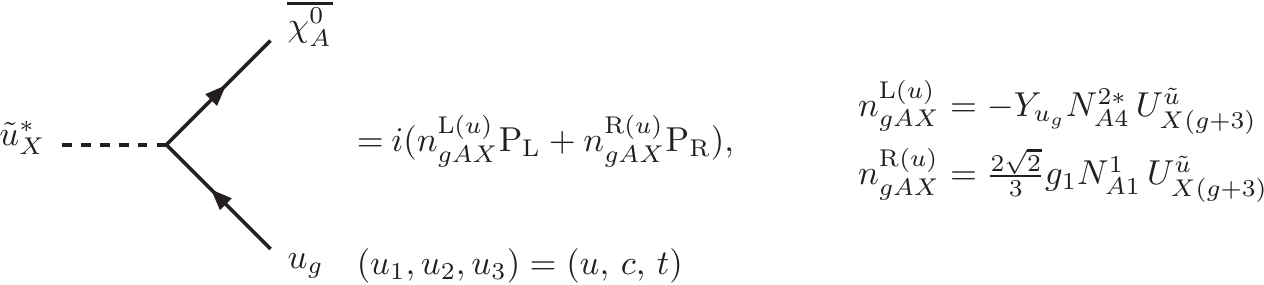}
\vspace{1.5cm}\\
\includegraphics[scale=1.]{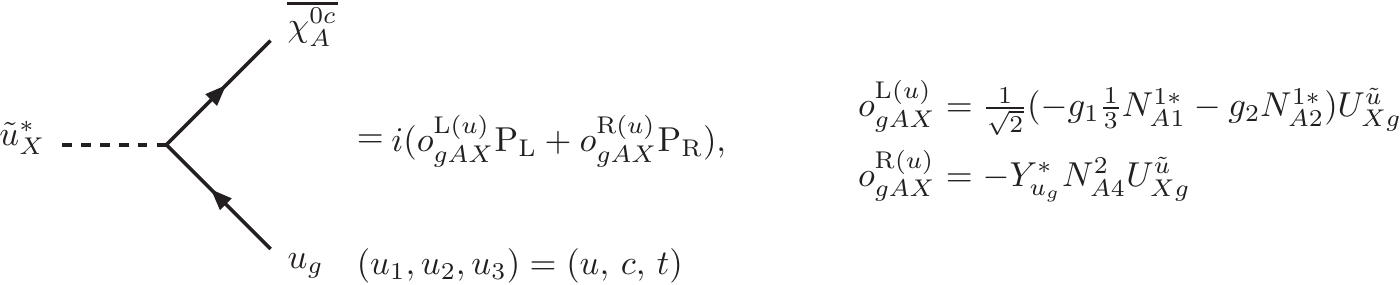}
\vspace{1.5cm}\\
\includegraphics[scale=1.]{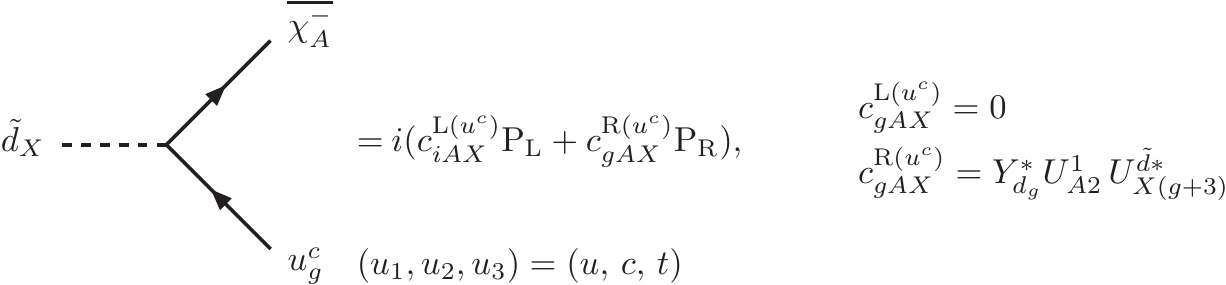}
\vspace{1.5cm}\\
\includegraphics[scale=1.]{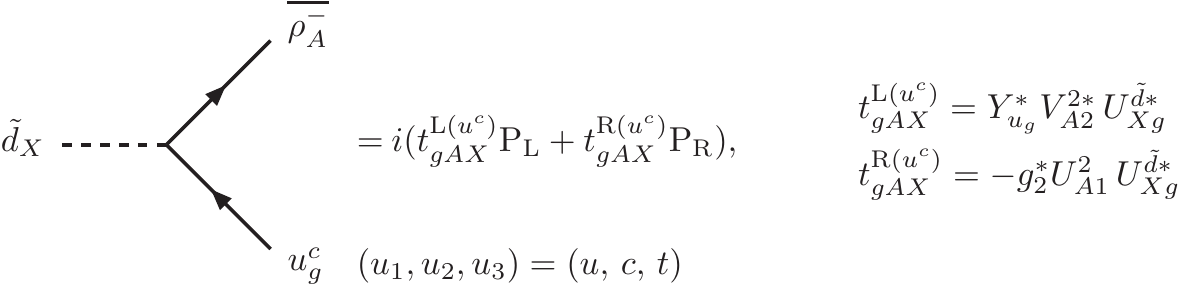}

\subsection*{Down--quarks--squarks--neutralinos/charginos}
\vspace{2cm}
\includegraphics[scale=1.]{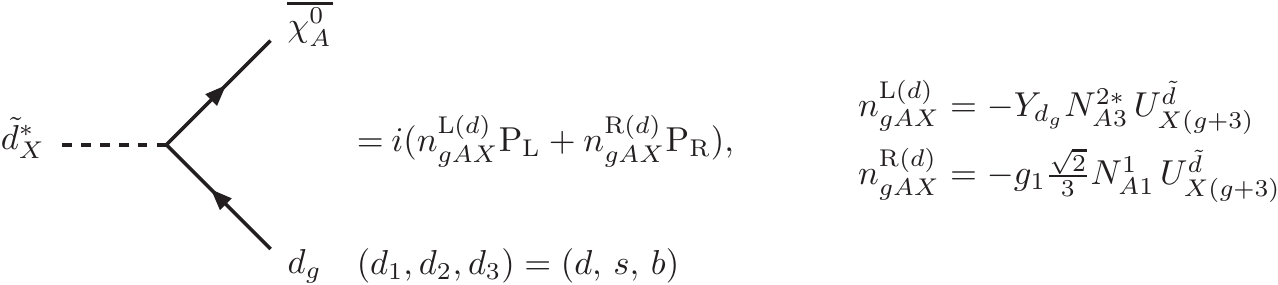}
\vspace{1.5cm}\\
\includegraphics[scale=1.]{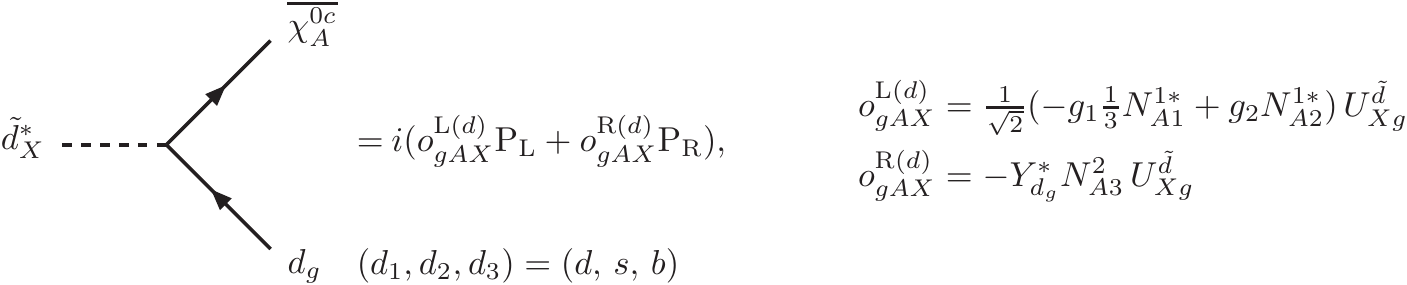}
\vspace{1.5cm}\\
\includegraphics[scale=1.]{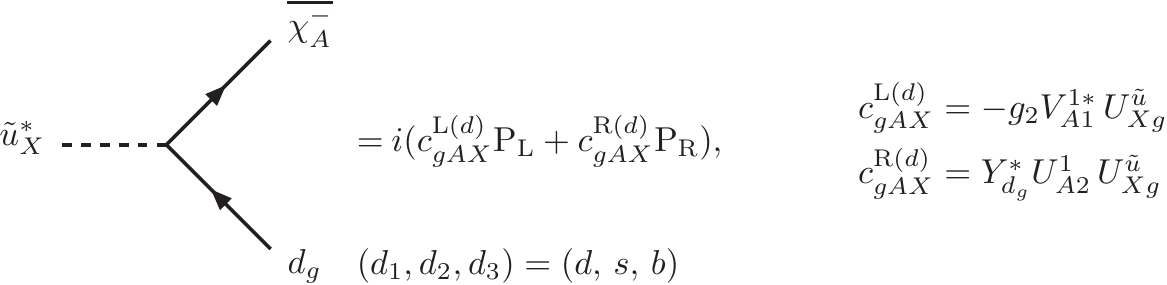}
\vspace{1.5cm}\\
\includegraphics[scale=1.]{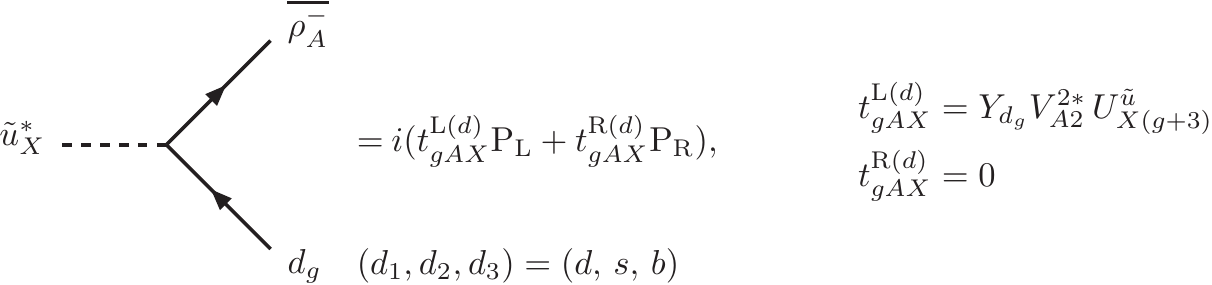}

\subsection*{$Z$--Neutralinos/charginos}
\vspace{2cm}
\includegraphics[scale=1.]{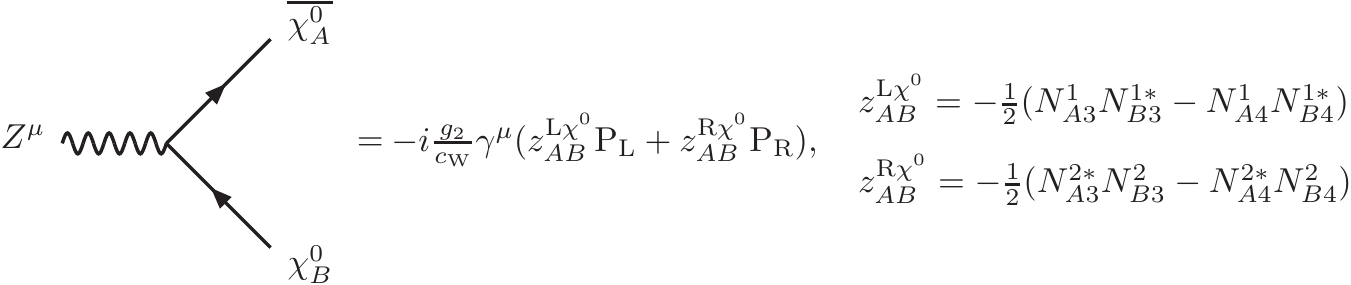}
\vspace{1.5cm}\\
\includegraphics[scale=1.]{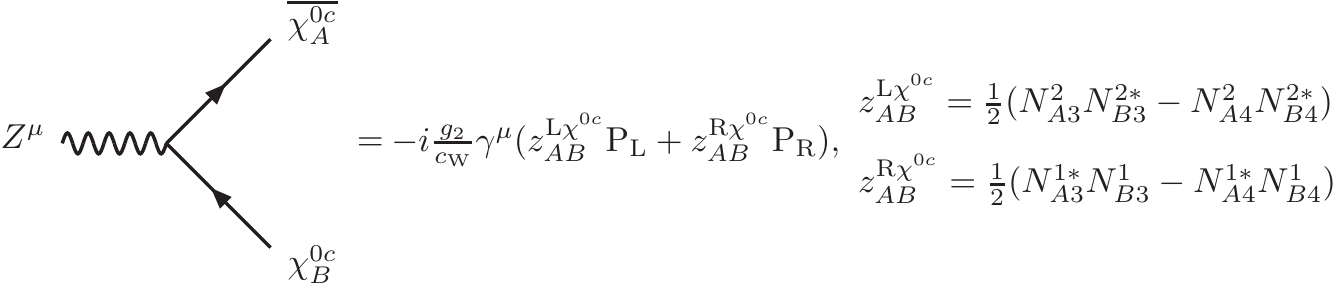}
\vspace{1.5cm}\\
\includegraphics[scale=1.]{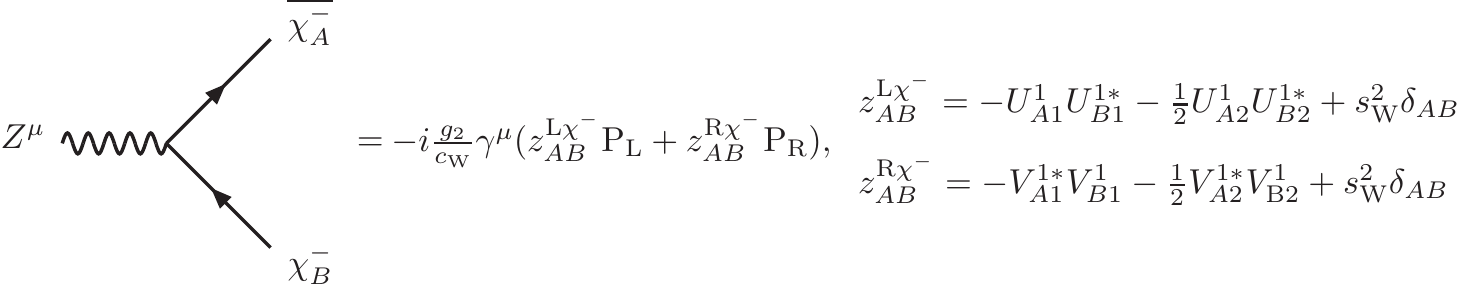}

\subsection*{$Z$--fermions/sfermions}
\vspace{2cm}
\includegraphics[scale=1.]{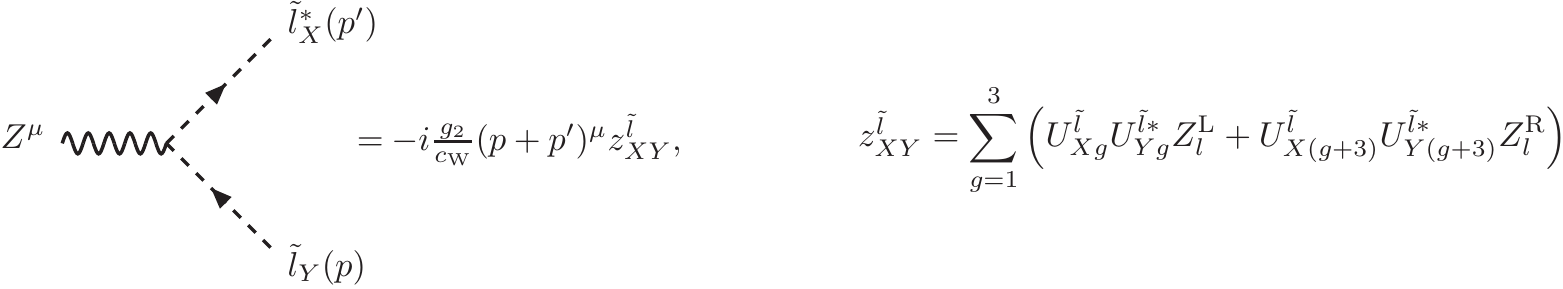}
\vspace{1.5cm}\\
\includegraphics[scale=1.]{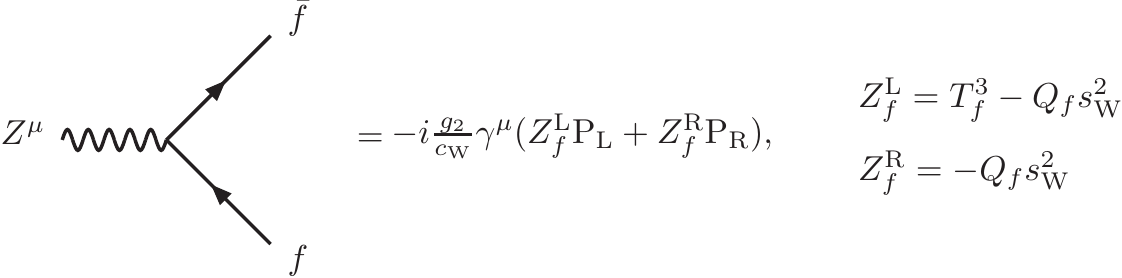}

\bibliographystyle{JHEP}
\bibliography{biblio}

\end{document}